\documentclass{aa} 
\usepackage{comment}
\usepackage{natbib}
\usepackage{multirow}
\usepackage{booktabs}
\usepackage{graphicx}
\graphicspath{{Fig/}}
\usepackage{menukeys}
\usepackage{array, multirow, bigdelim, makecell, booktabs} 

\DeclareUnicodeCharacter{2212}{-}
\usepackage{txfonts}
 
\newcommand{\cht}{CH$_{2}$ }
\newcommand{\chtp}{CH$_{3}^{+}$ }
\newcommand{\trans}{$4_{04} - 3_{13}$ }
\newcommand{\kms}{km~s$^{-1}$}

\makeatletter
\renewcommand*\aa@pageof{, page \thepage{} of \pageref*{LastPage}}
\makeatother
\usepackage{hyperref}
\hypersetup{colorlinks=true, allcolors=blue}
\begin{document}

   \title{Hunting for the elusive methylene radical}

    \author{A. M. Jacob
          \inst{1}\thanks{Member of the International Max Planck Research School (IMPRS) for Astronomy and Astrophysics at the Universities of Bonn and Cologne}
          \and
          K. M. Menten\inst{1}
          \and
          Y. Gong\inst{1}
          \and
          P. Bergman\inst{2}
          \and 
          M. Tiwari\inst{1,3}
          \and
          S. Br\"{u}nken\inst{4}
          \and
          A.O.H. Olofsson\inst{2}
          }

   \institute{Max-Planck-Institut f\"{u}r Radioastronomie, Auf dem H\"{u}gel 69, 53121 Bonn, Germany
   \and
      Department of Space, Earth and Environment, Chalmers University of Technology, Onsala Space Observatory, 43992 Onsala, Sweden
      \and 
      University of Maryland, Department of Astronomy, College Park, MD 20742-2421, USA
      \and 
      Radboud University, Institute for Molecules and Materials, FELIX Laboratory, Toernooiveld 7, 6525 ED Nijmegen, The Netherlands\\        \email{ajacob@mpifr-bonn.mpg.de}}

   \date{Received November 13, 2020; accepted December 21, 2020}
   \titlerunning{Hunting the elusive Methylene radical}
   \authorrunning{A. Jacob et al.}
 
  \abstract
  {The ${N_{K_{\text{a}}K_{\text{c}}}=\,}$\trans transitions of ortho-CH$_{2}$ between 68 and 71~GHz were first detected toward the Orion-KL and W51~Main star-forming regions. Given their high upper level energies (225~K) above the ground state, they were naturally thought to arise in dense, hot molecular cores near newly formed stars. However, this has not been confirmed by further observations of these lines and their origin has remained unclear. 
  Generally, there is a scarcity  of observational data for CH$_2$ and, while it is an important compound in the astrochemical context, its actual occurrence in astronomical sources is poorly constrained.}
  {In this work, we aim to investigate the nature of the elusive \cht emission, address its association with hot cores, and examine alternative possibilities for
  its origin. Owing to its importance in carbon chemistry, we also extend the search for \cht lines by observing an assortment of regions, guided by the hypothesis that the observed \cht emission is likely to arise from the hot gas environment of photodissociation regions (PDRs).}
  {We carried out our observations first using the Kitt Peak 12~m telescope to verify the original detection of \cht toward different positions in the central region of the Orion Molecular Cloud 1. These were followed-up by deep integrations using the higher angular resolution of the Onsala 20~m telescope. We also searched for the ${N_{K_{\text{a}}K_{\text{c}}}=2_{12}-3_{03}}$ transitions of para-\cht between 440--445~GHz toward the Orion giant molecular cloud complex using the APEX 12~m telescope. We also obtained auxiliary data for carbon recombination lines with the Effelsberg 100~m telescope and employing archival far infrared data.}
  {The present study, along with other recent observations of the Orion region reported here, rule out the possibility of an association with gas that is both hot and dense. We find that the distribution of the \cht emission closely follows that of the [C{\small II}] 158~$\mu$m emission, while \cht is undetected toward the hot core itself. The observations suggest, rather, that its extended emission arises from hot but dilute layers of PDRs and not from the denser parts of such regions as in the case of the Orion Bar. This hypothesis was corroborated by comparisons of the observed \cht line profiles with those of carbon radio recombination lines (CRRLs), which are well-known PDR tracers. In addition, we report the detection of the 70~GHz fine- and hyperfine structure components of ortho-\cht toward the W51~E, W51~M, W51~N, W49~N, W43, W75~N, DR21, and S140 star-forming regions, and three of the ${N_{K_{\text{a}}K_{\text{c}}}=}$~\trans fine- and hyperfine structure transitions between 68-71~GHz toward W3~IRS5. While we have no information on the spatial distribution of CH$_2$  in these regions, aside from that in W51, we again see a correspondence between the profiles of \cht lines and those of CRRLs. We see a stronger \cht emission toward the extended H{\small II} region W51~M rather than toward the much more massive and denser W51 E and N regions, which strongly supports the origin of \cht in extended dilute gas. We also report the non-detection of the $2_{12}-3_{03}$ transitions of para-\cht toward Orion. Furthermore, using a non-LTE radiative transfer analysis, we can constrain the gas temperatures and H$_{2}$ density to $(163 \pm 26)$~K and ($3.4\pm0.3$)$\times10^{3}~$cm$^{-3}$, respectively, for the 68--71~GHz ortho-\cht transitions toward W3~IRS5, for which we have a data set of the highest quality. This analysis confirms our hypothesis that CH$_{2}$ originates in warm and dilute PDR layers. Our analysis suggests that for the excitation conditions under the physical conditions that prevail in such an environment, these lines are masering, with weak level inversion. The resulting amplification of the lines' spontaneous emission greatly aids in their detection.  }
   {}
   

   \maketitle
%
\section{Introduction} \label{sec:intro}

The methylene radical, CH$_{2}$, is of considerable astrophysical interest given that it is both produced and destroyed at an early stage in the sequence of ion-molecule reactions that govern interstellar chemistry \citep[see][and references therein]{Godard2014}. Theoretical models of the diffuse interstellar clouds in the line of sight (LOS) toward $\zeta~$Per by \citet{black1978models} and \citet{van1986comprehensive} as well as equilibrium models of dense clouds in Orion by \citet{prasad1980model} have predicted \cht to possess abundances similar to, if not greater than, that of CH. Both species are primarily formed by the dissociative recombination of the methyl ion \chtp with an electron and both are  destroyed in dense clouds via reactions with atomic oxygen to form HCO and HCO$^{+}$, which are chemical species that act as building blocks for more complex interstellar molecules. \citet{Vejby1997} computed the complete branching ratios of the different possible reaction products of the dissociative recombination reaction of CH$_{3}^+$ and found that \cht was the major reaction product with a branching ratio of 40\%. Additionally, \cht has been speculated to play an important role in the photo-dissociation sequence of methane (CH$_{4}$) in cometary ice mantles \citep{vanDishoeck1996}. Despite its importance and large predicted abundances, \cht is yet to be detected in comets and only a handful of detections have been made in the interstellar medium (ISM), unlike the ubiquitous CH. 

\begin{figure}
    \centering
    \includegraphics[width=0.48\textwidth]{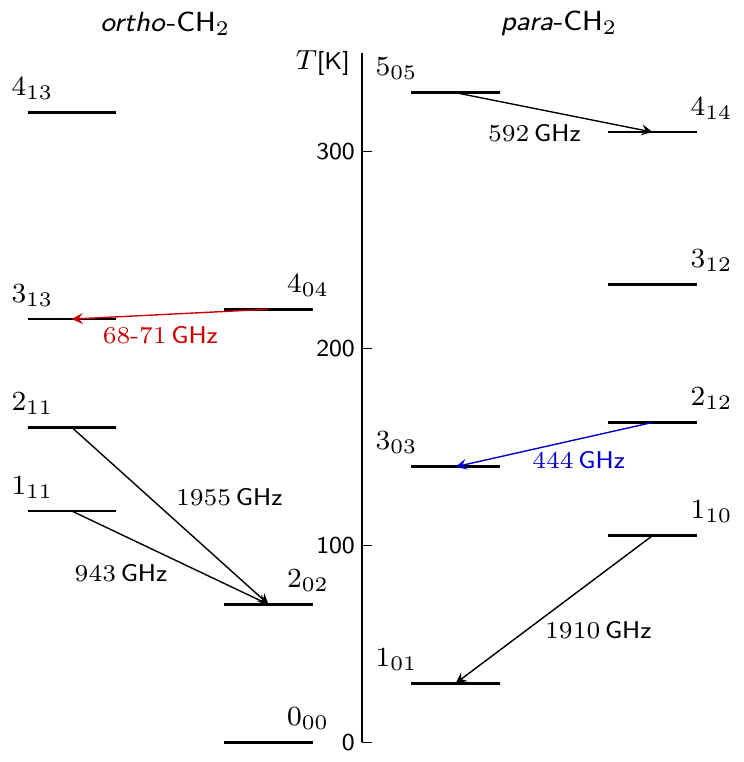}
    \caption{Ground state rotational energy level diagram of ortho-, and para-CH$_{2}$ up to an energy of 350~K, adapted from \citet{brunken2004high}. Transitions measured in the laboratory are labelled with their respective transitions indicated by arrows. Highlighted in red and blue are the transitions that are discussed in this paper. The corresponding fine- and hyperfine structure transitions are not depicted here as they would not be visible on this scale.}
    \label{fig:energy_level}
\end{figure}

Even laboratory measurements of this simple radical have proven to be difficult owing to its reactive nature and to its lightness and peculiar b-type selection-rules \citep{Michael2003, brunken2004high}. The latter two characteristics result in widely spaced energy levels particularly between the energetically lowest rotational states of CH$_{2}$. With only four rotational transitions below 1000~GHz (as shown in Fig.~\ref{fig:energy_level}), the rotational transitions of \cht are difficult to target in the laboratory, and even more so with astronomical observations, as they are either inaccessible from the ground or lie close to the edges of atmospheric windows. 

\cht was first unambiguously detected in the ISM by \citet{hollis1995confirmation} who identified the intrinsically strongest fine- and hyperfine-structure (HFS) transitions of the ${N_{K_{\text{a}}K_{\text{c}}}=}$~\trans multiplet of ortho-CH$_2$. The frequencies of the detected lines lie in the $68$–$71\,$GHz spectral range (${\approx\!4.3}$~mm), not far from the high frequency edge of the absorption band from atmospheric O$_2$, which effectively demarcates the radio- from the millimetre wavelength range. Their detections became possible after accurate frequencies had been measured in the laboratory by \citet{lovas1983laboratory}. These transitions arise from levels with energies of ${\sim 225}$~K above the ground state and were observed in emission toward the dense molecular `hot cores' associated with the Kleinmann-Low Nebula in Orion Molecular Cloud 1 (Orion-KL) and W51 Main. 

Following this work, several other astronomical searches for \cht have been published, but \citet{lyu2001search} reported the tentative detection of several of the electronic bands of \cht near 1410 and 1416~\AA{} toward HD~154368 and $\zeta$ Oph in absorption. However, recently \citet{Welty2020} were only able to publish an upper limit for \cht absorption near 1397~\AA{} from the relatively dense molecular core of the translucent interstellar cloud in the LOS to HD 62542. Their upper limits on the \cht\ column density are six and three times lower, respectively, than the nominal values given by \citet{lyu2001search} for HD~154368 and $\zeta$ Oph.

A breakthrough came with the detection of far-infrared (FIR) absorption lines from low-lying energy levels of both of CH$_2$'s spin isomeric forms, the ortho- and para-CH$_2$, which were discovered toward the intense dust continuum emission of the SgrB2~(M) and W49~N high-mass star-forming regions (SFRs) by \citet{polehampton2005far} from the data taken with the Long Wavelength Spectrometer \citep[LWS,][]{clegg1996iso} aboard the Infrared Space Observatory (ISO); see also \citet{polehampton2007lws}. Absorption was not only found at the systemic LSR velocities of the SFRs, but also at the velocities of diffuse and translucent interstellar clouds intervening along the lines of sight. These detections were enabled by accurate frequency measurements provided by the Cologne molecular spectroscopy laboratory and astrophysics group \citep{brunken2004high}. 
These supra-terahertz absorption lines not only represent the first detection of \cht in low excitation states, but they also yield reliable total \cht column densities, ${N(\text{CH}_{2}})$, of $(7.5\pm1.1)\times10^{14}~\text{cm}^{-2}$, which are consistent with chemical model predictions for diffuse clouds.

In this paper, we report the detection of the ${N_{K_{\text{a}}K_{\text{c}}}= }$\trans transitions of \cht toward nine SFRs -- seven of which \cht is detected for the first time and attempt to address questions regarding the origin of its emission. In addition, we also report the non-detection of the para-CH$_{2}$, ${N_{K_{\text{a}}K_{\text{c}}}= 3_{03}-2_{12}}$ transitions between 440--445~GHz toward the Orion molecular cloud, which were observed with the Atacama Pathfinder Experiment (APEX) 12~m sub-millimetre telescope.

\section{\texorpdfstring{CH$_{2}$}{CH2} spectroscopy } \label{sec:ch2_transition}

For \cht ($^3B_1$), which has a spin angular momentum, $S$, of unity, a state with rigid body angular momentum quantum number, $N$, splits into three levels with the total angular momentum quantum numbers, $J$, which have values of $N+1$, $N$, or $N-1$. Each of these rotational levels ($N_{K_{\text{a}}K_{\text{c}}}$ displayed in Fig.~\ref{fig:energy_level}) splits into three fine-structure levels due to spin-spin and spin-rotation interactions. Furthermore, \cht contains two identical protons, each with a nuclear spin, $I_{\text{H}} = 1/2$, and, consequently, has two nuclear-spin isomers: ortho ($I_{\text{tot}}=1$) and para ($I_{\text{tot}}=0$). With a total hydrogenic nuclear spin $I_{\text{tot}}$ of unity, each of the three fine-structure levels of ortho-CH$_{2}$ further splits into three hyperfine structure (HFS) levels with total angular quantum number, $F$, of $J-1$, $J$, and $J+1$ for $J\neq0$, while para-CH$_{2}$ states with $I_{\text{tot}} = 0$ do not have any HFS splitting.

\section{Ortho-\texorpdfstring{CH$_2$}{CH2}  emission at 4.3~mm -- A summary of prior results}\label{sec:ocht_summaryobs}
The \trans rotational transitions of ortho-\cht lie in a rarely observed spectral window, close to the 53--66~GHz atmospheric O$_{2}$ band and comprise three main fine-structure components $J\!=\!5\rightarrow\!4, 4\!\rightarrow\!3$, and $3\!\rightarrow\!2$ at 68.37, 70.68, and 69.01~GHz, respectively (which show additional HFS splitting). The corresponding frequencies and spectroscopic parameters of all the studied HFS transitions are summarised in Table~\ref{tab:freq}. In addition to the  HFS lines listed, with regard to the ortho-CH$_2$ rotational transitions, for every $J\rightarrow J-1$ fine structure transition, there is an additional pair of HFS lines with $\Delta F = 0$. The Einstein A-coefficients for these lines are smaller than those of the lines listed in Table~\ref{tab:freq} with $\Delta F = \Delta J = 1$ by factors between ${\approx 5}$ and 30\footnote{The CDMS contains a full listing of the CH$_2$ transitions under: \url{https://cdms.astro.uni-koeln.de/cgi-bin/cdmssearch?file=c014501.cat}}. While these lines were covered in our band pass, none were detected; for root mean square (rms) noise levels, see Appendix~\ref{appendix:ch2_nondetections}. Detected thus far in emission toward hot cores, the fine-structure lines of o-\cht and their corresponding HFS lines tabulated in Table~\ref{tab:freq} are naively expected to be weakly populated given the physical conditions that prevail in these regions. Intrigued by the fact that these transitions are observed in emission, \citet{Dagdigian2018} carried out simple non-LTE radiative transfer calculations using the one-dimensional escape probability code, RADEX \citep{van2007computer}. The negative excitation temperatures that they obtained suggest that the $4_{04}$ energy level is selectively enhanced by population inversion and that all three fine-structure transitions show weak maser emission. However, questions regarding the source of their emission or the association of these lines with the hot cores could not be definitively answered. 
\begin{table*}
    \centering
     \caption{Spectroscopic parameters of the rotational transitions of CH$_{2}$ studied in this work.}
    \begin{tabular}{ccrcrr}
    \hline \hline
         \multicolumn{2}{c}{Transition} & \multicolumn{1}{c}{Frequency} & \multicolumn{1}{c}{$A_\text{E}$} & \multicolumn{1}{c}{$S_\text{ij}$} & \multicolumn{1}{c}{$E_\text{u}$}\\
         $J^{\prime} - J^{\prime \prime}$ & $F^{\prime} - F^{\prime \prime}$ & \multicolumn{1}{c}{[MHz]} & \multicolumn{1}{c}{1$\times10^{-6}$[s$^{-1}$]} & & \multicolumn{1}{c}{[K]}\\
         \hline
         \multicolumn{5}{c}{$N_{K_{\text{a}}K_{\text{c}}}  = 2_{12}-3_{03}$ (para-CH$_{2}$)} \\
         \hline
         3-4 & -- & 444825.666 & 59.8
         & 0.928 & 155.97 \\
         2-3 & -- & 439960.991 & 64.3
         & 1.353 & 156.27\\
         1-2 & -- & 444913.930 & 70.1
         & 0.631 & 155.85\\
         \hline 
         \multicolumn{5}{c}{$N_{K_{\text{a}}K_{\text{c}}}~$= \trans (ortho-CH$_{2}$)} \\
         \hline 
         5-4 & 6-5 & 68371.278 & 0.216 &  2.329 \\
             & 5-4 & \tablefootmark{**}68375.875 & 0.208 &  1.892 & 224.22 \\
             & 4-3 & 68380.873 & 0.206 &  1.533 \\ 
            
             \hline
         4-3 
             & 3-2 & 70678.633 & 0.214 &  1.077 \\
            
             & 4-3 & \tablefootmark{**}70679.543 & 0.210 &  1.413 & 224.76 \\
             & 5-4 & 70680.720 & 0.223 &  1.842 \\ \hline 
         3-2 
             
             & 2-1 & 69007.179 & 0.171 &  0.691  \\
             & 3-2 & \tablefootmark{**}69014.202 & 0.181 &  1.023 & 224.15\\
             & 4-3 & 69019.187 & 0.204 &  1.480 \\ \hline
         
    \end{tabular}
   \tablefoot{The spectroscopic data - frequencies, Einstein $A$-coefficients ($A_\text{E}$) and line strengths ($S_\text{ij}$) were taken from \citet{Ozeki1995} and the Cologne Database for Molecular Spectroscopy ~\citep{muller2005cologne}.\tablefoottext{**}{Indicates the central HFS transition, which was used to set the velocity scale in the analysis.} }
    \label{tab:freq}
\end{table*}

The Orion Kleinmann-Low (KL) nebula, one of the two regions in which  \citet{hollis1995confirmation} first detected ortho-\cht (hereafter o-CH$_{2}$), is the densest part of Orion Molecular Cloud 1 (OMC-1). In itself, it is a highly complex region, which encompasses: (1) the eponymous `hot core', that is, a compact and very dense, hot \textgreater120~K) region that is surrounded by more extended dense molecular gas; (2) Orion South (here, Orion S), another hot core that, in contrast to the more famous one in KL \citep{Zapata2011}, harbours an embedded young stellar object (YSO) that was first identified as the NH$_3$ emission peak S6 \citep{Batrla1983} and by SiO emission, indicating outflow activity \citep{Ziurys1987}, which conclusively proves the presence of an embedded YSO, probably of intermediate mass. In a complex arrangement, Orion S lies in front of an extension of the H{\small II} region and the photodissociation region (PDR) associated with the Orion Nebula (M42) \citep{Mangum1993}; (3) A high density part of this PDR presents the prominent Orion Bar \citep{Walmsley2000}. To make things complicated, the o-\cht spectra toward Orion-KL obtained by \citet{hollis1995confirmation} using the NRAO\footnote{The National Radio Astronomy Observatory is operated by Associated Universities, Inc., under contract by the National Science Foundation.} Kitt Peak (KP) 12~m radio telescope in Arizona, with a full width at half maximum (FWHM) beam width of 86$^{\prime\prime}$, covers several of these components, leaving the origin of the observed \cht emission unclear. Given the high energies above the ground state of these lines' (upper-level energies of ${\approx\! 224}$~K), it was a natural conclusion to attribute the hot core to the source of their emission in Orion, whose molecular material was known to be characterised by temperatures \textgreater120~K or even higher values \citep[see][and references therein.]{Hermsen1985}.  

In a series of attempts to elucidate the location and size of the \cht emission in Orion, we compared the KP 12~m telescope results with data taken with two larger telescopes, namely, the Institut de Radioastronomie Millim\'etrique (IRAM) 30~m and the Green Bank telescope (GBT) operated by the NRAO. We scale the emission expected to be observed with the different telescopes by the inverse of the beam filling factor, that is, $({\theta_{\text{S}}^2+\theta_{\text{B}}^2})/\theta_{\text{S}}^2$.
For the Orion-KL hot core, we estimate a FWHM source size, $\theta_S$, of $12^{\prime\prime}$, from the Very Large Array (VLA) imaging of the NH$_3$ $J,K = (3,3)$ line \citep{Pauls1983}, whose energy levels are 127~K above the ground. For the FWHM beam diameters, $\theta_{\text{B}}$, of the KP 12~m, and the IRAM 30~m telescopes and the 100~m GBT, we assume $86^{\prime\prime}$, $35^{\prime\prime}$ and $12^{\prime\prime}$, respectively at ${\sim\!70}$~GHz. Compared to the KP 12~m telescope, this should result in 5.5 and 26 times higher main-beam brightness temperatures, $T_{\text{mb}}$, for the IRAM 30~m telescope and the GBT, respectively.

In 2004, we carried out observations of the 70.68~GHz fine-structure component of \cht using the IRAM 30~m telescope toward the Orion-KL region. Remarkably, we did not detect the o-\cht $J\!=\!4\rightarrow\!3$ line toward Orion-KL down to an rms noise level of 25~mK in the $T_\text{mb}$ scale, smoothed to a common velocity resolution of 0.66~km~s$^{-1}$. 
If the o-\cht emission was indeed shown to arise from the famous hot core, then we would expect the line to have been detected with a ${T_\text{mb}\sim\!140}$~mK with the IRAM 30~m telescope.

Moreover, the \cht transitions also remained undetected in the 67--93.6~GHz spectral line survey presented by \citet{frayer2015gbt} toward Orion-KL using the GBT, at an rms noise level of ${\sim\!61~}$mK at 68~GHz and $36~$mK at 70~GHz. The 4~mm receiver on the GBT (FWHM ${\sim\!12^{\prime\prime}}$) was pointed directly at the position of the Orion-KL hot core. Scaling here also the KP 12~m telescope data by the beam filling factor for a compact source with a FWHM of $12^{\prime\prime}$ in both the KP 12~m telescope and GBT, we expect the $T_{\text{mb}}$ measured by the GBT to be ${\sim\!0.65}$~K for the $J\!=\!5\rightarrow\!4$ transition of o-\cht at 68~GHz, that is, $26$ times higher than that measured using the KP 12~m telescope. The non-detections of these \cht transitions using the IRAM 30~m telescope and the GBT
suggest that o-\cht is simply not present
within the dense Orion-KL hot core itself and that, instead, its emission arises from a more diffuse (and presumably also warm) surrounding medium. This is not surprising as chemical models \citep{black1978models, prasad1980model, Lee1996} have previously predicted high abundances of o-\cht in both diffuse regions as well as in intermediate density gas layers at the edges of dense clouds. This motivated our search for hot o-\cht in Orion among regions outside of the hot core. 

The remainder of this paper is structured as follows. In Sect.~\ref{sec:observations}, we describe all our observations of CH$_2$, as  well as those of carbon radio recombination lines (CRRLs). We present the data in Sect.~\ref{sec:results}. Here, we start out by quite extensively describing our attempts to clarify the origin of the 68--71 GHz $N_{K_{\text{a}}K_{\text{c}}}=\,$\trans o-\cht emission lines in several diverse environments spread over ${\sim\!0.2}$~pc in OMC-1 and the PDR associated with the Orion Nebula.

Next, we briefly summarise our negative  results for the $N_{K_{\text{a}}K_{\text{c}}}=2_{12}-3_{03}$ sub-millimetre wavelength transitions of para-\cht (hereafter p-CH$_{2}$) that have frequencies between 440 and 445~GHz. 
After this, we present the results of our observations of the 68--71 GHz o-\cht line toward other sources,  which also contain PDRs, and compare them with the CRRL data.   
This is followed-up by our analysis of the \cht data in Sect.~\ref{sec:discussion}, where we describe our chemistry modelling and radiative transfer analysis. Finally, in Sect.~\ref{sec:conclusions} we discuss the main conclusions drawn from our results and provide a brief summary of the observational state of the art and future prospects of studies of interstellar CH$_2$ in Sect.~\ref{sec:outlook}.

\section{Observations}\label{sec:observations}
Motivated by the above considerations, we searched for o-\cht in hot media with intermediate densities near the envelopes of hot cores, that is, PDRs, which form at the interface between H{\small II} regions and dense molecular clouds and are characterised by the density of the cloud and the strength of the far-ultraviolet (FUV) ($6~\text{eV} < h\nu < 13.6~\text{eV}$) radiation field. In order to investigate the possible scenario in which the observed o-CH$_{2}$ emission arises from PDRs, we first re-examined the fine-structure transitions of o-\cht at 68 and 70~GHz and their associated HFS transitions using the KP 12~m telescope toward a number of positions in the OMC-1. Then we followed-up with observations carried out with the new receiver covering the 4~mm band of the Onsala 20~m telescope \citep{belitsky2015new}. 

The majority of the additional sources toward which we carried out our search (and not located in Orion) represent well-known H{\small II} regions and giant molecular cloud (GMC) cores showing active star formation. We targeted positions corresponding to peaks identified by a previous mapping of other continuum emission and molecular lines. The observed positions coincide with bona fide H{\small II} regions and the substantial distances ($\ge\!2$~kpc) imply that their associated PDRs have little or no angular offset from the fully ionised gas and are also covered in our $52^{\prime\prime}$ FWHM single-dish beam. Our study is further supplemented with observations of p-CH$_{2}$ at 444~GHz made using the APEX 12~m telescope and of CRRLs made using the Effelsberg 100~m telescope. In the following sections, we describe the technical aspects of these observations.

\subsection{KP 12~m telescope}
 The observations of Orion-KL were carried out between October and November 2005 (project id: 5029) and between January and April 2006 using the 3~mm receiver of the KP 12~m telescope. The dual channel single sideband system was operated in dual polarisation mode, using the Millimetre Auto Correlator (MAC) as backend, which provided a spectral resolution of 391~kHz over 8192 channels, spanning an effective bandwidth of 300~MHz. Thus, two frequency setups were required, centred on 68.370 and 70.679~GHz to cover each of the individual $5-4$ and $4-3$ HFS triplets, respectively. The observed positions lie in the vicinity of the hot core region and include the hot core itself (Orion-KL in the following), as well as the Orion S\footnote{We pointed at a position that is at an offset of $(-5, +12)^{\prime\prime}$ relative to that used by \citet{Tahani2016} for their line survey, which is negligible given the KP 12~m's $86^{\prime\prime}$ FWHM beam.} a position between Orion-KL and S (here referred to as KL/S), the so-called radical-ion peak (RIP) located $4^{\prime}$ north of Orion-KL \citep{ungerechts1997chemical}, and the Orion Bar, a neighbouring PDR. This Orion Bar position was neither covered by the beam of the KP 12~m, nor that of the IRAM 30~m telescopes in previous observations. Moreover, the 68-71~GHz \cht transitions were also not covered by the line survey carried out by \citet{Cuadrado2016} between 80--360~GHz using the IRAM 30~m telescope toward the same position. In our observations, the ${\sim 86^{\prime \prime}}$ beam of the KP 12~m telescope was centred on the source coordinates tabulated in Table~\ref{tab:source_coordinates_line_params}. The spectra were calibrated using a main beam efficiency of 0.64. The resultant spectra were then subsequently processed using the GILDAS-CLASS software\footnote{Software package developed by IRAM, see \url{https://www.iram.fr/IRAMFR/GILDAS/} for more information regarding GILDAS packages.} \citep{Pety2005} and up to a second order polynomial baseline was removed.

\begin{table*}
     
     \caption{Summary of o-\cht observations using the KP 12~m, and Onsala 20~m telescopes toward the Orion pointing positions and their corresponding line parameters.}
     \begin{tabular}{llllcrrrrll}
          \hline \hline
         Source & \multicolumn{1}{c}{$\alpha_{{\rm J2000}}$}  & \multicolumn{1}{c}{$\delta_{{\rm J2000}}$} & \multicolumn{1}{c}{Line} & \multicolumn{1}{c}{$\upsilon_\text{sys}$}& \multicolumn{1}{c}{$\upsilon_\text{LSR}$} & \multicolumn{1}{c}{$\Delta \upsilon$} &  \multicolumn{1}{c}{$T_\text{mb}$\tablefootmark{a}}  & rms\tablefootmark{b} &  S/N\tablefootmark{c} & (S/N)$_{\text{tot}}$\tablefootmark{d}\\
                   &   &   & \multicolumn{1}{c}{[MHz]} & \multicolumn{1}{c}{[km~s$^{-1}$]}  & 
                   \multicolumn{1}{c}{[km~s$^{-1}$]}  &\multicolumn{1}{c}{[km~s$^{-1}$]} & \multicolumn{1}{c}{[mK]}  & [mK]  & \\ \hline
                 \multicolumn{9}{c}{KP 12~m Telescope} \\
                 \hline
        Orion-KL & 05:35:14.10 & -05:22:26.54 & 70678.633 & 3.0--6.0 & 13.4(0.6) & 5.2(0.6) & 12.7(3.2) & 6.5 & 2.0 & 4.3\\
        & & & 70679.543 & & 9.5(0.6) & 5.2(0.6) & 16.6(4.2) & 6.5 & 2.5 \\
        & & & 70680.720 & & 4.5(0.6) & 5.2(0.6) & 21.7(5.4) & 6.5 & 3.3 \\
        Orion S  & 05:35:13.10 & -05:23:56.00 & 70678.633 & 6.5--8 & 12.4(0.4) & 5.2(0.5) & 20.1(3.2) & 7.0 & 2.8 & 6.1\\
        & & & 70679.543 & & 8.5(0.4) & 5.2(0.5) & 26.4(4.2) & 7.0 & 3.8 \\
        & & & 70680.720 & & 3.5(0.4) & 5.2(0.5) & 34.4(5.4) & 7.0 & 4.9 \\
        & & & 68371.278 & & -13.8(0.4) & 4.9(1.0) & 35.2(9.6) &  7.1 & 5.0 \\
        & & & 68375.875\tablefootmark{*} & & 7.6(0.4) & 5.2(0.9) & 38.3(8.6) & 7.0 & 5.4 \\
        & & & 68380.873 & & 28.8(0.4) & 4.8(0.8) & 28.7(7.0) & 7.0 & 4.1 \\
        Orion-KL/S  & 05:35:07.60  & -05:23:11.00 & 70678.633 &6--12 & 16.4(0.7) & 3.5(0.5) & 20.1(4.6)  & 7.2 & 2.7 & 3.3\\  
        & & & 70679.543 & & 12.5(0.7) & 3.5(0.5) & 26.4(6.0) & 7.2 & 3.6 \\
        & & & 70680.720 & & 7.5(0.7)  & 3.5(0.5) & 34.4(6.8) & 7.2 & 4.7 \\
        Orion RIP & 05:35:15.80  & -05:19:00.50 & 70678.633 & 8--10& 12.8(0.9) & 3.1(0.7) & 19.9(3.6) & 6.2 & 3.2 & 4.0\\
        & & & 70679.543 & & 9.0(0.9) & 3.1(0.7) & 26.1(4.7) & 6.2 & 4.2 \\
        & & & 70680.720 & & 4.0(0.9) & 3.1(0.7) & 34.0(6.1) & 6.2 & 5.5 \\
        Orion Bar & 05:35:22.80  &  -05:25:01.00 & 70678.633 & 9--10&  15.4(0.3) & 2.6(0.6) & 21.1(6.0) & 11.5 & 1.8 & 2.8\\
        & & & 70679.543 & & 11.5(0.3) & 2.6(0.6)& 27.7(8.0) & 11.5 & 2.4 \\
        & & & 70680.720 & & 6.5(0.3) & 2.6(0.6) & 36.1(10.3) & 11.5 & 3.1 \\
        & & & 68371.278 & & -11.4(0.3) & 2.1(0.5) & 33.1(11.6) & 7.8 & 4.2 \\
        & & & 68375.875\tablefootmark{*} & & 10.8(0.3) & 2.2(0.5) & 39.4(12.1)&  7.8 & 5.0 \\
        & & & 68380.873 & & 30.8(0.3) & 2.8(0.5) & 39.5(10.7) & 7.8 & 5.0\\ \hline
   
            \multicolumn{9}{c}{OSO 20~m Telescope} \\ 
            \hline
        Orion-KL/S (1) & 05:35:16.96 & -05:22:02.7 & 70679.633 & 6--12 &  19.4(1.8) & 4.5(1.2) & 23.8(7.5) & 14.2 & 1.7 & 2.2\\
        & & & 70679.543 & & 15.5(1.8) & 4.5(1.2) & 31.3(9.7) & 14.2 & 2.2\\
        & & & 70680.720 & & 10.5(1.8) & 4.5(1.2) & 40.8(12.8) & 14.2 & 2.8\\
        Orion-KL/S (6) & 05:35:24.96 & -05:22:32.7 & 70679.633 & 6--12 &  16.3(1.3) & 4.7(0.9) & 23.9(5.1) & 13.8 & 1.7 & 3.1 \\
        & & & 70679.543 & & 12.5(1.3) & 4.7(0.9) & 31.4(6.6) & 13.8 & 2.3\\
        & & & 70680.720 & & 7.5(1.3) & 4.7(0.9) & 40.8(8.8) & 13.8 & 3.0\\
        Orion Bar (2) & 05:35:22.80 & -05:25:01.0 & 70679.633 & 9--10 & 15.4(1.0) & 4.7(1.1)& 28.8(14.6) &  25.3 & 1.1 & 2.3 \\
        & & & 70679.543 & & 11.5(1.0) & 4.7(1.1) &  37.9(19.1) & 25.3 & 1.5 \\
        & & & 70680.720 & & 6.5(1.0) & 4.7(1.1) & 49.3(24.7) & 25.3 & 1.9 \\
        Orion Bar (5) &  05:35:20.81 & -05:25:17.1 & 70679.633 & 9--10 & 15.4(0.8) & 5.0(1.0) & 28.8(8.4) & 15.0 &  1.9 & 3.8\\
        & & & 70679.543 & & 11.5(0.8) & 5.0(1.0) & 37.9(11.4) & 15.0 & 2.5 \\
        & & & 70680.720 & & 6.5(0.8) & 5.0(1.0) & 49.3(14.5) & 15.0 & 3.3\\
                       \hline
   \end{tabular}
      \tablefoot{ \tablefoottext{a}{Peak main-beam brightness temperature derived from the integrated intensity of the detected line features.} 
     \tablefoottext{b}{The rms noise level on the $T_{\text{mb}}$ scale, quoted for a spectral resolution of 0.85~km~s$^{-1}$ and 0.97~km~s$^{-1}$ for observations made using the KP 12~m, and Onsala 20~m telescopes, respectively.}
     \tablefoottext{c}{The signal-to-noise ratio (S/N) with respect to the peak line temperature.}\tablefoottext{d}{The S/N of the integrated intensity calculated from the listed rms noise levels and line widths for the 70~GHz CH$_{2}$ line.}\tablefoottext{*}{Represents the HFS component used to determine the LSR velocity axes of the spectra shown in Fig.~\ref{fig:Orion_KP_spec}. For each HFS line, the velocities of the other HFS components are listed as they appear in the spectra shown in this figure.}}
     \tablebib{For the radial velocities, see \citet{Gong2015} and references therein.}
   \end{table*}

\addtocounter{table}{-1}   
\begin{table*}
     
     \caption{Continued.: Summary of o-\cht observations using the Onsala 20~m telescope and derived line parameters toward the other sources presented in this study.}
     \begin{tabular}{llllcrrrrll}
          \hline \hline
         Source & \multicolumn{1}{c}{$\alpha_{{\rm J2000}}$}  & \multicolumn{1}{c}{$\delta_{{\rm J2000}}$} & \multicolumn{1}{c}{Line} & \multicolumn{1}{c}{$\upsilon_{\text{sys}}$}& \multicolumn{1}{c}{$\upsilon_\text{LSR}$} & \multicolumn{1}{c}{$\Delta \upsilon$} &  \multicolumn{1}{c}{$T_\text{mb}$\tablefootmark{a}}  & rms\tablefootmark{b} &  S/N\tablefootmark{c} & (S/N)$_{\text{tot}}$\tablefootmark{d}  \\
                   &   &   & \multicolumn{1}{c}{[MHz]} & 
                   \multicolumn{1}{c}{[km~s$^{-1}$]}  & \multicolumn{1}{c}{[km~s$^{-1}$]}  &  \multicolumn{1}{c}{[km~s$^{-1}$]} & \multicolumn{1}{c}{[mK]}  & [mK]  & \\ \hline
   
             W3~IRS5 & 02:25:40.5 & 62:05:52 & 70678.633 & -39.0[1] & -35.1(0.5) & 7.3(0.9) & 22.9(6.4) &  9.8 & 2.3 & 5.0\\
             & & & 70679.543 & & -39.0(0.5) & 7.3(0.9) & 30.0(8.4) & 9.8 & 3.1\\
             & & & 70680.720 & &-44.0(0.5) & 7.3(0.9) & 39.1(9.3) & 9.8 & 4.0\\
             &  &  & 68371.278 & &-60.4(0.3) &  5.4(0.8) & 21.6(4.0) & 4.4 & 4.9\\
             & & & 68375.875\tablefootmark{*} & & -40.1(0.3) & 6.2(0.6) & 25.0(3.3)& 4.5 & 5.6 \\
             & & & 68380.973 & & -19.0(0.3) & 5.5(0.5) & 27.5(3.4)& 4.5 & 6.2\\
             &  &  & 69007.179\tablefootmark{e} & & -59.3(0.4) & 8.7(0.9) & 8.2(2.7) & 3.0 & 2.8 \\
             & & & 69014.202\tablefootmark{*} & & -39.0(0.4) & 6.9(1.3)& 13.3(2.8)& 3.0 & 4.4\\
             & & & 69019.187 & &-7.5(0.4) & 9.7(2.2) & 9.9(2.9)& 3.0 & 3.3 \\
             W51~E & 19:23:44.0 & 14:30:30 & 70679.633 &  +57.8[2] & +60.4(0.7) & 9.0(0.5) & 19.9(0.6) & 11.0 & 1.8 & 4.4  \\
             & & & 70679.543 & & +56.5(0.7) & 9.0(0.5) & 26.1(0.7) & 11.0 & 2.4 \\
             & & & 70680.720 & & +51.5(0.7) & 9.0(0.5) & 34.0(0.9) & 11.0 & 3.1\\
             W51~M & 19:23:42.0 & 14:30:36  & 70679.633 & +60.8[2] &  +63.4(0.5) & 9.2(1.1)&  24.0(6.1) & 9.3 & 2.6  & 4.5 \\
             & & & 70679.543 & & +59.5(0.5) & 9.2(1.1) & 31.3(7.2) & 9.3 & 3.4\\
             & & & 70680.720 & & +54.5(0.5) & 9.2(1.1) & 40.8(9.5) & 9.3 & 4.4\\
             W51~N & 19:23:40.0 & 14:31:10& 70679.633  & +60.8[2]
             & +64.1(0.7) & 9.2(0.7) & 12.4(6.2) & 9.2 &  1.3 & 3.4\\
             & & & 70679.543 & & +60.2(0.7) & 9.2(0.7) & 16.3(7.0) & 9.2 & 1.8 \\
             & & & 70680.720 & & +55.2(0.7) & 9.2(0.7) & 21.2(9.0) & 9.2 & 2.3 \\
             W49~N & 19:10:13.6 & 09:06:15 & 70679.633 & +10.0[3]& +11.8(0.8) & 9.2(1.0) &  23.6(0.8) & 11.5 & 2.0  & 4.0 \\ 
             & & & 70679.543 & & +8.0(0.8) & 9.2(1.0) & 31.0(10.1) & 11.5 & 2.7\\
             & & & 70680.720 & & +3.0(0.8) & 9.2(1.0) & 40.4(13.1) & 11.5 & 3.5\\
             W43 & 18:47:36.9 & -01:55:30 & 70679.633 & +89.4[4] & +92.8(1.1) & 9.2(0.8) & 11.2(4.3) & 10.2 & 1.1 & 3.3 \\ 
             & & & 70679.543 & & +88.5(1.1) & 9.2(0.8) & 14.7(5.4) & 10.2 & 1.4 \\
             & & & 70680.720 & & +83.5(1.1) & 9.2(0.8) & 19.1(7.3) & 10.2 & 2.0 \\
             DR21 & 20:39:02.0 & 42:19:42 & 70679.633 & $-3.0[5]$ & +1.9(0.7) & 9.2(0.7) & 7.5(2.3) & 5.4 & 1.4  & 4.1  \\
             & & & 70679.543 & & -2.0(0.7) & 9.2(0.7) & 9.8(3.0) & 5.4 & 1.8  \\
             & & & 70680.720 & & -7.0(0.7) & 9.2(0.7) & 12.8(3.8) & 5.4 & 2.4  \\
             W75~N & 20:38:36.5 & 42:37:35 &70679.633 & +9.0[5] & +13.4(0.8) & 2.4(0.1) & 10.4(2.0) & 5.3 & 2.0 & 3.5  \\ 
             & & & 70679.543 & & +9.5(0.8) & 2.4(0.1) & 13.7(2.6) & 5.3 & 2.6 \\
             & & & 70680.720 & & +4.5(0.8) & 2.4(0.1) & 17.9(3.4) & 5.3 & 3.4 \\
             S140 & 22:19:11.5 & 63:17:47 & 70679.633 & $-8.5$[6] & -7.1(0.8) & 4.7(0.6) & 26.1(10.4) & 16.7 & 1.6 & 3.0\\ 
             & & & 70679.543 & & -11.0(0.8) & 4.7(0.6) & 34.3(13.7) & 16.7 & 2.1 \\
             & & & 70680.720 & & -16.0(0.8) & 4.7(0.6) & 44.6(17.8) & 16.7 & 2.7 \\
                        \hline

   \end{tabular} 
   \tablefoot{ \tablefoottext{a}{Peak temperature derived from the integrated intensity of the detected line features.} 
     \tablefoottext{b}{The rms noise level on $T_{\text{mb}}$ scale, quoted for a spectral resolution of 0.97~km~s$^{-1}$ for observations made using the Onsala 20~m telescope.} \tablefoottext{c}{The signal-to-noise ratio (S/N) with respect to the peak line temperature.} \tablefoottext{d}{The S/N of the integrated intensity calculated from the listed rms noise levels and line widths for the 70~GHz CH$_{2}$ line.}
     \tablefoottext{e}{The reported intensity for this component accounts for contributions from the blended NS $J,F=3/2,3/2 \rightarrow 1/2,1/2$ line at 69017.995~MHz. This line was modelled using the NS $J,F=3/2,5/2 \rightarrow 3/2,1/2$ HFS line at 69002.890~MHz, which is also covered in the same bandpass.}\tablefoottext{*}{Represents the HFS component used to determine the LSR velocity axes of the spectra shown in Fig.~\ref{fig:W3_6869_finestructure}. For each HFS line, the velocities of the other HFS components are listed as they appear in the spectra shown in this figure.} }
     \tablebib{For the radial velocities: [1]~\citet{Imai2000}; [2]~\citet{Parsons2012}; [3]~\citet{Jackson1994}; [4]~\citet{Bally2010}; [5]~\citet{dickel78}; [6]~\citet{Bally2002}.}
   \label{tab:source_coordinates_line_params}
\end{table*}
\subsection{Onsala 20~m telescope}
In April and May 2019, we observed the ${N_{K_{\text{a}}K_{\text{c}}}}$ = \trans transitions of o-\cht using the 4~mm receiver on the Onsala 20~m telescope \citep{Walker2016} (project id: O2018b-07) and followed-up with further observations in  January and February 2020. The new 4~mm receiver, equipped with a cooled dual-polarisation high electron mobility transistor (HEMT) amplifier, was tuned to a frequency of 69.52~GHz, such that all three fine-structure components (with a maximum separation of 2.3~GHz) and their respective HFS components could be observed simultaneously, while leaving enough baseline on either side of the 4~GHz IF bandpass. The FWHM beam width at this frequency (69~GHz) was measured to be 52$^{\prime\prime}$. The observations were carried out in dual beam switch mode with a beam throw of 10.$^{\prime}$5. Using the 4~GHz bandwidth provided by the OSA (Onsala Spectrometer A), fast Fourier Fourier Transform spectrometer (FFTS) backend, with a spectral resolution of 76.294~kHz, we achieved a velocity resolution of 0.3~km~s$^{-1}$. Timely pointing and focus accuracy checks were performed by observing nearby stellar SiO ($\varv =1, J =2 \rightarrow 1$) masers. 

The intensity calibration was done every 12 minutes using the standard chopper-wheel method, whereby the second order chopper-wheel correction term was calculated and applied following \citet{ulich1976absolute}. We express the intensity scale of our spectra in units of $T_{\text{mb}}$, by assuming a main beam efficiency of 0.55 (on average), while the velocity scale is given with respect to the local standard of rest. Similarly to the KP 12~m data, the calibrated spectra obtained using the Onsala 20~m telescope were further analysed using the CLASS software. However, we see signatures of a regular standing-wave pattern with a frequency of ${\sim\!16}~$MHz, which are likely associated with the enclosing radome structure whose reflective properties worsen at the lower end of the 4$\,$mm band. This is particularly so for the observations toward Orion, whose transit at the Onsala 20~m telescope is much lower in comparison to that at the KP 12~m telescope. We corrected for contributions from the standing-wave features by using a standard standing-wave removal method based on  a Fast Fourier Transform analysis. The resulting spectra are then box-smoothed to velocity bins of ${\sim\! 1}$~km~s$^{-1}$ and polynomial baselines up to the third order were subtracted. 

\subsection{APEX 12~m telescope\label{sec:apexobservations}}
The fine-structure components of the $N_{K_{\text{a}}K_{\text{c}}}= 2_{12}-3_{03}$ transition of p-\cht were observed in 2013 August\footnote{Project id: M-091.F-0040-2013}, using the high frequency channel of the sideband separating (2SB), dual frequency band First Light APEX  Submillimetre Heterodyne Receiver, FLASH \citep[hereafter FLASH-460,][]{Heyminck2006} on the APEX 12~m sub-millimetre telescope\footnote{APEX is a collaboration between the Max-Planck-Institut fur Radioastronomie, the European Southern Observatory, and the Onsala Space Observatory.} \citep{Gusten2006}. The FWHM beam size at 443~GHz is 14$^{\prime\prime}$. The bandpass was selected such that we covered the 444~GHz transitions in the upper sideband alongside the $^{13}$CO $J\!=\!4\rightarrow\!3$ transition at 439.088~GHz, which made for an excellent reference for monitoring the calibration during the observations. We observed the same Orion-KL, Orion S, and Orion RIP positions as discussed above, along with two positions in the Orion Bar, corresponding to the CO and HCN abundance peaks \citep[see ][and references therein]{Nagy2015}. The FFTS, providing a 4~GHz bandwidth in each sideband, was used as backend for these observations to achieve a spectral resolution of 76.3~kHz. The data was subsequently reduced and processed using the CLASS software with a forward efficiency of 0.95 and a main beam efficiency of 0.60. Polynomial baselines up to a second order were removed and the subsequently obtained spectrum was box-smoothed to channel widths of 1~km~s${}^{-1}$. Each pointing position was integrated on, for a total time of $\sim\!23$~mins, except for the Orion S position, toward which we carried out deeper integration\footnote{Project id: M-091.F-0045-2013}, for a total time of 12.7~hours.

\subsection{Effelsberg-100~m telescope}
We performed CRRL measurements (project id: 08-19) toward those targets with successful detections of CH$_{2}$, in position-switching mode with the S20mm receiver of the 100-m telescope at Effelsberg, Germany\footnote{The 100-m telescope at Effelsberg is operated by the Max-Planck-Institut f{\"u}r Radioastronomie (MPIFR) on behalf of the Max-Planck Gesellschaft (MPG).}, on 2019 5--6 July and 22--23 August. The S20mm receiver is a double-beam and dual-polarisation receiver operating in the frequency range between 12--18 GHz. This range contains $\Delta n = 1$, `$\alpha$', radio recombination lines from H, He and C with principal quantum numbers, $n$, between 80 and 72; see Appendix \ref{appendix:summary_RRL}. Frequencies of RRLs from all three species can be calculated following the prescriptions of \citet{LilleyPalmer1968}, or simply retrieved from the Splatalogue database\footnote{\url{https://splatalogue.online//}}.

For our analysis, we only use data from the central beam of the S20 receiver. The FFTS \citep[e.g.][]{Klein2012} data serve as backend, each of which consists of 65536 channels. For the observations toward the different Orion positions and W3~IRS5, we used a total bandwidth of 300~MHz with a channel width of 4.6~kHz, corresponding to a velocity spacing of 0.09~km~s$^{-1}$ at 15~GHz. Our observations of the other targets utilised a total bandwidth of 5~GHz and a channel width of 38.1 kHz, corresponding to a velocity spacing of 0.76~km~s$^{-1}$ at 15~GHz. The focus was adjusted using observations of strong continuum sources at the beginning of each observing session. Pointing observations were carried out roughly every two~hours toward strong continuum sources nearby. The pointing accuracy was found be less than $5^{\prime\prime}$. NGC~7027 was used as the flux calibrator, and the flux calibration accuracy is estimated to be within $\sim$10\%. The observations took about 20~hours in total. The FWHM beam width is 48$^{\prime\prime}$ at 15~GHz. This angular resolution allows for a meaningful comparison with the \cht transitions observed using the Onsala telescope at ${\sim\!69}$~GHz (with a FWHM of $52^{\prime\prime}$). The main beam efficiency is about 0.65 and the typical system temperature is about 15~K. The data reduction was once again performed using the CLASS software. 

\section{Results}\label{sec:results}
\subsection{Line profiles }\label{subsec:line_profiles}
The lines of the HFS triplet of the ${J=4\rightarrow3}$ fine structure transition of CH$_{2}$ overlap with one another because of their close frequency spacing, $\Delta \nu$, of 0.88 and 1.18~MHz, corresponding to 3.7 and 5.0 \kms, respectively. This blending of the individual components with one another broadens the observed profile. While this greatly aids in the detection of this component, it does not reveal accurate line properties, such as the intrinsic line width. On the other hand, with a frequency separation of 9.6 and 12~MHz, respectively, the individual HFS components of the ${J=5\rightarrow4}$ and ${J=3\rightarrow2}$ transitions of CH$_{2}$ near 68~GHz and 69~GHz, respectively, are well resolved. However, the ${F=4\rightarrow3}$ HFS component of the latter at 69.019187~GHz is contaminated by emission from the low-lying ($E_\text{u} = 3.3~$K) NS ${J,F=3/2,3/2\rightarrow1/2,1/2}$ line at 69.017995~GHz. Therefore, the HFS-resolved o-\cht line profiles are modelled by simultaneously fitting individual Gaussian profiles to the observed HFS lines using the CLASS software. The relative contribution of the emission from the NS transition present in the strongest HFS component of the ${J=3\rightarrow2}$ line, is modelled by scaling its intensity with that of the ${F=5/2\rightarrow3/2}$ transition of NS at 69.002890~GHz covered in the same bandpass, by their relative line strengths under the assumption of local thermal equilibrium (LTE). 

For the blended ${J=4\rightarrow3}$ component of o-\cht, we modelled the observations by using the line widths derived from the ${J=5\rightarrow4}$ transitions near 68~GHz, which are unaffected by blending and contamination, and thereby we were able to reveal the true shape of the CH$_{2}$ emission profile. While the thus determined line width can be used to model the line profiles of the other fine structure lines, this line is unfortunately not detected toward a majority of our sources above a 3$\sigma$ noise level. The rms noise levels we obtained for the different sources are listed in Appendix~\ref{appendix:ch2_nondetections}. For those sources toward which the ${J=5\rightarrow4}$ fine structure line is not detected, the HFS of the ${J=4\rightarrow3}$ transition is decomposed\footnote{This analysis was carried out using Python packages numpy and scipy \citep{numpy}.} by minimising the mean square error between the modelled fit and the observations over several iterations covering a range of line widths typically between 2 and 10~km~s$^{-1}$ in steps of 0.23~km~s$^{-1}$. This `empirical' model simultaneously fits the three HFS lines using Gaussian profiles with: (1) positions determined by the velocity separation of the HFS lines relative to the 70.679543~GHz component; (2) one common line width; and (3) individual line intensities estimated by scaling the peak temperature by the line strength of each HFS line. Since the typical positions are well-known for a given source and the line intensities can be constrained from the observations, the line width is the only free parameter. Varying the line width in each iteration and comparing the mean square error between the observed line profile and that of the combined HFS fit, the scheme converges for the line width returning the least error. The errors in the line parameters are determined from the covariance matrix and depend on the rms noise of the system. There are additional systematic uncertainties in the line width, which are caused by the low signal-to-noise ratio (S/N) in these spectra. A summary of the derived line parameters is given in Table~\ref{tab:source_coordinates_line_params}.  

\subsection{o-\texorpdfstring{CH$_{2}$}{CH2} in Orion}\label{subsec:results_orion}

\begin{figure*}
\includegraphics[width=0.95\textwidth]{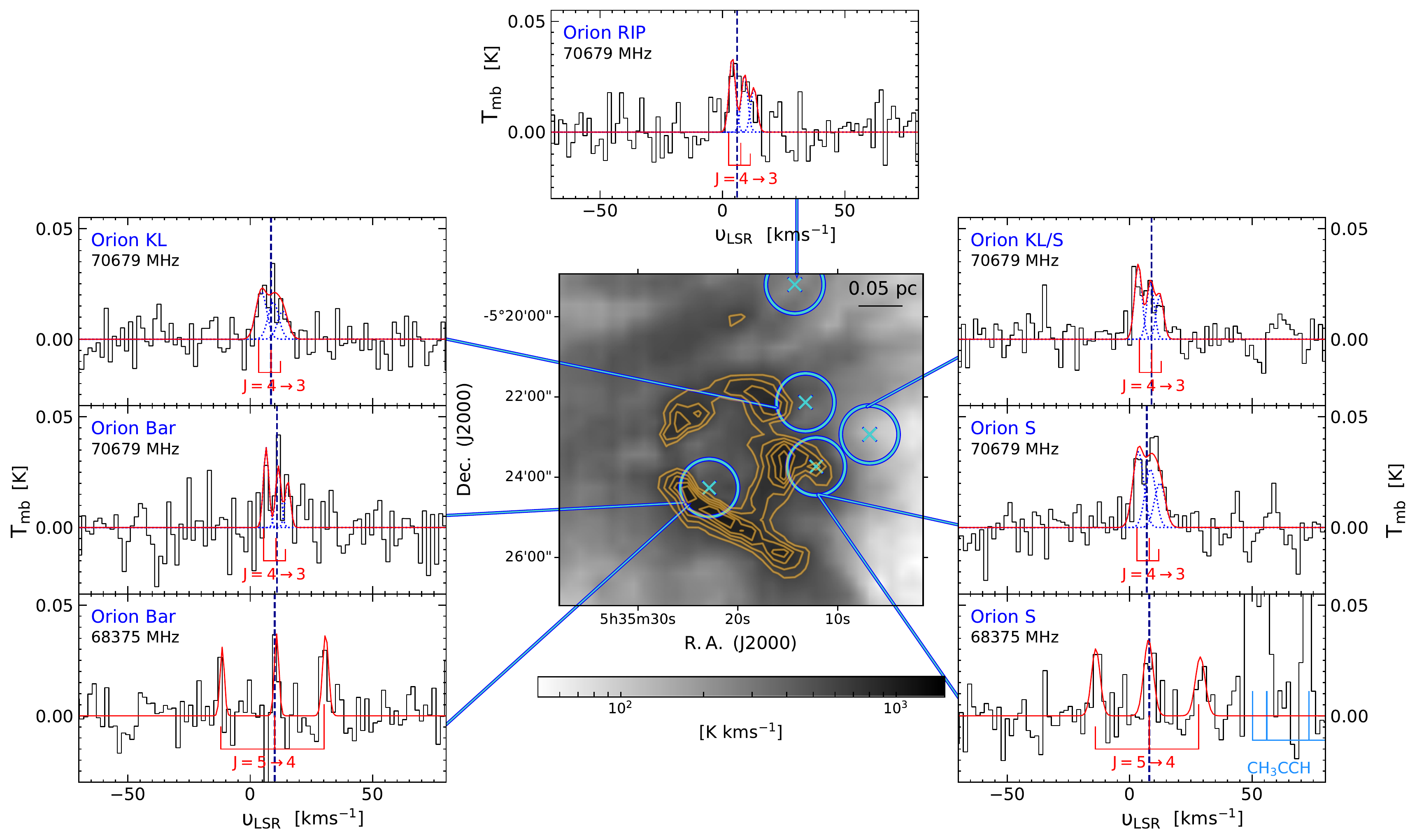}
\caption{ Integrated intensity map of the [C{\small II}] 158~$\mu$m emission overlaid with integrated intensity contours (orange) from 600 to 1000~K~km~s$^{-1}$ in steps of 100~K~km~s$^{-1}$ toward OMC-1, adapted from \citet{pabst2019disruption}. The positions observed using the KP 12~m telescope are marked by blue crosses, while the blue circles represent the KP beam. The corresponding calibrated and baseline subtracted o-\cht spectra are displayed alongside the map. The individual fits to the HFS components are displayed by dotted blue curves while solid red curves displays the combined fit. The positions and relative intensities of the HFS components are marked below the spectra. The dashed blue line indicates the systemic velocities of the sources which lie between +7 and +11~km~s$^{-1}$. In light blue we also mark the CH$_{3}$CCH lines covered in the 68375~MHz spectrum toward Orion S. The velocity scale is set by the $F^{\prime}- F^{\prime\prime} = J$--$J-1$ HFS line (see Table~\ref{tab:freq}).}
\label{fig:Orion_KP_spec}
\end{figure*}

The spectra observed toward the different Orion positions with the KP 12~m telescope are displayed in Fig.~\ref{fig:Orion_KP_spec}. We detect the ${J=4\rightarrow3}$ fine-structure component of the \trans transition of \cht near 70~GHz toward all the Orion pointing positions at systemic velocities between 8 and 10~km~s$^{-1}$. A second observational setup was used toward the Orion S and Bar positions, with the band centred on the ${J=5\rightarrow4}$ fine-structure components at 68~GHz. We were able to detect the well-separated HFS components of the 68~GHz lines toward both positions at a ${\gtrsim 5\sigma}$ level. In the spectrum of the 68~GHz component toward Orion S, we also detect part of the ${K=0~\text{to}~3}$ ladder of the ${J=4\rightarrow3}$ transition of methyl acetylene, that is, CH$_{3}$CCH near 68.3649~GHz. 

While the measured line intensity toward Orion-KL is consistent with that of the original detection by \citet{hollis1995confirmation}, we find the line intensities derived toward the other Orion positions to be of very comparable, if not greater, strengths than the strongest emission being seen toward the Orion S position. The successful detection and higher \cht intensities observed toward regions outside of the hot core once again brings into question the association of the \cht emission with the hot core. Moreover, we detect \cht emission toward an intermediate position between Orion-KL and S, which we name Orion KL/S and toward which we would not expect any \cht emission if it were confined to the hot core. The detection of this high-lying \cht transition toward all the Orion positions with comparable line strengths, indicates an extended emission component of \cht arising from a rather dilute, but hot ISM component and also suggests that this molecule is simply not found within the hot core. The compatibility of the intrinsic velocity of the observed \cht emission toward Orion~KL with that of the extended ridge component, $\upsilon_{\rm LSR}$ = 8--10~km~s$^{-1}$ \citep[see][and references therein]{Gong2015}, which represents the typical velocities of the ambient gas present in this region and further hints at the extended nature of the \cht emission. Since these observations were once again carried out using the KP 12~m telescope with a large beam size, the resulting signal may certainly contain contributions from neighbouring regions located within the beam. 

In an attempt to address the question of where the \cht emission arises from, we studied the emission characteristics of this molecule at several positions within the Orion complex, using the Onsala 20~m telescope, which has a beam size of 54$^{\prime\prime}$ at 69~GHz. These results are discussed in Sect.~\ref{sec:CH2_closerlook}. For convenience, the pointing positions chosen for our study are grouped into two, namely, the Orion KL/S and Orion Bar regions, respectively. The different positions are classified based on their proximity to either the hot core, Orion-KL, or to the PDR, the Orion Bar. 

\subsection{A closer look at the source of \texorpdfstring{CH$_{2}$}{CH2} emission}\label{sec:CH2_closerlook}
\subsubsection*{The Orion KL/S region}
\begin{figure*}
\sidecaption 
\includegraphics[width=0.95\textwidth]{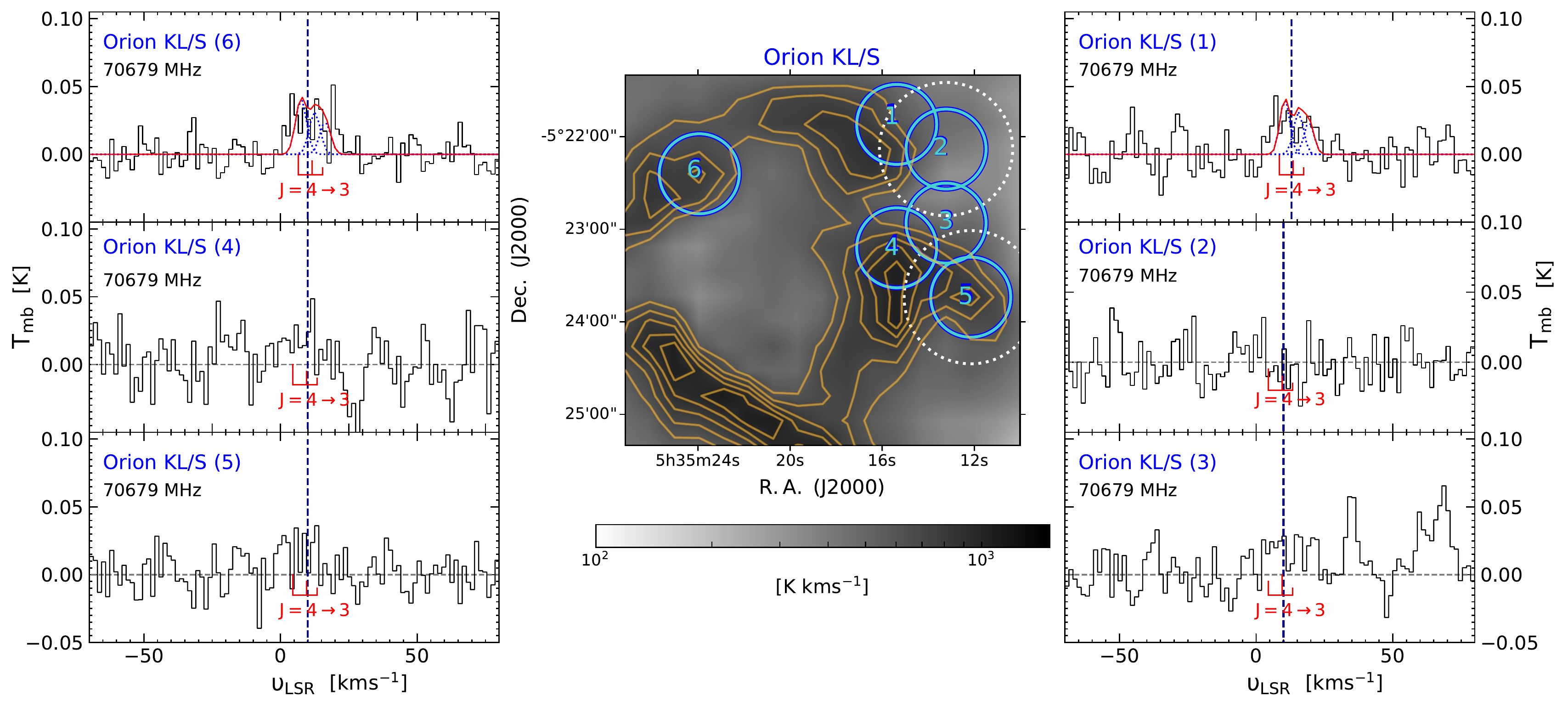} \quad
\includegraphics[width=0.95\textwidth]{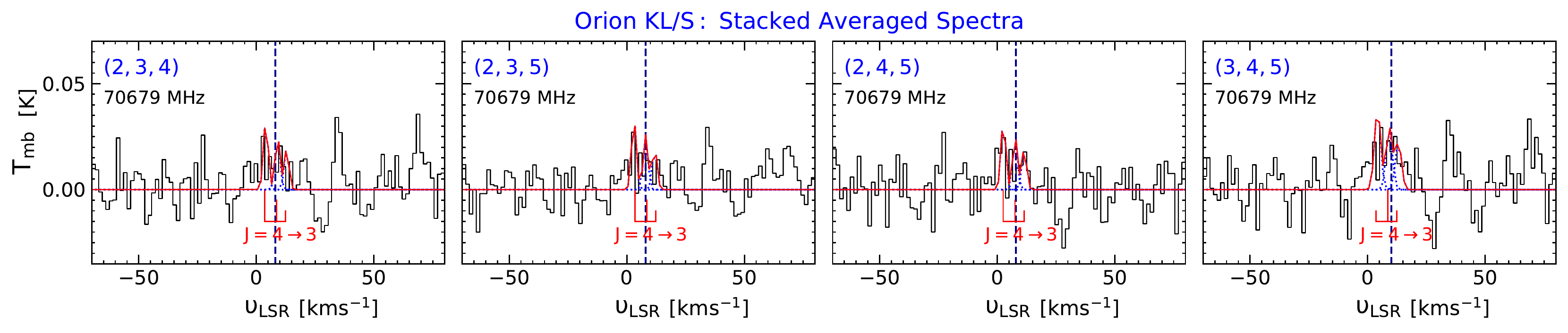}
\caption{Top: Same as Fig.~\ref{fig:Orion_KP_spec} but for o-\cht observations made using the Onsala 20~m telescope toward the Orion KL/S region. Blue circles represent the beam size of the Onsala 20~m telescope centred at the different pointing positions and the dotted white circle marks the KP beam at the Orion-KL and Orion S positions marked in Fig.~\ref{fig:Orion_KP_spec}.
Bottom: Stacked and averaged o-\cht spectra obtained by combining the Orion KL/S pointing positions (2), (3), (4), and (5), taken three at a time. The line fits are scaled using the line parameters obtained toward Orion KL/S positions (1) and (6).} 
\label{fig:OrionKLS}
\end{figure*}

\begin{figure}
    \centering
    \includegraphics[width=0.49\textwidth]{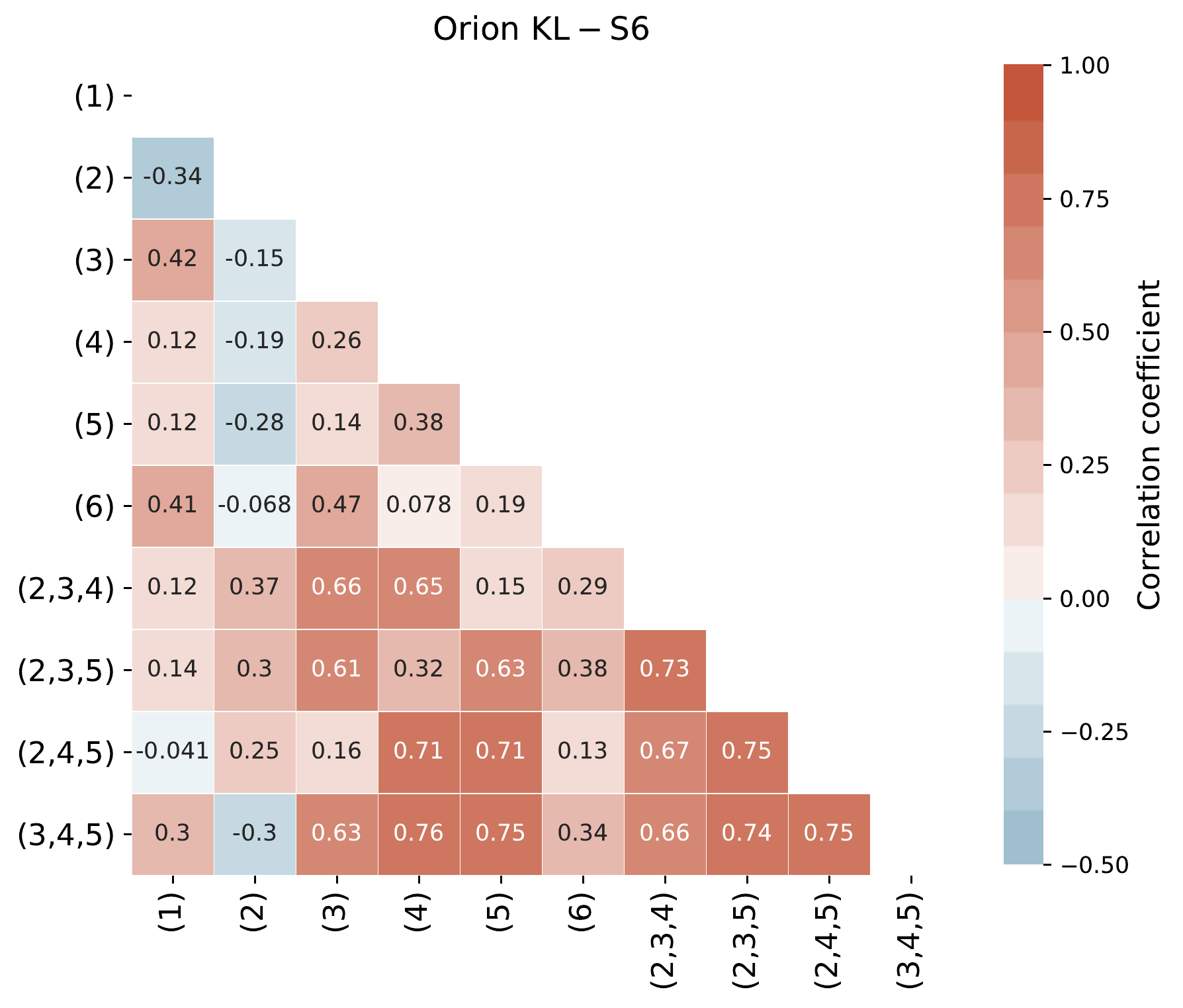}
    \caption{Pearson product-moment correlation coefficients between the integrated o-\cht intensities, observed toward the individual and combined pointing positions in the Orion KL/S region.}
    \label{fig:line_correlation_coeffs_OKL}
\end{figure}

In addition to the nominal Orion-KL and Orion S positions used in our KP observations (hereafter referred to as Orion KL/S positions (2) and (5)), we also observe four positions corresponding to the massive O7V star $\theta^{1}$~Ori~C, position (4), which is responsible for the ionisation of the Orion Nebula, [C{\small II}] peak positions (1) and (6) selected from \citet{pabst2019disruption} and a position between the Orion-KL and S regions, position (3). In Fig.~\ref{fig:OrionKLS}, we display the calibrated and baseline-subtracted spectra observed toward the different Orion KL/S positions. We detect \cht emission toward the two [C{\small II}] peak positions marked (1) and (6), at a ${\sim\!3\sigma}$ level with an average rms noise of 14~mK for the two spectra. However, we see no clear signal toward the other Orion KL/S positions, above an rms noise level of 22~mK on average. For the positions toward which we do not clearly detect \cht, we stack and then average the spectra (taken three at a time) in order to reduce the spectral noise and investigate our suspicion that the \cht emission is extended. The resulting spectra are displayed in the bottom panel of Fig.~\ref{fig:OrionKLS}. Upon scaling the fit results obtained toward positions (1) and (6), we see that all four independent combinations of the stacked and averaged profiles hint at the presence of a signal at velocities close to the systemic velocity of the source. The strongest \cht signals are observed in the profiles that result from the combination of positions (2, 3, 5) and (3, 4, 5). This suggests that the \cht emission is weakest at positions (2) and (4) which correspond to the Orion-KL hot core and the $\theta^{1}$~Ori~C positions, respectively. Cross-correlating the intensities integrated over a velocity range from $-10$ to +20~km~s$^{-1}$ (which is roughly the velocity interval over which we expect to observe \cht signatures) obtained toward the individual, and combined sets of pointing positions amongst each other, we find that positions (2) and (4) show negative correlations toward almost all other sources. The negative and/or weak correlations (tending to zero) shown by these components (see Fig.~\ref{fig:line_correlation_coeffs_OKL}) suggest that there is no association between them and reveal that \cht is likely not present within the hot core. The correlations we present were computed using the Pearson product moment correlation coefficient. This coefficient describes the strength of the linear relationship between each pair of spectra by using the standard deviation of each data set and the covariance between them. The underlying assumption made in this calculation is that the data follow a Gaussian distribution. For comparison, we also computed the correlation strengths based on non-parametric statistics by using the Spearman correlation coefficient. While the relative strengths produced by both methods are different, they both reproduce the same monotonic trends in their correlations. Since the aim of this analysis is to simply distinguish the nature of the correlation, namely positive or negative, in all subsequent analyses that make use of correlation coefficients, we use the standard Pearson coefficient. 

The detection of \cht toward the hot core initially discussed in \citet{hollis1995confirmation} and also in this work, can now be explained through its successful detection in the Orion KL/S (1) position. Offset from the nominal Orion KL position by (40.2\arcsec, 24.8\arcsec), a part of Orion KL/S (1) is covered by the KP 12~m beam. This resolves the observational discrepancies between the KP 12~m telescope, the GBT, and the IRAM 30~m telescope (detailed in Sect.~\ref{sec:ocht_summaryobs}) and indicates that the \cht emission is associated with gas layers similar to those traced by [C{\small II}]. Sensitivity is of course another important factor. As discussed earlier, given its extended emission, we would expect to observe \cht with a similar line strength using the different telescopes. The non-detections using the IRAM 30~m telescope and GBT therefore exclude the possibility that the \cht emission originates in compact regions like hot cores.

\subsubsection*{The Orion Bar region}

\begin{figure*}
\sidecaption 
\includegraphics[width=\textwidth]{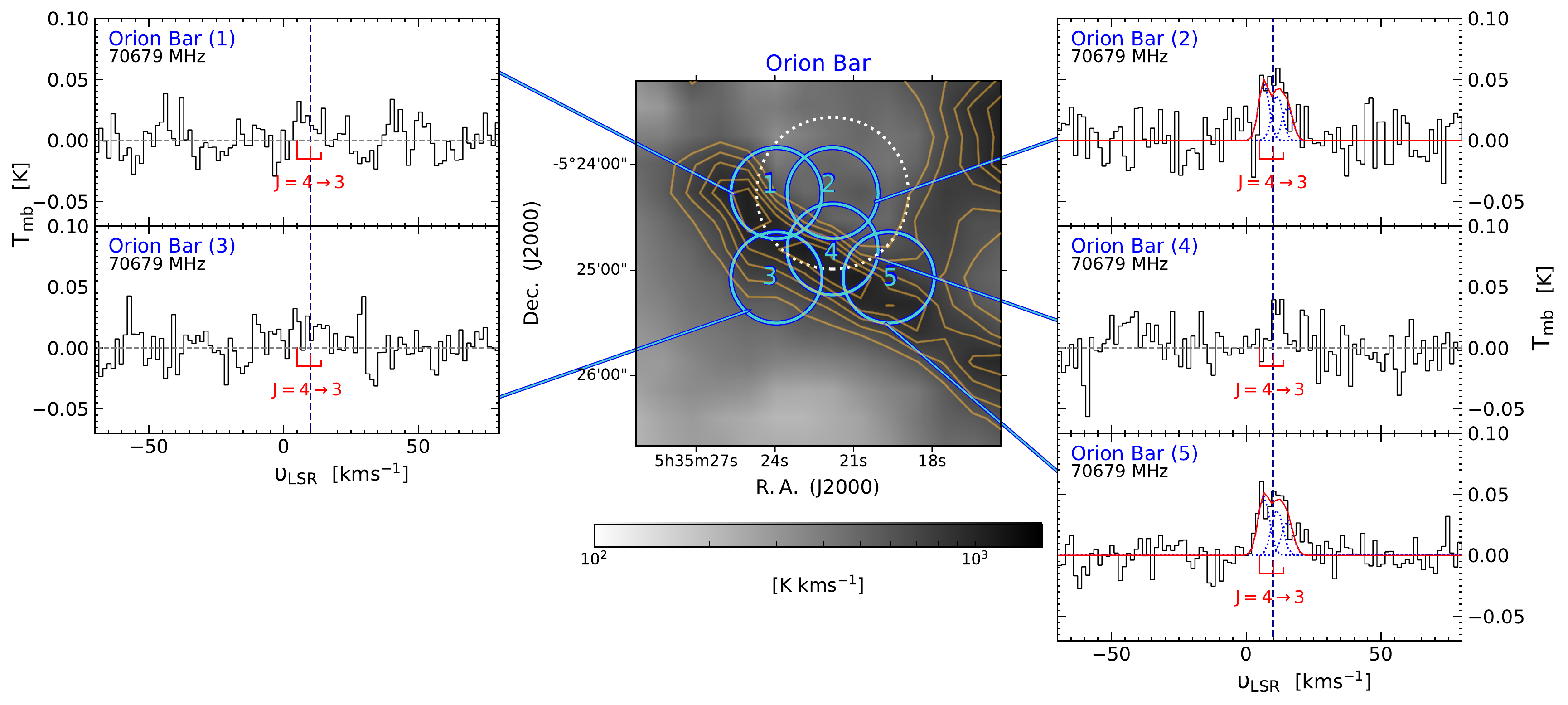} \quad 
\includegraphics[width=\textwidth]{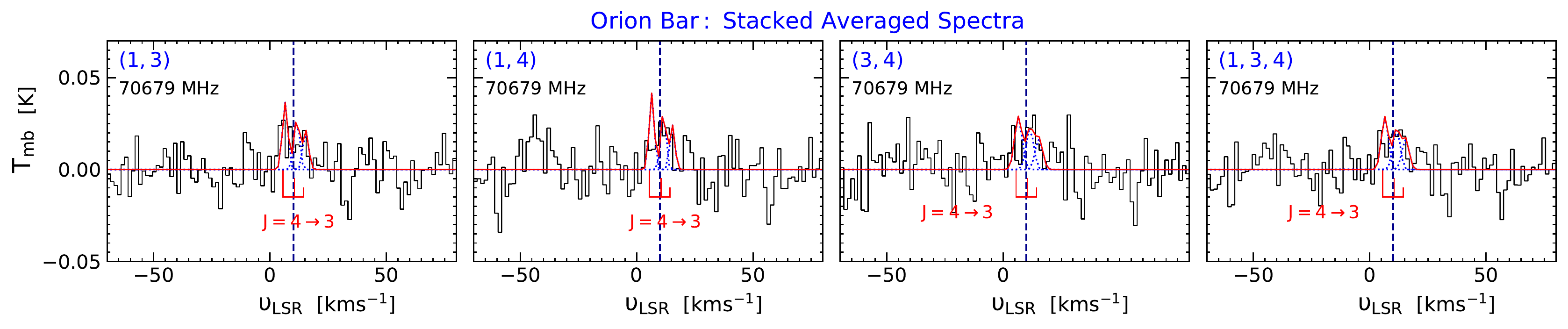}
\caption{Top: Same as Fig.~\ref{fig:OrionKLS} but for o-\cht observations made using the Onsala 20~m telescope toward the Orion Bar region. The blue circles represent the beam size of the Onsala 20~m telescope centred at the different pointing positions and the dotted white circle marks the KP beam at the Orion Bar position marked in Fig.~\ref{fig:Orion_KP_spec}. Bottom: Stacked and averaged o-\cht spectra obtained by combining the Orion KL/S pointing positions (1), (3), and (4), two at a time and all together. The line fits are scaled using the line parameters obtained toward Orion Bar positions (2) and (5).} 
\label{fig:OrionBar}
\end{figure*}

In addition to the Orion Bar position previously observed using the KP 12~m telescope, which corresponds to the peak of the CO emission [hereafter known as Orion Bar position (2)], we carried out observations toward four additional positions. They correspond to positions at or near the emission peaks of HCN, CF$^+$, C$_{2}$H, and CO$^+$ and are labelled (1), (3), (4), and (5), respectively, as displayed in Fig.~\ref{fig:OrionBar}. The positions were selected on the basis of previous studies by \citet{neufeld2006discovery}, \citet{Stoerzer1995}, \citet{Cuadrado2015}, and \citet{Nagy2015} and references therein. CH$_{2}$ emission is detected at a 2.3 and 4$\sigma$ level toward the Orion Bar nominal position (2) and the C$_{2}$H emission peak at position (5) at an rms noise level of 25 and 15~mK, respectively. We do not detect any appreciable signal from the other Orion Bar positions, even toward positions (1) and (3) at rms noise levels down to 17~mK. 

\begin{figure}
    \centering
    \includegraphics[width=0.49\textwidth]{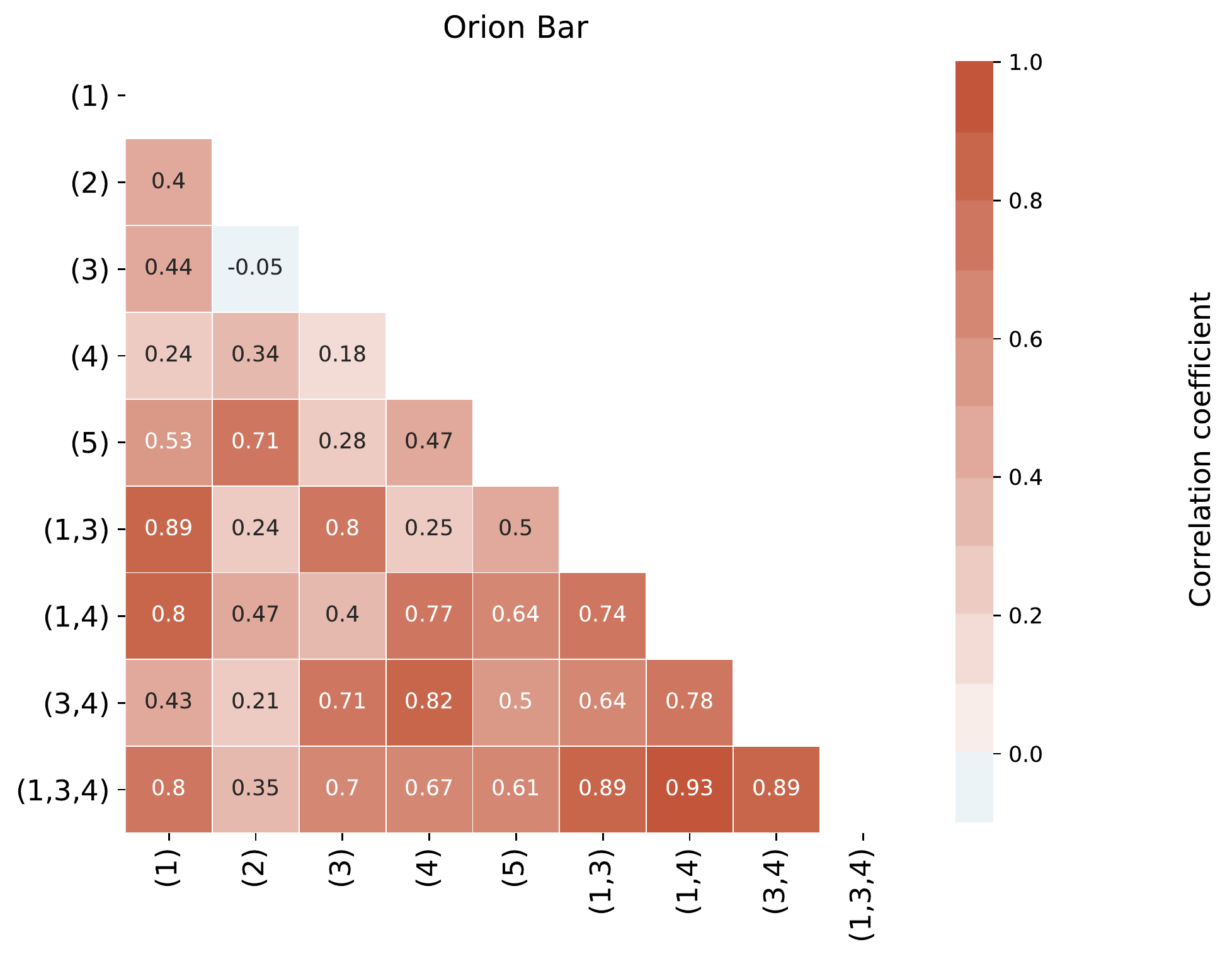}
    \caption{Pearson product-moment correlation coefficients between the integrated o-\cht intensities, observed toward the individual and combined pointing positions in the Orion Bar region.}
    \label{fig:line_correlation_coeffs_OBar}
\end{figure}

Similar to the pointing positions toward the Orion KL/S region for which we do not detect \cht emission, we stack and average the positions with non-detections in the Orion Bar region. The combined profiles of the independent pairs between positions (1), (3), and (4), as well as that obtained when considering all three together, are displayed in the bottom panel of Fig.~\ref{fig:OrionBar}. Scaling the fit parameters obtained from position (2), we find that there is a weak indication of \cht emission present in all the combinations. Carrying out a correlation analysis amongst the different Orion Bar positions we do not see any anti-correlations. However, from the correlation matrix presented in Fig.~\ref{fig:line_correlation_coeffs_OBar} we infer that position (3) has the weakest correlation coefficients particularly with that of positions (1), (4), and (5). 

The observed \cht emission in the Orion Bar suggests that the molecule's abundance decreases as we move away from the ionisation front (near the H{\small II} region) and toward the molecular clouds deeper within the PDR. Observations of HF emission by \citet{Kavak2019} across the Orion Bar centred near the CO$^+$ peak reveals a similar morphology. These authors were able to show that the bulk of the HF emission peaked in a region separating the H$_{2}$ and [C{\small II}] emission from the molecular emission in the denser clumps or close to the ionisation front. A direct comparison between the \cht and HF emissions is difficult because both sets of observations were carried out toward different positions in the Bar. However, since emission from both species has been observed near the Orion Bar CO$^+$ emission peak, we can compare the \cht line profile with that observed for HF at this position (this corresponds to position (2) in \citet{Kavak2019} and position (5) in this work). Both species show peak emission at velocities and line widths that are consistent within the error bars, between 10--10.5~km~s$^{-1}$ and 4.4--5~km~s$^{-1}$, respectively. From their line emission survey over the entire range of frequencies offered by the Herschel Heterodyne Instrument for the Far Infrared (HIFI) under the HEXOS HIFI key guaranteed time project, \citet{Nagy2015HEXOS} inferred typical line widths between 2 and 3~km~s$^{-1}$ toward the Orion Bar CO$^+$ peak, with the exception of a handful of species including HF. The broader line width of HF was reconciled as being due to its association with diffuse gas present in the inter-clump regions of Orion Bar. Therefore, it is conceivable that o-\cht is also likely to arise from a similar cloud population tracing dilute but hot ($T_\text{kin} = 120$~K) gas, unlike most of the other species studied by \citet{Nagy2015HEXOS} and others, including \citet{Cuadrado2015, Cuadrado2017}, which trace gas densities between 10$^{5}$ and 10$^{6}$~cm$^{-3}$. However, it is difficult to extend such a comparison toward the other sources in our study because HF is typically observed in absorption toward the envelope of these molecular clouds even showing several absorption components along the LOS. 

\subsection{\texorpdfstring{p-CH$_{2}$}{CH2} in Orion}\label{subsec:results_orion-pcht}
The $N_{K_{\text{a}}K_{\text{c}}}=2_{12}-3_{03}$ transitions of p-\cht between 440 and 445~GHz that have upper level energies of 156~K. Lying within a sub-millimetre window, they are accessible from high-mountain sites but have gone undetected thus far. Observations of these p-\cht lines could potentially aid our understanding of CH$_{2}$'s excitation. 
We do not detect any sign of the 444~GHz p-\cht transitions toward the different Orion positions given in Sect.~\ref{sec:apexobservations} above an rms noise level of 77~mK on average, for a spectral resolution of 1~km~s$^{-1}$. Moreover, even after a deeper integration toward the Orion S position, which resulted in an rms level of 8~mK, we did not detect any signatures of p-CH$_{2}$. The Einstein $A$-coefficients and hence, the critical densities of the 444~GHz p-CH$_{2}$ lines are two orders of magnitude larger than that of the 70~GHz o-\cht lines. With critical densities on the order of $2\times10^{7}~$cm$^{-3}$ the non-detection of these lines is no surprise and consistent with our finding that \cht exclusively resides in hot, but low-density regions.  

\subsection{\texorpdfstring{CH$_{2}$}{CH2} in other sources}\label{subsec:results_othersources}
The results of our observations discussed thus far point to the origin of the o-\cht emission in regions of intermediate gas densities in the envelopes of hot cores, probing PDR layers, rather than the hot cores themselves.  
In order to confirm the association of the observed o-\cht emission with PDRs, we have also searched for o-\cht emission in: (1) other well-known SFRs that harbour PDRs, (2) (proto-) planetary nebulae (PNe) that are surrounded by molecule-rich envelopes that resemble the composition of PDRs and (3) supernovae remnants (SNRs). Their coordinates are listed in Tables~\ref{tab:source_coordinates_line_params},~\ref{tab:detection_limits}, and~\ref{tab:detection_limits2}, alongside their assumed centroid LSR velocities and, for each line group, the rms noise levels. 

We successfully detected the blended $J\!=\!4\rightarrow\!3$ transitions of o-\cht in emission toward W51~M, N, and E, W49~N, W3~IRS5, W43, W75~N, DR21, and S140 at the systemic velocities of these sources. The resulting spectra are displayed in Figs.~\ref{fig:w51_panel}--\ref{fig:other_sources}. In the following sections, we discuss the observed characteristics of this line blend for select sources in more detail.

\subsection*{W51}
Given that W51~Main~(M) was one of the original targets toward which \citet{hollis1995confirmation} first detected CH$_{2}$, we re-observed this position in order to first verify their detection and then carried out observations toward two luminous condensations harboring high-mass YSOs present in this region, W51~North (N), also known as W51 IRS~2, and W51(E). 

The W51 cloud complex, lying in the Sagittarius spiral arm at a distance of 5.4~kpc from the Sun \citep{sato2010trigonometric}, is one of the best-studied SFRs in our Galaxy. W51 E and N are the two active and presumably youngest centres of activity, hosting the ultracompact (UC) H{\small II} regions e1--e8 and d, respectively \citep[][and references therein]{Ginsburg2017}, infrared and sub-millimetre continuum emission \citep{Thronson1979, jaffe1984massive}, H$_{2}$O and OH masers \citep{genzel1981proper}, and knots of hot NH$_{3}$ ($J,K$) = (3,3) line-emitting gas \citep{ho1983vla, Goddi2015, Ginsburg2017}, all of which are signposts of active star formation within cores with masses of $\gtrsim 10^{4}~\text{M}_{\odot}$ each. 
W51~N and E are separated by a projected distance of $1\rlap{.}^{\prime\prime}9$~pc, with W51~M about halfway in between. M shows more extended FIR and radio continuum emission than E and N \citep{Moon1994, Thronson1979}.  

\begin{figure*}
    \includegraphics[width=1.02\textwidth]{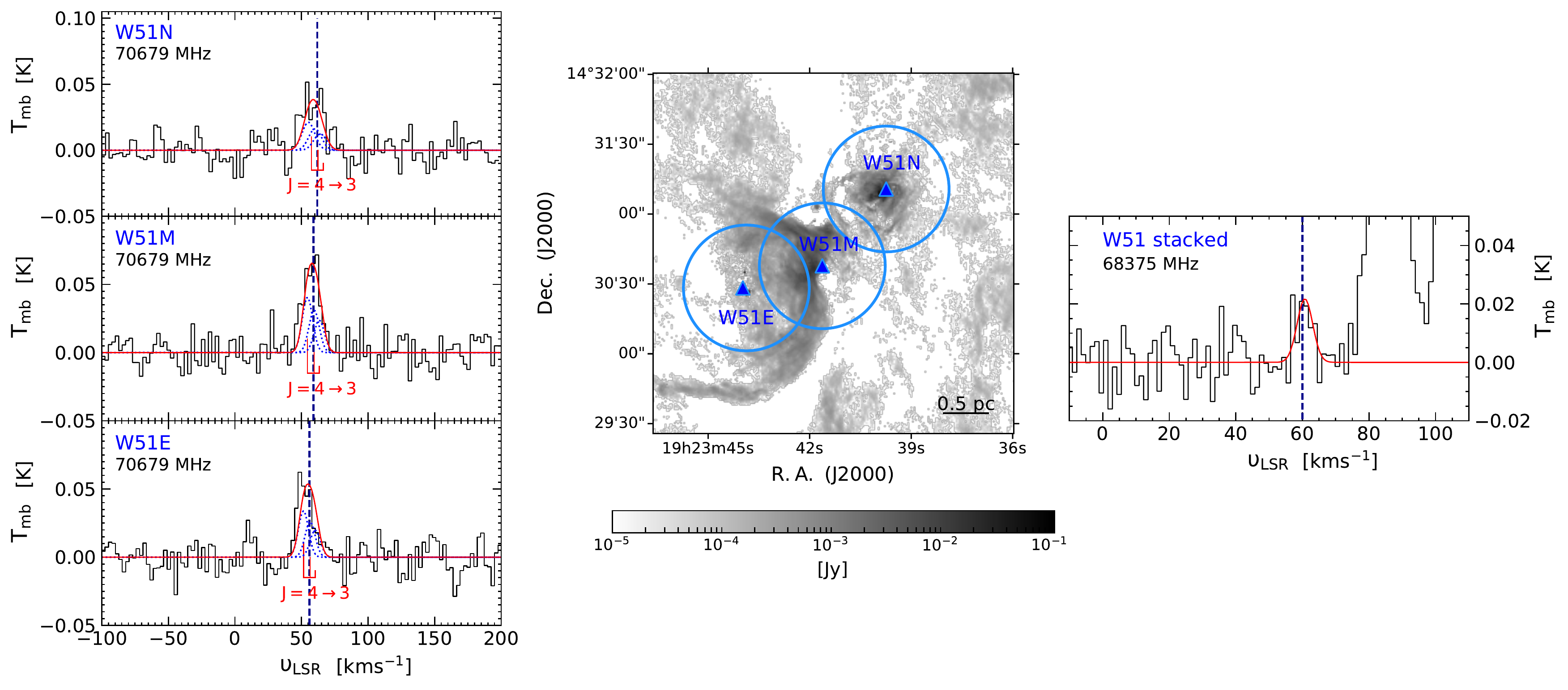}
    \caption{ Left: Spectra of the o-\cht $J\!=\!4\rightarrow\!3$ fine-structure level toward (top to bottom) W51~N, W51~M, and W51~E observed using the Onsala 20~m telescope.  The individual fits to the HFS components are displayed by dotted blue curves while solid red curves display the combined fit. The positions and relative intensities of the corresponding, blended HFS components are displayed below. The dashed blue line indicates the systemic velocities of the sources which lie ${\sim\!+60}~$km~s$^{-1}$.
    Centre: Ku-band image (14.5~GHz, tracing ionised gas) of the W51 region obtained using a combination of JVLA B and D arrays taken from \citet{ginsburg2016toward}, is used to indicate the different Onsala pointing positions. The distance between the different positions is greater than half the beam size of $27^{\prime\prime}$ at $69\,$GHz. Right: The position- and HFS-stacked spectrum of the 68~GHz \cht lines. The strong emission at $\varv >\!77$~km~s$^{-1}$ represents contamination from a CH$_3$CCH line (see Sect.~\ref{subsec:results_orion}).}
    \label{fig:w51_panel}
\end{figure*}

We are able to unambiguously detect the $J=4\rightarrow3$ fine-structure component of the \trans transition of o-\cht near 70~GHz toward all three positions at the source intrinsic velocities between 57 and 62~km~s$^{-1}$ (see Fig.~\ref{fig:w51_panel}). While the emission profiles toward each of the three positions are comparable, the strongest emission arises from the W51~M region, toward which \citet{hollis1995confirmation} pointed the KP 12~m telescope. This is not surprising, given that the radio emission (and that from its PDR, too) has a larger angular size than that from E and N, resulting in a larger beam filling factor. 
Therefore, the fact that we see stronger \cht emission toward the extended H{\small II} region W51~M than toward the much more massive and denser W51~E and N regions, strongly supports an origin of \cht\ in extended dilute gas. 

The observed line profile of the 70~GHz component toward W51~M is comparable to that reported by \citet{hollis1995confirmation} however, we do not detect distinct emission from the HFS lines corresponding to the $J\!=\!5\rightarrow\!4$ and $J\!=\!3\rightarrow\!2$ fine-structure transitions near 68~GHz, and 69~GHz, toward any of the three positions above an average rms noise level of 20~mK. Therefore, in an attempt to recover an average measurement, we stacked the observed spectra toward the different W51 pointing positions by aligning their frequency scales. Due to the contamination from strong NS emission, it is difficult to disentangle the HFS features corresponding to the 69~GHz transition. However, the HFS lines of the 68~GHz component are weakly visible. In order to gauge the true nature of this emission, we further stacked each of the individual HFS lines by aligning their velocity scales. This stacking exercise revealed a 3.5~$\sigma$ detection of this line with a width of 9.1~km~s$^{-1}$ and a peak line temperature of 23~mK. The line width determined from this 68~GHz component is consistent with the value of the intrinsic line width derived from the iterative HFS decomposition of the 70~GHz lines. The spectrum resulting from our stacking analysis is displayed alongside the 70~GHz lines in Fig.~\ref{fig:w51_panel}. This highlights the fact that the non-detection of the 68~GHz and 69~GHz transitions of \cht in our observations is primarily due to a sensitivity issue.

\subsection*{W3 IRS5}

\begin{figure}
     \centering
    \includegraphics[width=0.4\textwidth]{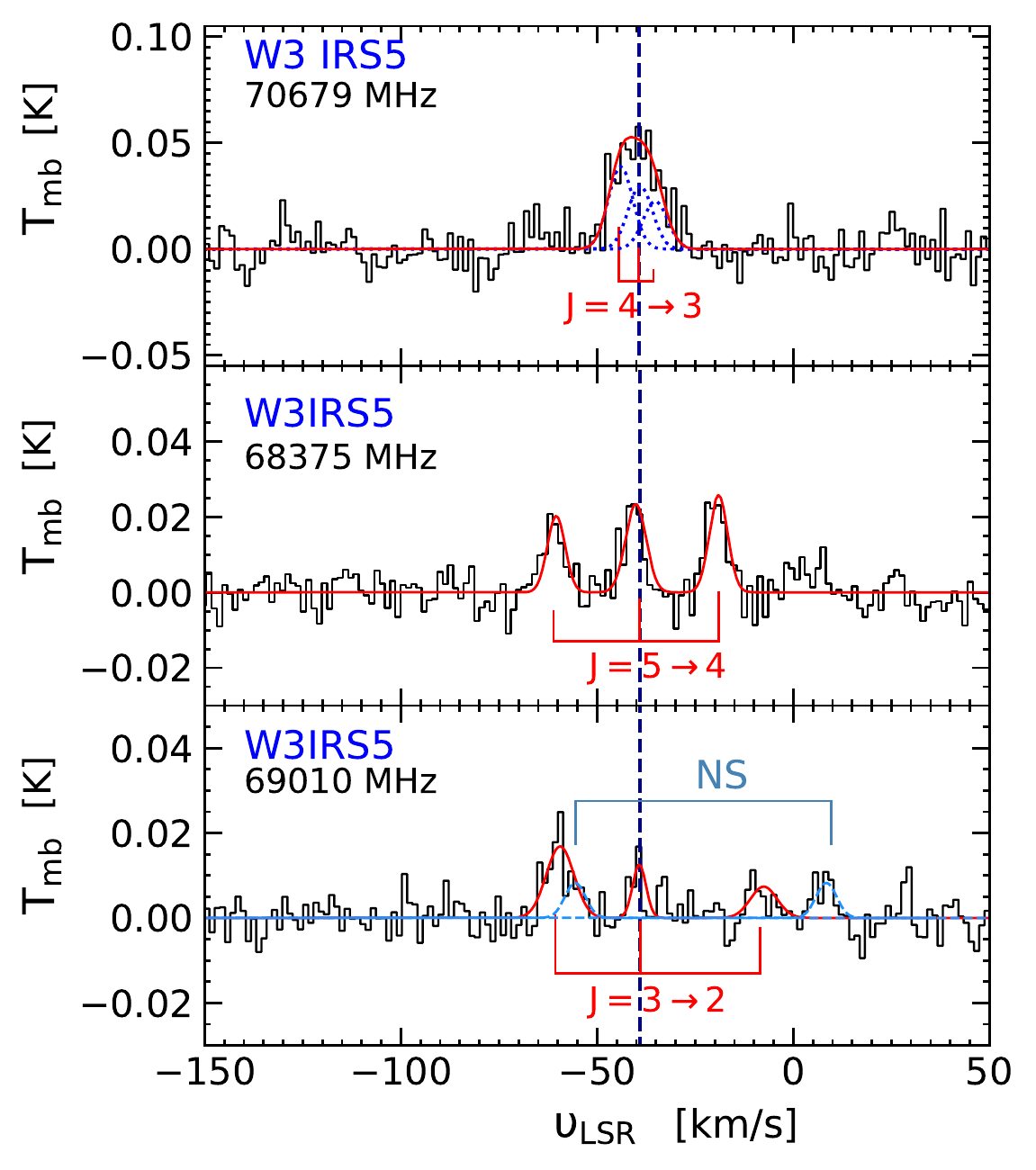}
    \caption{Spectra showing the HFS transitions corresponding to the ${J\!=\!4\rightarrow\!3}$ (top), ${J\!=\!5\rightarrow\!4}$ (middle) and ${J\!=\!3\rightarrow\!2}$ (bottom), fine-structure levels observed using the Onsala 20~m telescope toward W3~IRS5. The individual fits to the HFS components are displayed by red curves. The NS contamination near 69019.180~MHz that blends with the ${J,F = 3,4 \rightarrow 2,3}$ o-\cht HFS line at 69017.995~MHz is marked in light blue. The relative contribution of the NS feature is modelled using the nearby NS $F=5/2\rightarrow3/2$ transition at 69002.890~MHz and is displayed in light blue. The velocity scale is set by the $F^{\prime}- F^{\prime\prime} = J$--$J-1$ HFS line (see Table~\ref{tab:freq}).}
    \label{fig:W3_6869_finestructure}
\end{figure}

The W3 IRS5 cluster system has a well known double IR source at the centre of an embedded cluster of a few hundred low mass stars \citep{Megeath1996}. Located in the W3-Main region at a distance of $2.1$~kpc \citep{megeath2008low, navarete2019} and a total luminosity of $2\times10^{5}~\mathrm{L}_{\odot}$ \citep{campbell1995high}, it is considered to be at the early stages of star formation. 
We detect the blended HFS transitions of o-\cht near 70~GHz toward this region in emission, centred around -40~km~s$^{-1}$ (with a S/N $>5$), a typical velocity found for molecular lines in this region \citep{Dickel1980} (see Fig.~\ref{fig:W3_6869_finestructure}). Akin to the observations toward the different W51 positions, we did not achieve noise levels that are low enough to clearly detect the 68~GHz, and 69~GHz fine-structure lines. However, by stacking our data with a deep integration spectrum of a bandpass covering 67.3--69.8~GHz, obtained by one team member during the course of a different study using the OSO 20$\,$m telescope toward this source, we were able to detect both the 68~GHz and 69~GHz transitions of o-\cht and their respective HFS lines as well. 

We find the HFS transitions of the 68~GHz fine-structure line to be well-resolved with peak temperatures between 21 and 27~mK and line widths of 5.3--6~km~s$^{-1}$, which are comparable with one another, as well as the HFS-stacked 68~GHz component, toward the combined W51 positions.
As discussed in Sec.~\ref{subsec:line_profiles}, the strongest HFS component of the 69~GHz line blends with the $F=3/2\rightarrow1/2$ transition of NS. Assuming that the intensity of the blended NS line at 69.017995~GHz scales with that of the $F=5/2\rightarrow3/2$ transition of NS at 69.002890~GHz by 0.36, based on their Einstein $A$-coefficients and upper level degeneracies, we can decompose the relative contribution of the NS line from that of o-\cht under the assumption that they are optically thin for conditions of LTE. The o-\cht spectra for both the 68~GHz, and 69~GHz transitions are displayed in Fig.~\ref{fig:W3_6869_finestructure} along with the modelled fit to the contamination from NS.

\subsection*{Other sources}
\begin{figure*}
\centering
\includegraphics[width=0.85\textwidth]{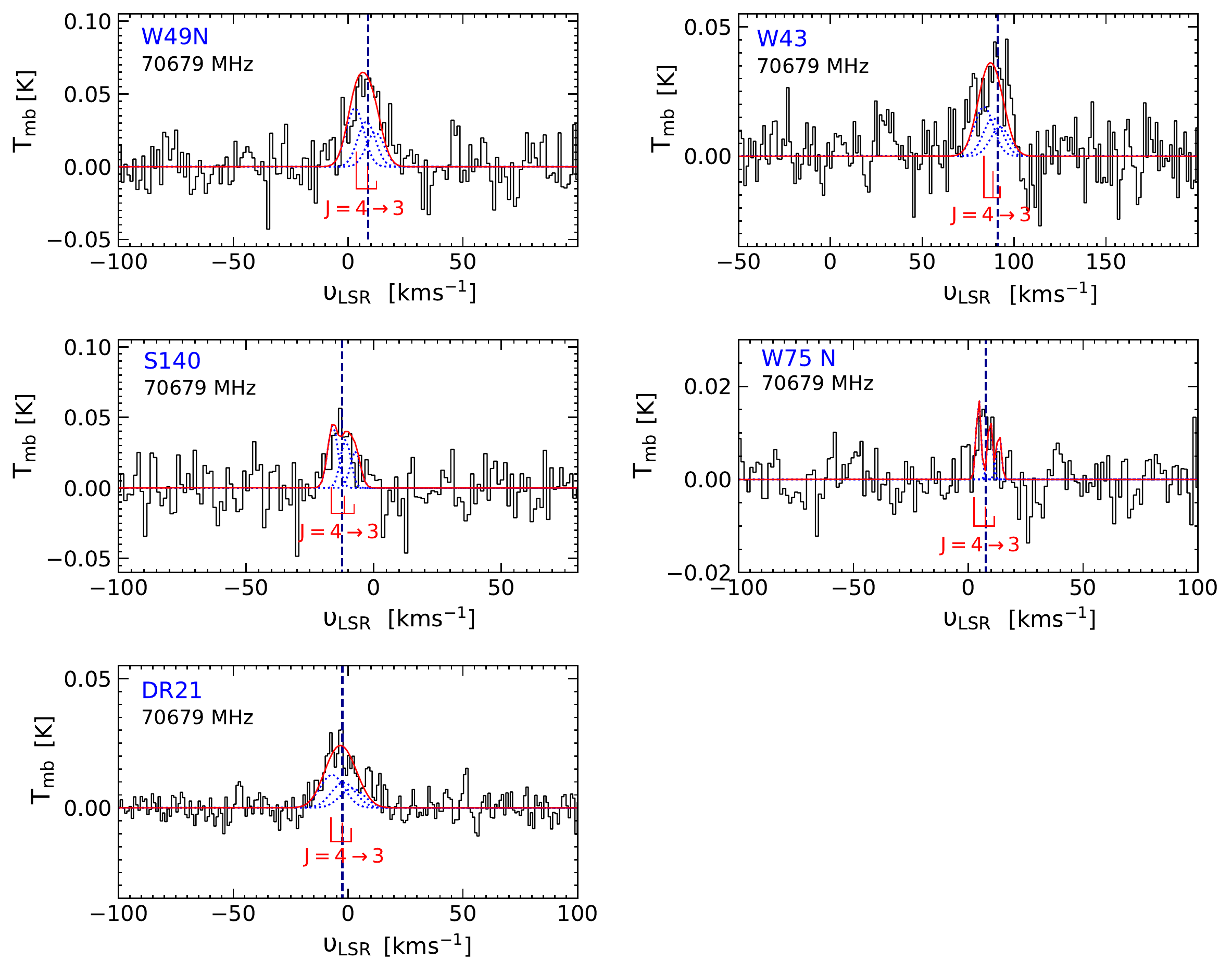}
\caption{Spectra of the o-\cht $J=4\rightarrow3$ fine-structure level toward (clockwise from the top-left) W49~N, W43, W75~N, DR21, and S140 observed using the Onsala 20~m telescope. The individual fits to the HFS components are displayed by dotted blue curves while solid red curves display the combined fit. The positions and relative intensities of the corresponding, blended HFS components are displayed below. The dashed blue line in each panel indicates the systemic velocities of the sources.}
\label{fig:other_sources}
\end{figure*}

In addition to the results presented in the above sections, we successfully detected o-\cht emission from the HFS blended $J = 4\rightarrow3$ transition near 70~GHz, toward W49~N, W43, W75~N, DR21, and S140 at a $\geq\!3\sigma$\ level. The observed spectra are displayed in Fig.~\ref{fig:other_sources} and the results are tabulated in Table~\ref{tab:source_coordinates_line_params}.

At first glance, it appears surprising that the observed line intensities in the nearby Orion region, which is a distance of ${\approx\!400}$~pc \citep{Menten2007, Kounkel2017}, are comparable to the values we find for the other more distant regions, which are between 1.4~kpc (DR21) to 11~kpc (W49~N) away from the Earth. If all sources (including Orion) were unresolved and had an identical intrinsic size and line luminosity, then ultimately all of their emission would be detected in our beam and their measured intensities would scale with the beam-filling factor. In that case, emission from Orion would be by far the strongest. In contrast, our results indicate that the \cht emission from Orion is very extended and has a very low surface brightness. This means that in contrast to the other more distant sources, in the case of Orion, our beam only samples a small portion of the \cht emission, resulting in a low intensity in some PDR positions, or even a non-detection in others, although \cht may be present over much of the volume of the PDR.

\subsection{Comparison with carbon radio recombination lines}\label{subsec:comparison_with_crrl}
The detection of o-\cht toward SFRs strongly suggests that \cht resides in all cases, as in Orion, in warm intermediate-density regions, namely PDRs, which surround the denser, fully molecular material harbouring embedded YSOs. Indeed, the HFS-decomposed line profiles of the observed \cht spectra show LSR velocities and line widths that are similar to those of previously observed prime PDR tracers, namely low frequency CRRLs and neutral atomic carbon lines; see, for example, \citet{Heiles1996, Wyrowski1997, Roshi2006} and \citet{Jakob2007}. 

Having spatial distributions, line widths, and radial velocities consistent with an origin in the neutral gas close to the C$^{+}$/C/CO transition layer, CRRLs are particularly useful tools for probing the physical conditions and kinematics of these regions \citep{Hoang-Binh1974, natta1994carbon, Salas2019}. Since the properties derived from the observed CRRL line profiles reflect the physical conditions of the PDR, coupling our observations of o-\cht with ancillary CRRL data will help us to constrain the origins of the observed o-\cht emission.

In the following analysis, we compare the profiles of HFS-stacked 70~GHz lines with those observed for CRRLs. Because of the small velocity (frequency) separation between the HFS lines corresponding to the 70~GHz fine-structure transitions, they appear to be blended, which makes it difficult to simply stack them. For this reason, we use the modelled results from the individual HFS fits for stacking. The expected rms noise of the stacked and averaged spectrum is then added back to the modelled HFS-stacked profile. The different steps involved in this exercise are detailed and illustrated in Appendix~\ref{appendix:hfs_stacking}.

In the following, as a basis for a CH$_{2}$/CRRL comparison, we use a PDR model to explore the abundance versus visual extinction ($A_{v}$) profiles of \cht, ionised and neutral carbon (C$^+$ and C$^0$, respectively), and other species. We then compare the model results with observational constraints derived from the line profiles. 

\begin{figure*}
\sidecaption 
    \includegraphics[width=12cm]{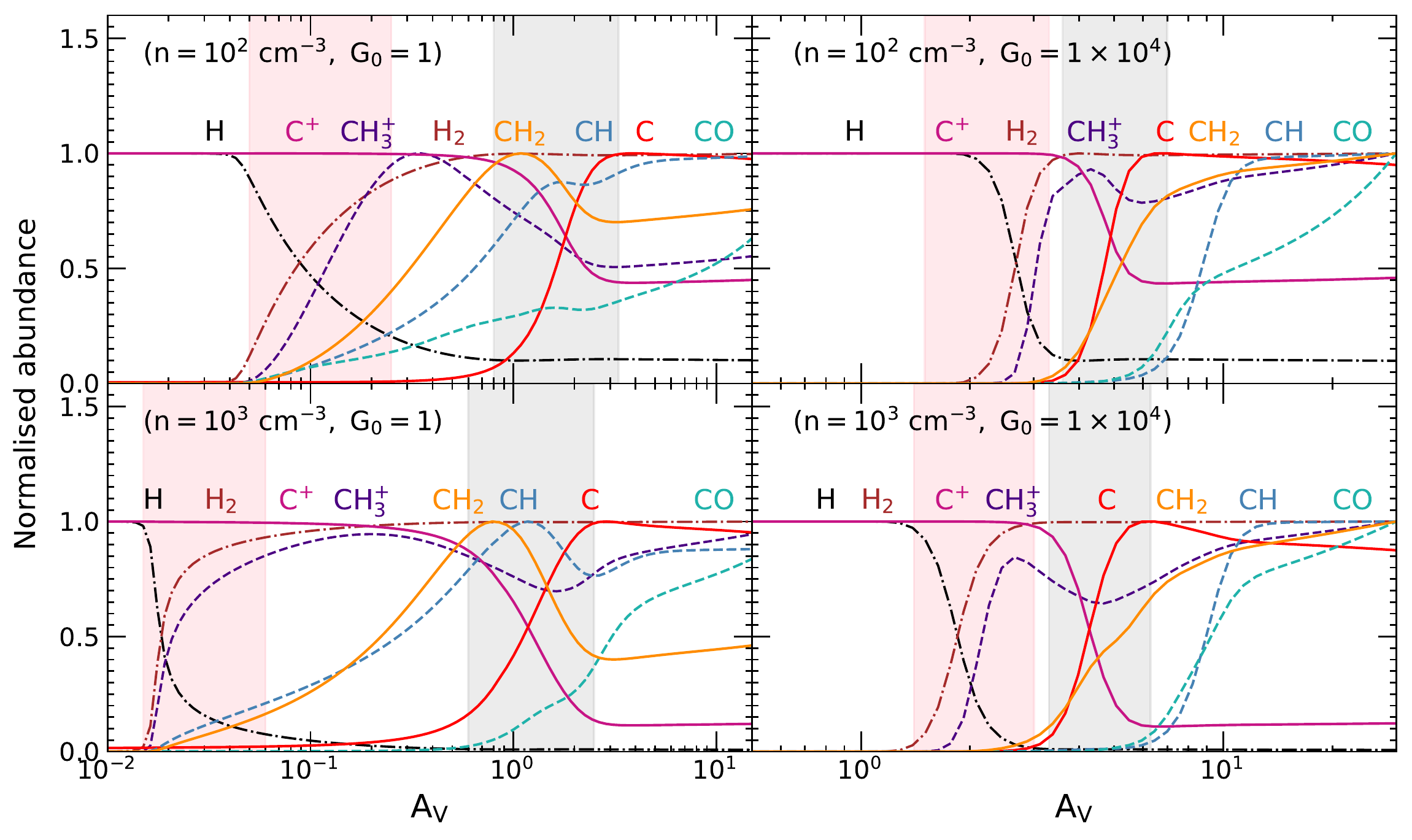}
    \caption{Variation of the gas-phase abundances of fundamental species in carbon chemistry, normalised by their respective peak abundances as a function of visual extinction, $A_{v}$. The cloud parameters for the different models are as follows: $n_{\text{H}} = 10^{2}~$cm$^{-3}$, $G_{0}=1$ (top-left), $n_{\text{H}} = 10^{2}~$cm$^{-3}$, $G_{0}=1\times10^{4}$, $n_{\text{H}} = 10^{3}~$cm$^{-3}$, $G_{0}=1,$ and $n_{\text{H}} = 10^{3}~$cm$^{-3}$, $G_{0}=1\times10^{4}$ (bottom-right). All the models presented here use a constant cosmic-ray ionisation rate of ${\zeta_{\text{H}} = 2.2\times10^{-16}~}$s$^{-1}$. The pink and grey shaded regions in each panel highlight the transition layer between H-H$_{2}$ and C$^{+}$-C, respectively.}
    \label{fig:chemical_mdl}
\end{figure*}

 Our models were created using a simple Python-based PDR code, PyPDR \citep{Bruderer2019}. The code computes chemical abundances by evolving chemical rate equations iteratively, by utilising a pure gas-phase, time-dependent chemical network (except for H$_{2}$), containing 30 species, including CH$_{2}$. The input conditions used for the model are UV radiation field, $G_{0}$, in Habing units, the total gas density, $n_{\text{H}}$, and the cosmic-ray ionisation rate, $\zeta_{\text{H}}$. We studied two sets of models with gas densities $n_{\text{H}}$ of $10^{2}$ and $10^{3}~$cm$^{-3}$, each of which was exposed to a UV radiation field of 1 and $10^{4}$. All four models were exposed to the same cosmic-ray ionisation rates, fixed at $\zeta_{\text{H}} = 2.2\times10^{-16}~$s$^{-1}$, which corresponds to the typical value expected in diffuse and translucent clouds \citep{indriolo2015herschel,neufeld2016chemistry,Jacob2020Arhp}. The resulting (normalised) abundances of relevant carbon-chain species are displayed in Fig.~\ref{fig:chemical_mdl} as a function of $A_{v}$. In both models with $G_{0} =1$, the CH$_{2}$ abundance peaks in the transition layer from C$^+$ to C with its distribution peaking at ${A_{v} \lesssim 1}$. The models with a higher UV radiation field show the CH$_{2}$ abundance distribution to follow more closely that of neutral atomic carbon. From both sets of modelled results, it is clear that \cht traces gas layers between the dissociation front (which marks the transition from H$~\rightarrow~$H$_{2}$) and the molecular cloud, tracing gas layers where C$^{+}$~$\rightarrow ~$C and CO is not the main reservoir of carbon. Overall, the modelled results are consistent with our premise that both \cht and ~CRRLs are probing similar cloud layers.

\begin{figure*}
    \centering
    \includegraphics[width=0.45\textwidth]{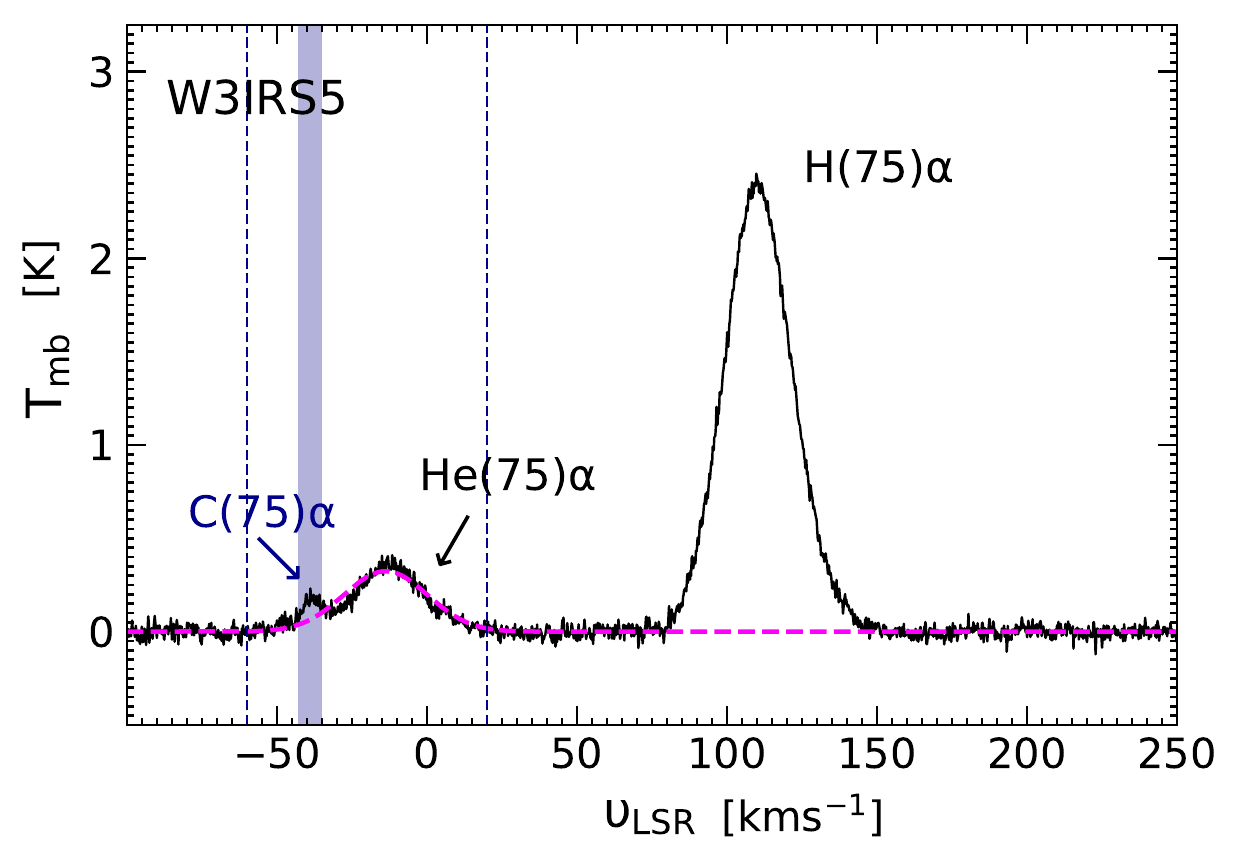} \quad 
    \includegraphics[width=0.45\textwidth]{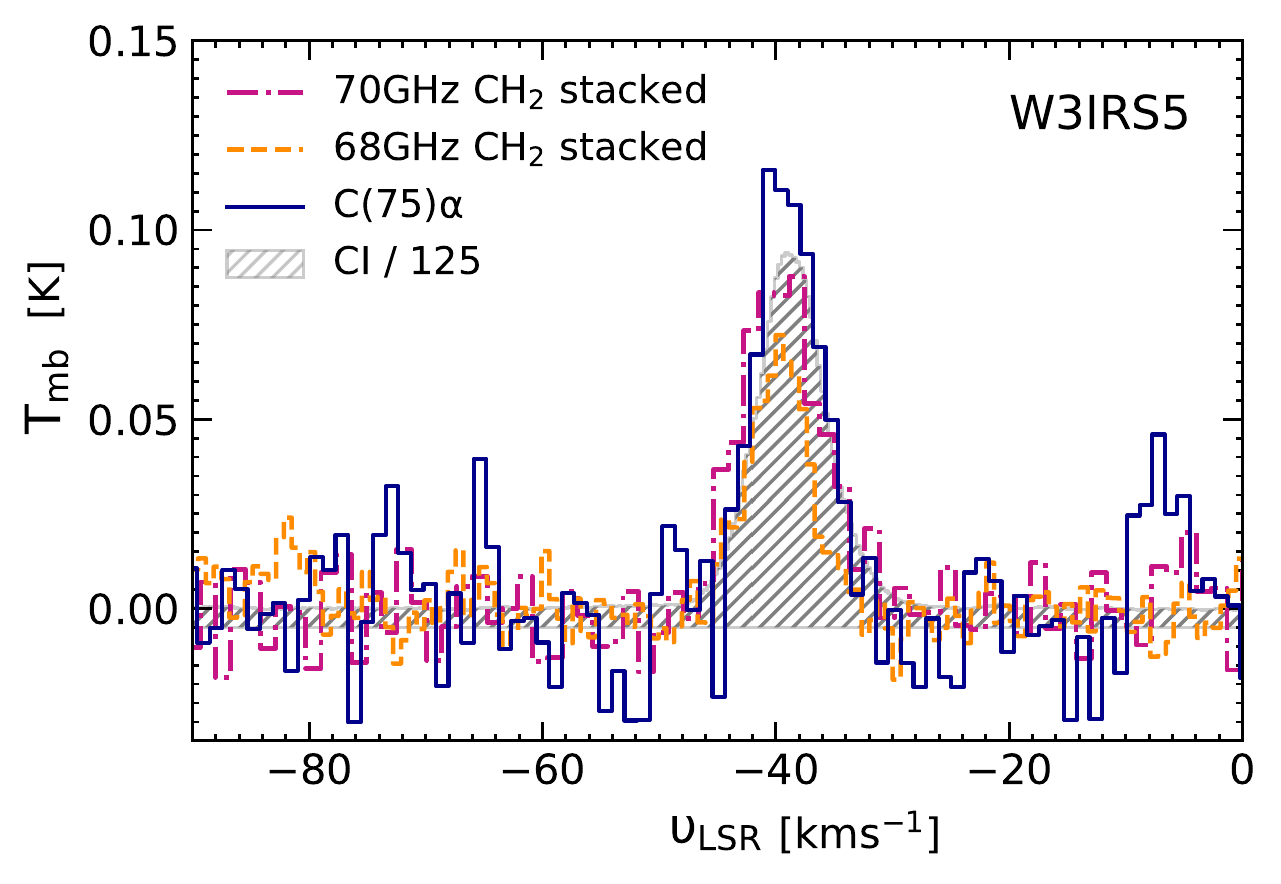}
    \caption{Left: Observed H$\alpha$, He$\alpha$, and C$\alpha$ transitions for $n=75$ toward W3~IRS5. The LSR velocity scale is given with respect to the C$\alpha$ line at 15.82~GHz. The pink dashed curve represents the fit to the He RRL and the shaded purple region confines the CRRL. Right: Decomposed CRRL profile in blue alongside the 68~GHz and 70~GHz HFS-stacked CH$_{2}$ line profiles displayed by the dashed orange and dashed-dotted violet curves, respectively. The stacked profile of the 70~GHz CH$_{2}$ transition was obtained from the HFS decomposition model. The hatched grey regions display the line profile of the $^{3}P_{1}-{}^{3}P_{0}$ transition of C{\tiny I} at 492.160~GHz scaled down by a factor of 125 on the $T_{\text{mb}}$ scale. }
    \label{fig:W3IRS5_RRL}
\end{figure*}

In Appendix~\ref{appendix:summary_RRL}, we present a complete summary of the observed CRRLs; however, as an example, we display the observed recombination line spectrum toward W3~IRS5 in Fig.~\ref{fig:W3IRS5_RRL}. Comparing the narrow line profile of the ~CRRL with that of the corresponding broader (${\Delta \upsilon \gtrsim 30~}$km~s$^{-1}$) H, and He RRLs, it is clear (as has been known for a long time) that the observed CRRLs do not arise from the hot ionised gas of the H{\small II} region but, rather, from the periphery of these regions, next to the neutral gas, namely, in PDRs. We separate the narrow line profiles of the CRRL transitions from those of the broader He RRLs, with which they are blended, by subtracting a Gaussian fit to He RRL from the observed spectra. The residual line profile represents the relative contribution of the CRRL which is then used in our analysis. As mentioned in Sect.~\ref{sec:observations}, toward some of the sources, the observational setup we used covers several RRLs with principal quantum numbers, $n$, ranging from 72 to 80 for $\Delta n= 1$. Since we expect the line properties of consecutive HRRLs to be similar, we can estimate calibration uncertainties by comparing the peak temperatures of the different HRRLs covered in our setups for each individual source. On average, we find the line strengths to vary by 23~\% at most. Figure~\ref{fig:W3IRS5_RRL} shows that the centroid velocities and line widths of the CRRLs match the values of the HFS stacked CH$_{2}$ line emission. This is in agreement with the gas-phase chemistry revealed by our simple PDR models and strongly suggests that the two species trace the same gas. 

It has been shown by \citet{Salgado2017}, that the line widths of high frequency CRRLs (with $n<100$) are dominated by thermal (Doppler) broadening, while Lorentzian profiles best describe low frequency CRRLs (with $n>100$) because of collisional and/or radiation broadening. Therefore, considering the effect of line broadening as being only due to the random thermal motion of particles in the gas and non-thermal effects or turbulence, the observed line widths can be expressed as follows,
\begin{equation} 
    \Delta \upsilon = \sqrt{\Delta\upsilon_{\text{th}}^2 + \Delta\upsilon_{\text{nth}}^2} = \sqrt{\frac{k_{\text{B}}T_{\text{kin}}}{m_\text{C}} + {\left<\Delta\upsilon_{\text{nth}}\right>_{\text{rms}}^2}}
    \label{eqn:velocity_broadening}
\end{equation} 
where, $T_\text{kin}$ is the gas temperature, $k_{\text{B}}$ is the Boltzmann constant,  $m_{\text{C}}$ is the mass of the carbon atom, and ${\left<\Delta\upsilon_{\text{nth}}\right>_{\text{rms}}}$ is the root mean-square measure of the turbulent velocities. Furthermore, based on the premise that the CRRLs and o-\cht lines trace the same gas layers, we assume that both species are impacted by the same turbulent flows and will hence have the same turbulent widths, $\left(\Delta\upsilon_{\text{nth}}\right)_{\text{CRRL}} = \left( \Delta\upsilon_{\text{nth}}\right)_{\text{CH}_{2}}$. Additionally, since the observed line width of the CH$_{2}$ emission is comparable to that of the CRRL (see Fig.~\ref{fig:W3IRS5_RRL}), when re-arranging Eq.~\ref{eqn:velocity_broadening} to equate the non-thermal components, it is clear that the o-\cht gas temperature is simply proportional to that of the CRRL, scaled by the ratio of their masses.

From CRRL data alone, it would be difficult to accurately determine the physical conditions of the C$^+$ region, for example, because of uncertainties involved in their excitation or, practically speaking, simply because of the low intensities of these lines and their blending with the stronger He RRLs. These issues can be overcome by comparing the line intensities of CRRLs with those of the FIR fine-structure line of ionised carbon at 158~$\mu$m. This approach has been employed by several studies to constrain the electron density, and temperature of C$^+$ layers. Typically, the temperatures cover a range of values from $1000$~K near the H{\small II} region to ${\sim\!100~}$K at the outer boundaries of this C$^+$ layer. Modelling this layer toward the W3 region, \citet{Sorochenko2000A} computed a value for the kinetic temperature, $T_{\text{kin}}$, of at most 200~K and an electron density of $n_{e}=54~$cm$^{-3}$. The PDR structure used in their analysis was adapted from \citet{Howe1991} and does not assume a homogeneous distribution of material but rather that it consists of dense clumps ($n_{\text{H}}\!\sim\!10^{5}~$cm$^{-3}$) embedded in a dilute medium ($n_{\text{H}}\!\leq\!300~$cm$^{-3}$) at $T_{\text{k}}\!\geq\!100$~K in order to be consistent with observations. Therefore, for the specific case of W3~IRS5 we can constrain the gas temperature to a value of at most 233~K using Eq.~\ref{eqn:velocity_broadening}.

\section{Discussion} \label{sec:discussion}
\subsection{Non-LTE radiative transfer analysis for \texorpdfstring{CH$_{2}$}{CH2}} \label{subsec:non-lte_analysis}
\begin{figure*} 
    \includegraphics[width=0.98\textwidth]{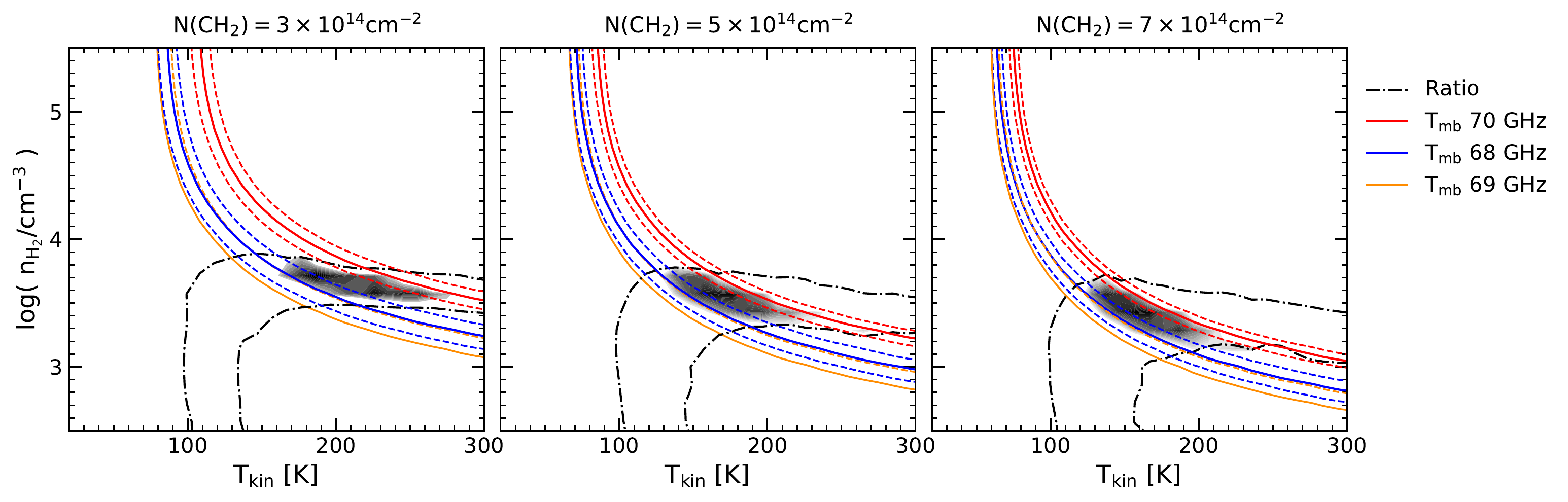}
    \caption{RADEX non-LTE modelling of o-CH$_{2}$ toward W3~IRS5. The solid and dashed- red, blue and orange curves represent the observed peak main-beam brightness temperatures of the 70~GHz, 68~GHz and 69~GHz, respectively, and their uncertainties. We note that for the 69~GHz o-\cht lines being contaminated by NS, we only represent the uncontaminated modelled intensity and an upper limit. The black dashed and dotted lines mark the limits of the brightness temperature ratio between the 70 and 68~GHz lines. The grey shaded region characterises $\chi^{2}_{\text{red}}$ values of $<11$ corresponding to a probability of 99.9~\% over the temperature-density plane for models with fixed column density values of $3\times10^{14}~\text{cm}^{-2}$ (left), $5\times10^{14}~\text{cm}^{-2}$ (centre), and $7\times10^{14}~\text{cm}^{-2}$ (right).}
    \label{fig:radex_dens-temp_mdl}
\end{figure*}

We perform non-LTE radiative transfer calculations using the statistical equilibrium radiative transfer code RADEX \citep{vanderTak2007}, for a uniform expanding sphere geometry under the large velocity gradient (LVG) approximation. The code computes level populations, line intensities, excitation temperatures, and optical depths as a function of the physical conditions specified as input, based on the escape probability formalism. Assuming that H$_{2}$ is the primary collisional partner of CH$_{2}$ in the ISM, we carry out our non-LTE analysis by using rate coefficients recently computed by \citet{Dagdigian2018} for collisions between \cht and Helium from which we obtain o-CH$_{2}$--H$_{2}$ collisional rate coefficients by scaling the rates by factor of 1.4, for all the fine-structure transitions among the lowest 22 energy levels of o-CH$_2$. By adopting a background temperature of 2.73~K, with a fixed line width as estimated from the intrinsic widths of the observed Onsala spectra, we run a grid of models with varying physical conditions, with the aim to constrain the gas densities, $n_{\text{H}_{2}}$, and kinetic temperatures, $T_{\text{kin}}$, to values that are consistent with the observed o-\cht emission or its upper limits. The models were computed over a temperature-density grid of size 500$\times$500, for $T_{\text{kin}}$ values between 20--300~K (constrained by the collisional data) and $n_{\text{H}_{2}}$ values in the range of 10--10$^{6}$~cm$^{-3}$. Given that we were only able to clearly detect all three sets of o-\cht HFS lines toward W3~IRS5, the non-LTE analysis is carried out specifically for this source. Using absorption spectroscopy, \citet{polehampton2005far} were able to (from their column density measurements) determine a [CH]/[CH$_{2}$] ratio through observations of both CH and CH$_{2}$ transitions near 150~$\mu$m and 107.7/127.6~$\mu$m, respectively, using the ISO-LWS. These authors obtained a [CH]/[CH$_{2}$] ratio of 2.7$\pm$0.5 for an ortho-to-para \cht ratio of 1.6, and a value closer to 3.7 for an ortho-to-para \cht ratio of 3 for the (systemic) +64~km/s velocity component toward Sgr~B2(M), values that are consistent with results obtained by \citet{Viti2000} -- whose models additionally take into account grain-surface chemistry. This corresponds to o-\cht abundances with respect to H$_{2}$ between $9.4\times10^{-9}$ and $1.6\times10^{-8}$ when scaled using the [CH]/[H$_{2}$] ratio of 3.5$\times10^{-8}$ as determined by \citet{Sheffer2008}. By using CH column densities determined toward W3~IRS5 by \citet{Wiesemeyer2018} from this radical's 150~$\mu$m ground-state transition observed using the GREAT instrument on board the Stratospheric Observatory For Infrared Astronomy (SOFIA) \citep{young2012early} and the above [CH]/[CH$_{2}$] ratio, we can constrain the column densities of o-CH$_{2}$ to about $(4.5\pm1.7)\times10^{14}~$cm$^{-2}$ for ortho-to-para CH$_{2}$ ratios of between 1.6 and 3. We therefore ran models in the temperature-density plane for fixed values of $N(\text{o-CH}_{2})$ at $3\times10^{14}$, $5\times10^{14}$ and $7\times10^{14}~$cm$^{-2}$, the results of which are displayed in Fig.~\ref{fig:radex_dens-temp_mdl}. The radiative transfer analysis is further simplified by assuming a beam filling factor of unity for an extended o-\cht cloud. From the results of each column density model, we see that the different o-\cht lines trace similar temperature and density conditions as indicated by their contours. We constrain these results by comparing them with the distribution of the line intensity ratio between the 70~GHz, and 68~GHz lines which are both free from contamination, in the temperature-density plane. The physical conditions that prevail in these regions are constrained based on the behaviour of $\chi^{2}_{\text{red}}$ across this parameter space. The $\chi^{2}_{\text{red}}$ value is computed across the entire grid as follows,
\begin{align} 
  \chi^{2} &= \sum {\left( T_{\text{r,obs}} - T_{\text{r,mod}} \right)^{2}}/{\sigma^{2}_{T_{\text{r,obs}}}}  ,\\
  \chi^{2}_{\text{red}} &= \chi^2/(n - 1), 
\end{align}
where $T_{\text{r,obs}}$ and $T_{\text{r,mod}}$ represent the observed and modelled line brightness temperatures on the $T_{\text{mb}}$ scale, $\sigma^{2}_{T_{\text{r,obs}}}$ represents uncertainties in the observed o-\cht spectra and $n$ is the number of degrees of freedom. The $\chi^{2}_{\text{red}}$ values were fit at a 99.9\% probability for two degrees of freedom. For each column density model, we find that the $\chi^{2}_{\text{red}}$ values show more variations with $T_{\text{kin}}$ than with $n_{\text{H}_{2}}$. In Table~\ref{tab:summary_radex}, we summarise the estimated range of gas temperatures and densities for the different models. Across the different models we derive $T_{\text{kin}}$ between (150--200)~K and ${n_{\text{H}_{2}}\!\sim\!3\times10^{3}~}$cm$^{-3}$ consistent with not only previous chemical models \citep{Lee1996} but also our hypothesis that o-\cht must arise in a hot but dilute media. The derived values of temperature are also consistent with that determined from the comparison with CRRLs, as discussed in Sect.~\ref{subsec:comparison_with_crrl}. Furthermore, with critical densities of the order of 10$^{5}$~cm$^{-3}$, the derived densities reveal that the observed 68--71~GHz o-\cht emission lines arise from sub-thermally excited gas.

We expect the non-LTE analysis toward the W51 pointing positions to reproduce physical conditions similar to those derived toward W3~IRS5 since the line intensities of both the 70~GHz and 68~GHz components of CH$_{2}$, and their corresponding line ratios are comparable toward both sources. The results of this analysis confirm our suspicions and clearly addresses why CH$_{2}$ is not widely detected in the Orion Bar PDR. From the non-LTE analysis presented here, we would expect the abundance of CH$_{2}$ to peak in regions with H$_{2}$ gas densities of ${\sim\!3\times10^{3}}~$cm$^{-3}$ and the Orion Bar PDR, as discussed earlier, simply traces gas of higher densities (\textgreater10$^{5}$~cm$^{-3}$). The dense Orion Bar is a spatially limited region in the much larger, on average, more diffuse PDR associated with the Orion Nebula, for which a density of $10^{5}~\text{cm}^{-3}$ is a representative value \citep{Cuadrado2015, Nagy2015HEXOS}. Because of the Orion Nebula's proximity (${\approx\!400}$~pc) this larger scale PDR is well resolved, even in our arc minute size beams. However, the other regions toward which we detect \cht emission are much further away and our observations sample the total extent of their large, low-density PDRs. This has already been discussed at the end of Sect.~\ref{subsec:results_othersources} and here finds support by our non-LTE calculations. Furthermore, the nominal hydrogen nucleus density we derive is a factor of ${\sim\!5}$--10 higher than the value of 500--1000~cm$^{-3}$ estimated by \citet{Welty2020}, who were only able to obtain upper limit for the \cht transition near 1397~\AA{} in the translucent cloud along the LOS toward HD 62542. 

\begin{table}
  \caption{Summary of RADEX results.}
    \centering 
    \begin{tabular}{cccc}
    \hline \hline
        Model & $N$(o-{CH}$_{2})$ & log$_{10}(n_{\text{H}_{2}}/{\rm cm}^{-3})$ & $T_{\text{kin}}$   \\
              & [cm$^{-2}$] &  &  [K]\\
              \hline 
              I & $3\times10^{14}$ & $3.63^{+0.10}_{-0.09}$& $202^{+33}_{-32}$ \\
              II & $5\times10^{14}$ & $3.57^{+0.11}_{-0.10}$& $163^{+27}_{-25}$\\
              II & $7\times10^{14}$ & $3.42^{+0.15}_{-0.14}$& $155^{+26}_{-25}$\\
              
              \hline 
    \end{tabular}
  
    \label{tab:summary_radex}
\end{table}

As for the excitation of these lines, similar to the results obtained by \citet{Dagdigian2018}, we find that over the very wide range of gas densities we modelled ($n_{\text{H}_{2}} = 10$--$10^{8}~$cm$^{-3}$), our calculations produce negative opacities (${\sim\!-10^{-2}}$) or line inversion at an excitation temperature of $-0.36$~K. We thus find all three of the fine-structure lines of the \trans transition of o-\cht seen in emission to be weak masers. From our RADEX models, we compute a ${\sim\!1\%}$ inversion in the population of the \trans fine structure transitions, where the percentage of population inversion for a two-level system is given by $\left(n_{\text{u}} - n_{\text{l}}\right)/\left( n_{\text{u}} + n_{\text{l}} \right)$ and $n_{\text{u}}$, and $n_{\text{l}}$ are the upper and lower energy level populations. Since the models produce masing conditions even in the absence of strong, external radiation fields and without line overlap considerations, it is clear that the observed masing effect in these high-lying lines is a robust phenomenon which preferentially populates the $N_{K_{a}K_{c}}$ = $4_{04}$ level over that of the $N_{K_{a}K_{c}}$ = $3_{13}$ level. The observed emission spectra may have contributions from the extended continuum background radiation in these regions as well as collisional pumping effects. While the weak level-inversion observed in these lines explains why they are detectable in the first place, the degree of population inversion itself is greatly dependent on the collisional rate coefficients.

We exclusively detect \cht within PDR cloud layers associated with H{\small II} regions and not in those associated with (P)PNe and SNRs, a few of which were also observed by us. 
The lack of detectable amounts of \cht in these types of objects can be attributed to the fact that \cht arises from dilute PDR layers, as implied by our non-LTE radiative transfer analysis. The PDRs surrounding (P)PNe are dense regions (${\sim 10^{5}-10^{6}}~{\rm cm}^{-3}$) as they form in the compressed inner layers of the remnant circumstellar envelopes of AGB stars \citep[see for example][]{Cox2002}. This argument explaining the non-detection of \cht in (P)PNe owing to the elevated  densities of their PDRs also holds true for the case of the SNRs for which upper limits are presented in this study. Our Onsala 20~m beam was pointed toward a dense molecular clump in IC~443 residing in the interaction zone of the SNR with a molecular cloud that has gas densities as high as 10$^{5}$~cm$^{-3}$ \citep{Dickman1992}, while for the dense molecular knots 
detected in high-$J$ CO lines toward Cas~A, gas densities of 10$^{6-7}~{\rm cm}^{-3}$ have been determined \citep{Wallstrom2013}.\\

\begin{figure}
    \centering
    \includegraphics[width=0.36\textwidth]{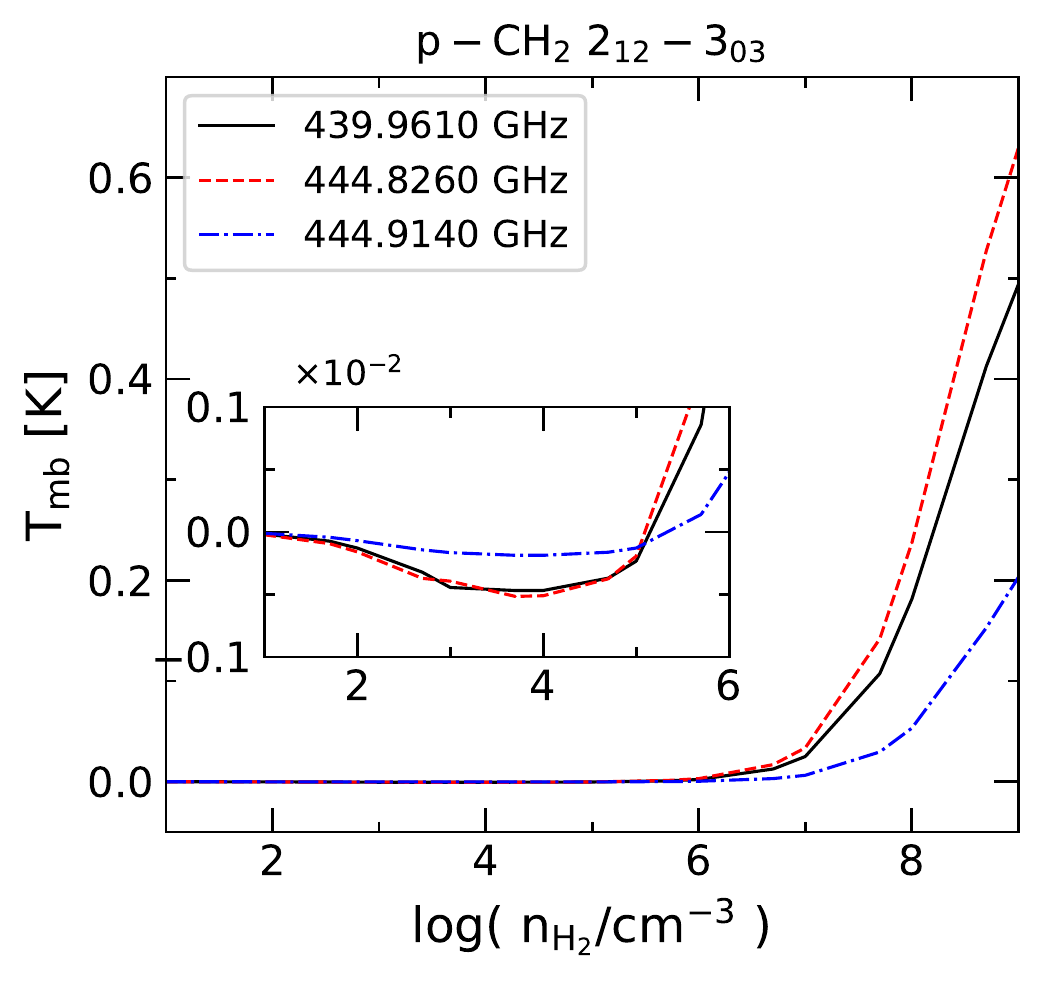}\quad
    \includegraphics[width=0.36\textwidth]{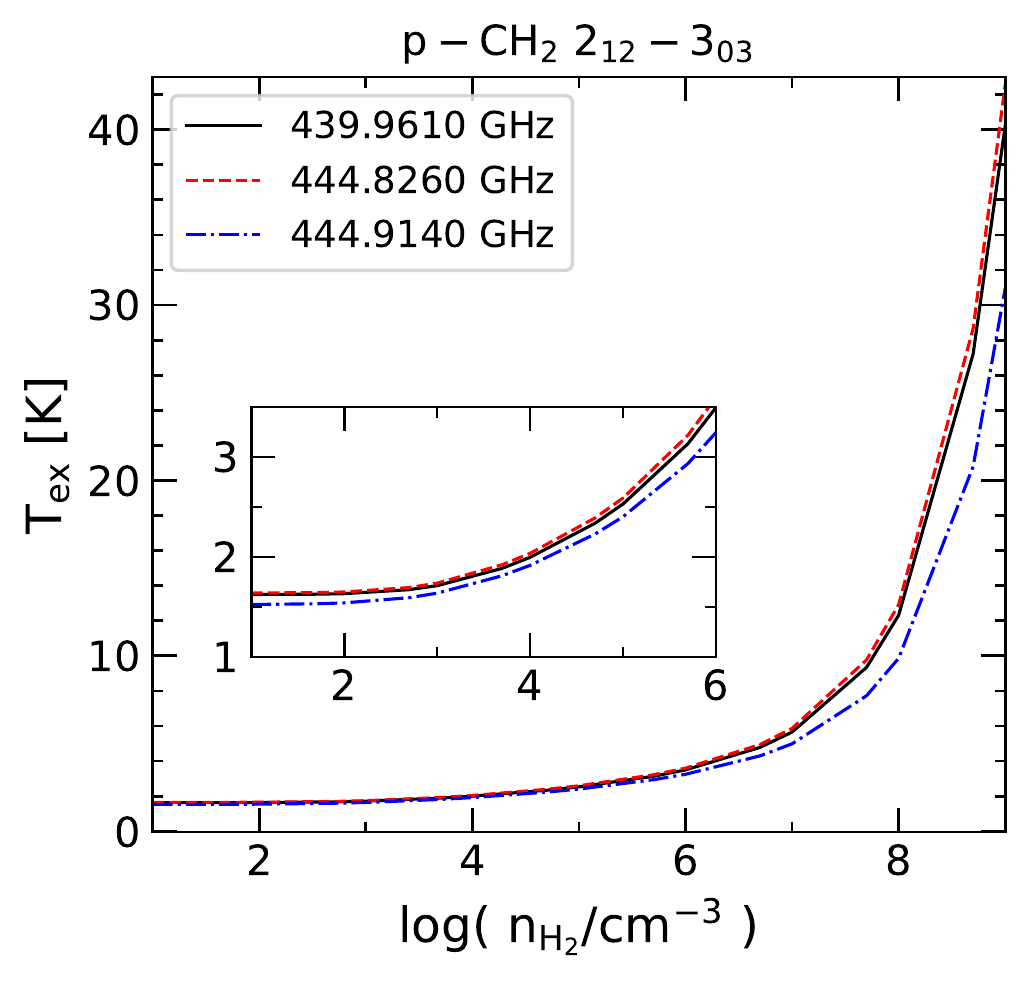}
    \caption{Main-beam brightness temperature ($T_{\text{mb}}$) and excitation temperature ($T_{\text{ex}}$) for the $2_{12}-3_{03}$ fine-structure lines of p-CH$_{2}$. The inset panels expand on the ($T_{\text{mb}}$) and ($T_{\text{ex}}$) values for gas densities between 10--10$^{6}~$cm$^{-3}$.}
    \label{fig:RADEX-444ghz}
\end{figure}

Using the physical conditions derived from these models and the rate coefficients calculated between $p$-CH$_{2}$--H$_{2}$, which are also based on calculations by \citet{Dagdigian2018}, we predict the brightness temperatures and excitation temperature for the 444~GHz p-\cht transitions. The model was run for a range of gas densities $n(\text{H}_{2})\!=\!10-10^{9}~$cm$^{-3}$ with a fixed CH$_{2}$ column density and gas temperature of $N$(p-{CH}$_{2})\!=\!5\times10^{14}~\text{cm}^{-2}$ (in the limit where $N$(p-{CH}$_{2}$) =  $N$(o-{CH}$_{2}$)) and $T_{\text{kin}} = 163~$K, respectively. The modelled results are displayed in Fig.~\ref{fig:RADEX-444ghz}. Up until the critical density of $2\times10^{7}~$cm$^{-3}$ is reached, the brightness temperatures reproduced by the models are low and even slightly negative, particularly for the range of densities derived from the 70~GHz o-\cht lines. The excitation temperature at these frequencies is also small and close to the background radiation temperature, which, for these models, is governed by the cosmic microwave background at 2.73~K. With excitation temperatures below 10~K up to densities of 10$^{8}~$cm$^{-3}$, the models suggest that it is highly unlikely to observe detectable amounts of p-CH$_{2}$ at 444~GHz and in absorption, if any at all.

\subsection{Comparison with atomic carbon}\label{sec:atomic_carbon}
We further compare the observed CH$_{2}$ and CRRLs with that of the neutral atomic carbon, C{\small I}, ${^{3}P_{1}-{}^{3}P_{0}}$ transition at 492.160~GHz for those sources toward which this transition is available, using ancillary data from the Herschel/HIFI archive\footnote{See, \url{http://archives.esac.esa.int/hsa/whsa/}}. These observations, published in \citet{Gerin2015}, were procured using the HIFI band 1a which provides a FWHM beam width of 43.1$^{\prime\prime}$ at these frequencies, which is comparable to the beam size of our \cht and CRRL observations. We find that the line ratio between the C{\small I}, and \cht 70~GHz emission to be almost a constant, with a value of 125:1 toward W3~IRS5, W51~M, and W49~N,  (see Fig.~\ref{fig:W3IRS5_RRL} and top panel of Fig.~\ref{fig:RRL_comparison_plots}). Furthermore, we see that toward W51~M and W49~N the C{\small I} emission shows two components along the LOS and, perhaps, the second component may also be present in \cht but has gone undetected because our observations do not attain the sensitivity necessary to detect this weaker component.

\begin{figure}
    \centering
    \includegraphics[width=0.49\textwidth]{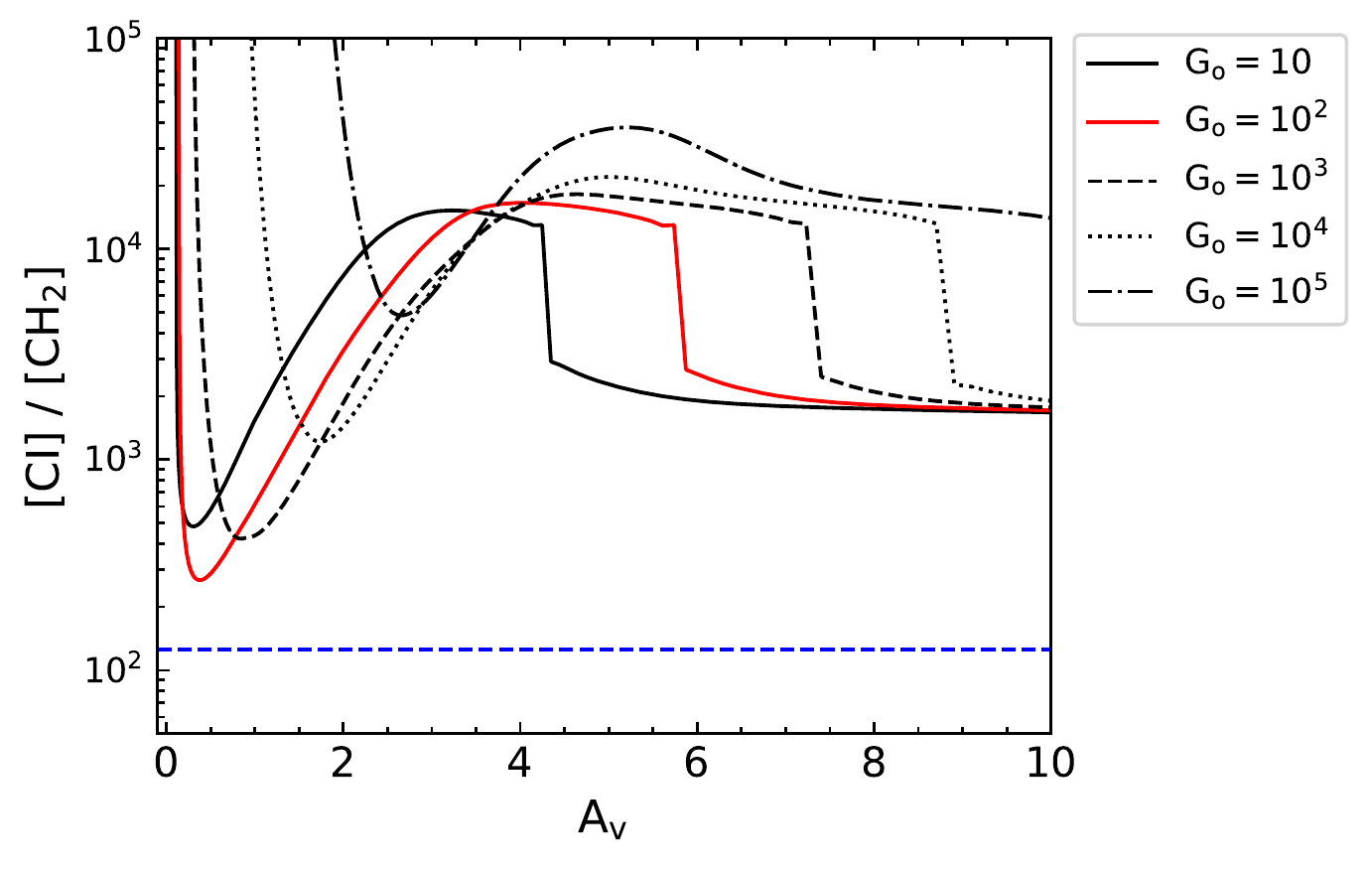}
    \caption{Variation in the [C{\tiny I}]/[CH$_{2}$] abundance ratio as a function of $A_{v}$. The different curves represent models at different values of $G_{\text{o}}$, as labelled. The dashed blue curve marks the observed [C{\tiny I}]/[CH$_{2}$] intensity ratio of 125.}
    \label{fig:revised_chemical_mdl}
\end{figure}

The similarity in the line intensity ratio between the emission of the C{\small I} and the 70~GHz \cht component toward these three sources strongly suggests that \cht is likely to be formed under similar physical conditions in these regions, in spite of their different levels of star-formation. Using the constraints on the physical conditions derived in Sect.~\ref{subsec:non-lte_analysis}, we revise the chemical models presented in Sect.~\ref{subsec:comparison_with_crrl}. We once again ran the PyPDR code, this time for the gas density and temperature values derived from our non-LTE analysis. In order to reproduce the observed [C{\small I}]/[CH$_{2}$] line intensity ratio of 125, we carried out the analysis for different models with varied values of $G_\text{o}$ at a fixed dust temperature of 50~K. None of the models were capable of reproducing the observed [C{\small I}]/[CH$_{2}$] line intensity ratio (see Fig.~\ref{fig:revised_chemical_mdl}). The model that most closely matched the observations gives us a [C{\small I}]/[CH$_{2}$] ratio of 268, and was that with a $G_{o} = 10^{2}$ in Habing units for values of $A_{v}$ between 0 and 1. However, overall the modelled line ratios are almost always overestimated; or rather, the \cht abundance is underestimated. This could either be because the underlying chemical network for CH$_{2}$ used in these models is incomplete or because the PDR model itself is too simplistic or a combination of both. Perhaps a more robust PDR model is required, namely, one which includes dust illumination by both an internal as well as an external heating source within the PDR structure. 

\section{Summary and conclusions}\label{sec:conclusions}
In this work, we present observations of the \trans transitions of o-\cht made with both the KP 12~m and Onsala 20~m telescopes. The former observations were prompted by the non-detection of the 68--71~GHz \cht lines toward the Orion-KL hot core using the IRAM 30~m telescope and the GBT. Our KP observations were therefore used, firstly, to verify the original detection of \cht toward the Orion-KL, and secondly, to address if the \cht emission is extended by investigating nearby positions in OMC-1 and its related H{\small II} region/PDR. These observations confirm the detection of \cht toward Orion-KL and reveal higher intensities of \cht toward PDR positions. The interpretation of these results clearly indicates that the \cht emission is extended but does not constrain their source of origin. 
We clarify the nature of the \cht emission using the smaller angular beam size of the Onsala 20~m telescope. These observations show that the distribution of \cht toward the Orion KL-S and Orion Bar regions coincides with  peaks of the [C{\small II}] emission. This is consistent with the results of simple PDR models which show that the \cht abundance peaks in gas layers where ionised carbon recombines to form neutral atomic carbon. As for the observations, we find that the line profiles of the \cht transitions are akin to those of CRRLs, which are well-known tracers of PDRs. Based on CH$_2$'s association with PDRs, we expanded our target sample to well known H{\small II} regions, which naturally have associated PDRs, and successfully detected the 70~GHz fine-structure line and its HFS components for the first time toward W51~E, W51~N, W49~N, W43, W75~N, DR21, and S140 as well as all three of the 68--71~GHz fine- and HFS components toward W3~IRS5. Stacking each of the HFS components of the 68~GHz transition in a previously averaged spectrum of the different W51 positions revealed a 3.5$\sigma$\ detection of the 68~GHz CH$_{2}$ line. Carrying out a non-LTE analysis, we constrained the physical conditions of the \cht clouds to gas densities and temperatures of ${\sim\!3\times10^{3}~}$cm$^{-3}$ and 150--200~K, respectively. The physical conditions probed are consistent with those that prevail in translucent clouds and are in agreement with the results of previous theoretical models by \citet{Lee1996}, who show the fractional abundances of \cht to peak at intermediate densities. While our non-LTE analysis was focused on W3~IRS5, the HFS-stacked and averaged pointing positions toward W51 show similar line intensities and therefore probe similar gas conditions. It is only with further, highly sensitive observations of the 68~GHz component of \cht can we verify if the physical properties of the gas probed by \cht are ubiquitous across sources. Our analysis of the excitation conditions reveal that these lines are weakly masing with low negative excitation temperatures. The weak level inversion amplifies the emission of these lines which may otherwise have remained undetectable. We also report the non-detection of the $2_{12}$--$3_{03}$ transitions of p-\cht toward Orion S at a 3$\sigma$ upper limit of 24~mK and at 0.23~K toward the other positions in Orion. 

Our results establish the \trans transition of CH$_{2}$ as tracing the  hot but low density component of the ISM present in PDR layers, with its abundance peaking at the edges of dense clouds. As an essential intermediary in interstellar carbon chemistry, it is important to carry out further observations of this molecule and particularly of its energetically lower transitions. We hope that the results of our study will help guide future searches for this radical in other sources at FIR and sub-millimetre wavelengths.

\section{Open questions and outlook}\label{sec:outlook}
The present work provides strong observational evidence for \cht in PDRs of relatively low density (several times $10^3$~cm$^{-3}$). This is supported by our radiative transfer modelling, which suggests temperatures \textgreater150~K for the CH$_2$-bearing material. While such densities are also found in translucent interstellar clouds, the temperatures in these clouds are thought to be much lower, 10--50(?)~K \citep{SnowMcCall2006}. Evidence for \cht in translucent, and maybe even diffuse interstellar clouds can also be found in the spectra of the FIR o- and p-\cht ground-state transitions toward Sgr~B2. In these, \citet{polehampton2005far} find, absorption not only at Sgr~B2's systemic velocity but also from clouds along the LOS to the Galactic centre region. Although these ISO/LWS data have a low spectral resolution, the fitted velocity components closely correspond to those of numerous mm-wavelength absorption lines that arise from such intervening clouds, whose physical conditions are still poorly constrained, but which are thought to consist of translucent or diffuse interstellar material \citep{Thiel2019}; however, for statistical reasons they are unlikely to be PDRs. This raises the question whether \cht can attain substantial abundance not only in PDRs, but in translucent (or even diffuse) clouds as well. The available evidence from electronic transitions discussed in Sect.~\ref{sec:intro} is so far inconclusive, but the procurement and analysis of further UV absorption data would be most interesting. 

As for extra-galactic observations, \citet{Muller2011} do not detect o-\cht in the line survey they carried out with the Australian Telescope Compact Array of the   7~mm window, corresponding to rest frequencies of 57--94~GHz along two sight lines through the red-shift $z=0.88582$ foreground galaxy that absorbs the continuum of the more distant gravitational lens-magnified blazar, PKS 1830−211. The typical densities and temperatures of the emitting gas in the foreground galaxy are a few times 10$^{3}$~cm$^{-3}$ and 80~K, respectively, conditions that are neither dilute enough nor hot enough to excite the CH$_{2}$ lines in question. Alternatively, their survey is still too shallow to detect the \cht lines. 

It is difficult to meaningfully constrain the \cht abundance just with HFS lines assigned to a single rotational level ($J = 5$) -- while emission from higher frequency rotationally excited sub-millimetre \cht\ transitions will remain undetectable because of their high critical densities. To compound this 
difficulty, the only accessible lines that are likely to be excited; the 69--71~GHz o-\cht lines, extensively discussed here, are ubiquitously weak and boosted into detectability by inversion, which makes their interpretation not quite so straightforward. Moreover, the validity of the conclusions drawn from these lines is heavily dependent on the quality of the collisional rate coefficients used in the radiative 
transfer modelling, which are currently only available for the CH$_2$--He system (and 
were scaled by us to apply to collisions with H$_2$). Recently, \citet{vdTak2020} 
have discussed the current status and future plans for the Leiden Atomic and Molecular Database (LAMDA). In this paper, they review, among others, the use of collisional rate coefficients for collisions of molecules with He and the usual practice of 
scaling to estimate H$_2$ rate coefficients. For the case of \cht they assign a low accuracy (factor of $\sim\!3$--4) for this scaling and conclude that 
CH$_2$, amongst other molecules, has `to be the object of new scattering studies considering H$_2$ as a projectile'. Calculating CH$_2$--H$_2$ rate coefficients would certainly be a significant  computational project. Given that in a practical sense, the 
complete interpretation of CH$_2$ emission in the ISM critically depends on these collisional rate coefficients, it is one that should be considered. 

Apart from the 68--71~GHz lines, we are only left with the UV and the FIR resonance lines. These also present a bleak picture as the former still await a convincing identification in the ISM. With regard to the FIR lines, unfortunately, neither the o- and nor the p-\cht ground state lines at 127.6 and 107.7~$\mu$m, respectively, were covered by the HIFI instrument on Herschel. Moreover, none of the modules of the GREAT instrument on SOFIA covers their wavelengths at present. Even if GREAT was able to do this (and it may do so in the future), detecting these lines might be difficult given their low opacities observed by ISO (with the low LWS resolution) of just 0.02 to 0.05 toward one of strongest FIR sources in the sky, Sgr~B2 (whereas for the 149~$\mu$m CH ground state, lines opacities of ${\sim\!0.4}$ are determined with the same instrument).
Given all these conditions, a full observational picture of \cht in an astrochemical context remains elusive.

\begin{acknowledgements}
The authors acknowledge support from Onsala Space Observatory for the provisioning of its facilities/observational support. The Onsala Space Observatory national research infrastructure is funded through Swedish Research Council grant No 2017-00648. We would like to thank the KP 12~m telescope operators for their help during the on-site and remote observations, and Abby Hedden for her invaluable assistance with the observations as well as Arnaud Belloche and Ed Polehampton. We thank Paul Dagdigian for a discussion on the collision rates. We would also like to thank the anonymous referee for a careful review of the article and valuable input. The authors would like to express their gratitude to the developers of the many C++ and Python libraries, made available as open-source software, in particular this research has made use of the NumPy \citep{numpy}, SciPy \citep{scipy} and matplotlib \citep{matplotlib} packages.  

\end{acknowledgements}

%
\bibliographystyle{aa} 
\bibliography{ref} 

\begin{thebibliography}{100}
\expandafter\ifx\csname natexlab\endcsname\relax\def\natexlab#1{#1}\fi

\bibitem[{{Bally} {et~al.}(2010){Bally}, {Anderson}, {Battersby}, {Calzoletti},
  {Digiorgio}, {Faustini}, {Ginsburg}, {Li}, {Nguyen Luong}, {Molinari},
  {Motte}, {Pestalozzi}, {Plume}, {Rodon}, {Schilke}, {Schlingman},
  {Schneider-Bontemps}, {Shirley}, {Stringfellow}, {Testi}, {Traficante},
  {Veneziani}, \& {Zavagno}}]{Bally2010}
{Bally}, J., {Anderson}, L.~D., {Battersby}, C., {et~al.} 2010, \aap, 518, L90

\bibitem[{{Bally} {et~al.}(2002){Bally}, {Reipurth}, {Walawender}, \&
  {Armond}}]{Bally2002}
{Bally}, J., {Reipurth}, B., {Walawender}, J., \& {Armond}, T. 2002, \aj, 124,
  2152

\bibitem[{{Batria} {et~al.}(1983){Batria}, {Wilson}, {Bastien}, \&
  {Ruf}}]{Batrla1983}
{Batria}, W., {Wilson}, T.~L., {Bastien}, P., \& {Ruf}, K. 1983, \aap, 128, 279

\bibitem[{{Belitsky} {et~al.}(2015){Belitsky}, {Lapkin}, {Fredrixon}, {Sundin},
  {Helldner}, {Pettersson}, {Ferm}, {Pantaleev}, {Billade}, {Bergman},
  {Olofsson}, {Lerner}, {Strandberg}, {Whale}, {Pavolotsky}, {Flygare},
  {Olofsson}, \& {Conway}}]{belitsky2015new}
{Belitsky}, V., {Lapkin}, I., {Fredrixon}, M., {et~al.} 2015, \aap, 580, A29

\bibitem[{{Black} {et~al.}(1978){Black}, {Hartquist}, \&
  {Dalgarno}}]{black1978models}
{Black}, J.~H., {Hartquist}, T.~W., \& {Dalgarno}, A. 1978, \apj, 224, 448

\bibitem[{{Bruderer}(2019)}]{Bruderer2019}
{Bruderer}, S. 2019, {PyPDR: Python Photo Dissociation Regions}

\bibitem[{Br{\"u}nken {et~al.}(2004)Br{\"u}nken, Michael, Lewen, Giesen, Ozeki,
  Winnewisser, Jensen, \& Herbst}]{brunken2004high}
Br{\"u}nken, S., Michael, E.~A., Lewen, F., {et~al.} 2004, Canadian Journal of
  Chemistry, 82, 676

\bibitem[{{Campbell} {et~al.}(1995){Campbell}, {Butner}, {Harvey}, {Evans},
  {Campbell}, \& {Sabbey}}]{campbell1995high}
{Campbell}, M.~F., {Butner}, H.~M., {Harvey}, P.~M., {et~al.} 1995, \apj, 454,
  831

\bibitem[{{Clegg} {et~al.}(1996){Clegg}, {Ade}, {Armand}, {Baluteau}, {Barlow},
  {Buckley}, {Berges}, {Burgdorf}, {Caux}, {Ceccarelli}, {Cerulli}, {Church},
  {Cotin}, {Cox}, {Cruvellier}, {Culhane}, {Davis}, {di Giorgio}, {Diplock},
  {Drummond}, {Emery}, {Ewart}, {Fischer}, {Furniss}, {Glencross},
  {Greenhouse}, {Griffin}, {Gry}, {Harwood}, {Hazell}, {Joubert}, {King},
  {Lim}, {Liseau}, {Long}, {Lorenzetti}, {Molinari}, {Murray}, {Naylor},
  {Nisini}, {Norman}, {Omont}, {Orfei}, {Patrick}, {Pequignot}, {Pouliquen},
  {Price}, {Nguyen-Q-Rieu}, {Rogers}, {Robinson}, {Saisse}, {Saraceno},
  {Serra}, {Sidher}, {Smith}, {Smith}, {Spinoglio}, {Swinyard}, {Texier},
  {Towlson}, {Trams}, {Unger}, \& {White}}]{clegg1996iso}
{Clegg}, P.~E., {Ade}, P.~A.~R., {Armand}, C., {et~al.} 1996, \aap, 315, L38

\bibitem[{{Cox} {et~al.}(2002){Cox}, {Huggins}, {Maillard}, {Habart},
  {Morisset}, {Bachiller}, \& {Forveille}}]{Cox2002}
{Cox}, P., {Huggins}, P.~J., {Maillard}, J.~P., {et~al.} 2002, \aap, 384, 603

\bibitem[{{Cuadrado} {et~al.}(2017){Cuadrado}, {Goicoechea}, {Cernicharo},
  {Fuente}, {Pety}, \& {Tercero}}]{Cuadrado2017}
{Cuadrado}, S., {Goicoechea}, J.~R., {Cernicharo}, J., {et~al.} 2017, \aap,
  603, A124

\bibitem[{{Cuadrado} {et~al.}(2015){Cuadrado}, {Goicoechea}, {Pilleri},
  {Cernicharo}, {Fuente}, \& {Joblin}}]{Cuadrado2015}
{Cuadrado}, S., {Goicoechea}, J.~R., {Pilleri}, P., {et~al.} 2015, \aap, 575,
  A82

\bibitem[{{Cuadrado} {et~al.}(2016){Cuadrado}, {Goicoechea}, {Roncero},
  {Aguado}, {Tercero}, \& {Cernicharo}}]{Cuadrado2016}
{Cuadrado}, S., {Goicoechea}, J.~R., {Roncero}, O., {et~al.} 2016, \aap, 596,
  L1

\bibitem[{{Dagdigian} \& {Lique}(2018)}]{Dagdigian2018}
{Dagdigian}, P.~J. \& {Lique}, F. 2018, \mnras, 473, 4824

\bibitem[{{Dickel} {et~al.}(1980){Dickel}, {Dickel}, {Wilson}, \&
  {Werner}}]{Dickel1980}
{Dickel}, H.~R., {Dickel}, J.~R., {Wilson}, W.~J., \& {Werner}, M.~W. 1980,
  \apj, 237, 711

\bibitem[{{Dickel} {et~al.}(1978){Dickel}, {Dickel}, \& {Wilson}}]{dickel78}
{Dickel}, J.~R., {Dickel}, H.~R., \& {Wilson}, W.~J. 1978, \apj, 223, 840

\bibitem[{{Dickman} {et~al.}(1992){Dickman}, {Snell}, {Ziurys}, \&
  {Huang}}]{Dickman1992}
{Dickman}, R.~L., {Snell}, R.~L., {Ziurys}, L.~M., \& {Huang}, Y.-L. 1992,
  \apj, 400, 203

\bibitem[{{Frayer} {et~al.}(2015){Frayer}, {Maddalena}, {Meijer}, {Hough},
  {White}, {Norrod}, {Watts}, {Stennes}, {Simon}, {Woody}, {Srikanth},
  {Pospieszalski}, {Bryerton}, {Whitehead}, {Ford}, {Mello}, \&
  {Bloss}}]{frayer2015gbt}
{Frayer}, D.~T., {Maddalena}, R.~J., {Meijer}, M., {et~al.} 2015, \aj, 149, 162

\bibitem[{{Genzel} {et~al.}(1981){Genzel}, {Downes}, {Schneps}, {Reid},
  {Moran}, {Kogan}, {Kostenko}, {Matveenko}, \& {Ronnang}}]{genzel1981proper}
{Genzel}, R., {Downes}, D., {Schneps}, M.~H., {et~al.} 1981, \apj, 247, 1039

\bibitem[{{Gerin} {et~al.}(2015){Gerin}, {Ruaud}, {Goicoechea}, {Gusdorf},
  {Godard}, {de Luca}, {Falgarone}, {Goldsmith}, {Lis}, {Menten}, {Neufeld},
  {Phillips}, \& {Liszt}}]{Gerin2015}
{Gerin}, M., {Ruaud}, M., {Goicoechea}, J.~R., {et~al.} 2015, \aap, 573, A30

\bibitem[{{Ginsburg}(2017)}]{Ginsburg2017}
{Ginsburg}, A. 2017, arXiv e-prints, arXiv:1702.06627

\bibitem[{{Ginsburg} {et~al.}(2016){Ginsburg}, {Goss}, {Goddi},
  {Galv{\'a}n-Madrid}, {Dale}, {Bally}, {Battersby}, {Youngblood}, {Sankrit},
  {Smith}, {Darling}, {Kruijssen}, \& {Liu}}]{ginsburg2016toward}
{Ginsburg}, A., {Goss}, W.~M., {Goddi}, C., {et~al.} 2016, \aap, 595, A27

\bibitem[{{Godard} {et~al.}(2014){Godard}, {Falgarone}, \& {Pineau des
  For{\^e}ts}}]{Godard2014}
{Godard}, B., {Falgarone}, E., \& {Pineau des For{\^e}ts}, G. 2014, \aap, 570,
  A27

\bibitem[{{Goddi} {et~al.}(2015){Goddi}, {Henkel}, {Zhang}, {Zapata}, \&
  {Wilson}}]{Goddi2015}
{Goddi}, C., {Henkel}, C., {Zhang}, Q., {Zapata}, L., \& {Wilson}, T.~L. 2015,
  \aap, 573, A109

\bibitem[{{Gong} {et~al.}(2015){Gong}, {Henkel}, {Thorwirth}, {Spezzano},
  {Menten}, {Walmsley}, {Wyrowski}, {Mao}, \& {Klein}}]{Gong2015}
{Gong}, Y., {Henkel}, C., {Thorwirth}, S., {et~al.} 2015, \aap, 581, A48

\bibitem[{{G{\"u}sten} {et~al.}(2006){G{\"u}sten}, {Nyman}, {Schilke},
  {Menten}, {Cesarsky}, \& {Booth}}]{Gusten2006}
{G{\"u}sten}, R., {Nyman}, L.~{\r{A}}., {Schilke}, P., {et~al.} 2006, \aap,
  454, L13

\bibitem[{Harris {et~al.}(2020)Harris, Millman, van~der Walt, Gommers,
  Virtanen, Cournapeau, Wieser, Taylor, Berg, Smith, Kern, Picus, Hoyer, van
  Kerkwijk, Brett, Haldane, Fernández~del Río, Wiebe, Peterson,
  Gérard-Marchant, Sheppard, Reddy, Weckesser, Abbasi, Gohlke, \&
  Oliphant}]{numpy}
Harris, C.~R., Millman, K.~J., van~der Walt, S.~J., {et~al.} 2020, Nature, 585,
  357–362

\bibitem[{{Heiles} {et~al.}(1996){Heiles}, {Koo}, {Levenson}, \&
  {Reach}}]{Heiles1996}
{Heiles}, C., {Koo}, B.-C., {Levenson}, N.~A., \& {Reach}, W.~T. 1996, \apj,
  462, 326

\bibitem[{{Hermsen} {et~al.}(1985){Hermsen}, {Wilson}, {Walmsley}, \&
  {Batrla}}]{Hermsen1985}
{Hermsen}, W., {Wilson}, T.~L., {Walmsley}, C.~M., \& {Batrla}, W. 1985, \aap,
  146, 134

\bibitem[{{Heyminck} {et~al.}(2006){Heyminck}, {Kasemann}, {G{\"u}sten}, {de
  Lange}, \& {Graf}}]{Heyminck2006}
{Heyminck}, S., {Kasemann}, C., {G{\"u}sten}, R., {de Lange}, G., \& {Graf},
  U.~U. 2006, \aap, 454, L21

\bibitem[{{Ho} {et~al.}(1983){Ho}, {Genzel}, \& {Das}}]{ho1983vla}
{Ho}, P.~T.~P., {Genzel}, R., \& {Das}, A. 1983, \apj, 266, 596

\bibitem[{{Hoang-Binh} \& {Walmsley}(1974)}]{Hoang-Binh1974}
{Hoang-Binh}, D. \& {Walmsley}, C.~M. 1974, \aap, 35, 49

\bibitem[{{Hollis} {et~al.}(1995){Hollis}, {Jewell}, \&
  {Lovas}}]{hollis1995confirmation}
{Hollis}, J.~M., {Jewell}, P.~R., \& {Lovas}, F.~J. 1995, \apj, 438, 259

\bibitem[{{Howe} {et~al.}(1991){Howe}, {Jaffe}, {Genzel}, \&
  {Stacey}}]{Howe1991}
{Howe}, J.~E., {Jaffe}, D.~T., {Genzel}, R., \& {Stacey}, G.~J. 1991, \apj,
  373, 158

\bibitem[{{Hunter}(2007)}]{matplotlib}
{Hunter}, J.~D. 2007, Computing in Science Engineering, 9, 90

\bibitem[{{Imai} {et~al.}(2000){Imai}, {Kameya}, {Sasao}, {Miyoshi}, {Deguchi},
  {Horiuchi}, \& {Asaki}}]{Imai2000}
{Imai}, H., {Kameya}, O., {Sasao}, T., {et~al.} 2000, \apj, 538, 751

\bibitem[{{Indriolo} {et~al.}(2015){Indriolo}, {Neufeld}, {Gerin}, {Schilke},
  {Benz}, {Winkel}, {Menten}, {Chambers}, {Black}, {Bruderer}, {Falgarone},
  {Godard}, {Goicoechea}, {Gupta}, {Lis}, {Ossenkopf}, {Persson},
  {Sonnentrucker}, {van der Tak}, {van Dishoeck}, {Wolfire}, \&
  {Wyrowski}}]{indriolo2015herschel}
{Indriolo}, N., {Neufeld}, D.~A., {Gerin}, M., {et~al.} 2015, \apj, 800, 40

\bibitem[{{Jackson} \& {Kraemer}(1994)}]{Jackson1994}
{Jackson}, J.~M. \& {Kraemer}, K.~E. 1994, \apjl, 429, L37

\bibitem[{{Jacob} {et~al.}(2020){Jacob}, {Menten}, {Wyrowski}, {Winkel}, \&
  {Neufeld}}]{Jacob2020Arhp}
{Jacob}, A.~M., {Menten}, K.~M., {Wyrowski}, F., {Winkel}, B., \& {Neufeld},
  D.~A. 2020, \aap, 643, A91

\bibitem[{{Jaffe} {et~al.}(1984){Jaffe}, {Becklin}, \&
  {Hildebrand}}]{jaffe1984massive}
{Jaffe}, D.~T., {Becklin}, E.~E., \& {Hildebrand}, R.~H. 1984, \apjl, 279, L51

\bibitem[{{Jakob} {et~al.}(2007){Jakob}, {Kramer}, {Simon}, {Schneider},
  {Ossenkopf}, {Bontemps}, {Graf}, \& {Stutzki}}]{Jakob2007}
{Jakob}, H., {Kramer}, C., {Simon}, R., {et~al.} 2007, \aap, 461, 999

\bibitem[{Jones {et~al.}(2001)Jones, Oliphant, Peterson, {et~al.}}]{scipy}
Jones, E., Oliphant, T., Peterson, P., {et~al.} 2001, {SciPy}: Open source
  scientific tools for {Python}

\bibitem[{{Kavak} {et~al.}(2019){Kavak}, {van der Tak}, {Tielens}, \&
  {Shipman}}]{Kavak2019}
{Kavak}, {\"U}., {van der Tak}, F.~F.~S., {Tielens}, A.~G.~G.~M., \& {Shipman},
  R.~F. 2019, \aap, 631, A117

\bibitem[{{Klein} {et~al.}(2012){Klein}, {Hochg{\"u}rtel}, {Kr{\"a}mer},
  {Bell}, {Meyer}, \& {G{\"u}sten}}]{Klein2012}
{Klein}, B., {Hochg{\"u}rtel}, S., {Kr{\"a}mer}, I., {et~al.} 2012, \aap, 542,
  L3

\bibitem[{{Kounkel} {et~al.}(2017){Kounkel}, {Hartmann}, {Loinard},
  {Ortiz-Le{\'o}n}, {Mioduszewski}, {Rodr{\'\i}guez}, {Dzib}, {Torres}, {Pech},
  {Galli}, {Rivera}, {Boden}, {Evans}, {Brice{\~n}o}, \& {Tobin}}]{Kounkel2017}
{Kounkel}, M., {Hartmann}, L., {Loinard}, L., {et~al.} 2017, \apj, 834, 142

\bibitem[{{Lee} {et~al.}(1996){Lee}, {Bettens}, \& {Herbst}}]{Lee1996}
{Lee}, H.~H., {Bettens}, R.~P.~A., \& {Herbst}, E. 1996, \aaps, 119, 111

\bibitem[{{Lilley} \& {Palmer}(1968)}]{LilleyPalmer1968}
{Lilley}, A.~E. \& {Palmer}, P. 1968, \apjs, 16, 143

\bibitem[{{Lovas} {et~al.}(1983){Lovas}, {Suenram}, \&
  {Evenson}}]{lovas1983laboratory}
{Lovas}, F.~J., {Suenram}, R.~D., \& {Evenson}, K.~M. 1983, \apjl, 267, L131

\bibitem[{{Lyu} {et~al.}(2001){Lyu}, {Smith}, \& {Bruhweiler}}]{lyu2001search}
{Lyu}, C.-H., {Smith}, A.~M., \& {Bruhweiler}, F.~C. 2001, \apj, 560, 865

\bibitem[{{Mangum} {et~al.}(1993){Mangum}, {Wootten}, \&
  {Plambeck}}]{Mangum1993}
{Mangum}, J.~G., {Wootten}, A., \& {Plambeck}, R.~L. 1993, \apj, 409, 282

\bibitem[{{Megeath} {et~al.}(1996){Megeath}, {Herter}, {Beichman}, {Gautier},
  {Hester}, {Rayner}, \& {Shupe}}]{Megeath1996}
{Megeath}, S.~T., {Herter}, T., {Beichman}, C., {et~al.} 1996, \aap, 307, 775

\bibitem[{{Megeath} {et~al.}(2008){Megeath}, {Townsley}, {Oey}, \&
  {Tieftrunk}}]{megeath2008low}
{Megeath}, S.~T., {Townsley}, L.~K., {Oey}, M.~S., \& {Tieftrunk}, A.~R. 2008,
  Handbook of Star Forming Regions, Volume I, 4, 264

\bibitem[{{Menten} {et~al.}(2007){Menten}, {Reid}, {Forbrich}, \&
  {Brunthaler}}]{Menten2007}
{Menten}, K.~M., {Reid}, M.~J., {Forbrich}, J., \& {Brunthaler}, A. 2007, \aap,
  474, 515

\bibitem[{{Michael} {et~al.}(2003){Michael}, {Lewen}, {Winnewisser}, {Ozeki},
  {Habara}, \& {Herbst}}]{Michael2003}
{Michael}, E.~A., {Lewen}, F., {Winnewisser}, G., {et~al.} 2003, \apj, 596,
  1356

\bibitem[{{Moon} \& {Koo}(1994)}]{Moon1994}
{Moon}, D.-S. \& {Koo}, B.-C. 1994, Journal of Korean Astronomical Society, 27,
  81

\bibitem[{{M{\"u}ller} {et~al.}(2005){M{\"u}ller}, {Schl{\"o}der}, {Stutzki},
  \& {Winnewisser}}]{muller2005cologne}
{M{\"u}ller}, H. S.~P., {Schl{\"o}der}, F., {Stutzki}, J., \& {Winnewisser}, G.
  2005, Journal of Molecular Structure, 742, 215

\bibitem[{{Muller} {et~al.}(2011){Muller}, {Beelen}, {Gu{\'e}lin}, {Aalto},
  {Black}, {Combes}, {Curran}, {Theule}, \& {Longmore}}]{Muller2011}
{Muller}, S., {Beelen}, A., {Gu{\'e}lin}, M., {et~al.} 2011, \aap, 535, A103

\bibitem[{{Nagy} {et~al.}(2015{\natexlab{a}}){Nagy}, {Ossenkopf}, {van der
  Tak}, {Choi}, {Bergin}, {Gerin}, {Joblin}, {R{\"o}llig}, {Simon}, \&
  {Stutzki}}]{Nagy2015HEXOS}
{Nagy}, Z., {Ossenkopf}, V., {van der Tak}, F., {et~al.} 2015{\natexlab{a}}, in
  IAU General Assembly, Vol.~29, 2254241

\bibitem[{{Nagy} {et~al.}(2015{\natexlab{b}}){Nagy}, {Ossenkopf}, {Van der
  Tak}, {Faure}, {Makai}, \& {Bergin}}]{Nagy2015}
{Nagy}, Z., {Ossenkopf}, V., {Van der Tak}, F.~F.~S., {et~al.}
  2015{\natexlab{b}}, \aap, 578, A124

\bibitem[{{Natta} {et~al.}(1994){Natta}, {Walmsley}, \&
  {Tielens}}]{natta1994carbon}
{Natta}, A., {Walmsley}, C.~M., \& {Tielens}, A.~G.~G.~M. 1994, \apj, 428, 209

\bibitem[{{Navarete} {et~al.}(2019){Navarete}, {Galli}, \&
  {Damineli}}]{navarete2019}
{Navarete}, F., {Galli}, P. A.~B., \& {Damineli}, A. 2019, \mnras, 487, 2771

\bibitem[{{Neufeld} {et~al.}(2006){Neufeld}, {Schilke}, {Menten}, {Wolfire},
  {Black}, {Schuller}, {M{\"u}ller}, {Thorwirth}, {G{\"u}sten}, \&
  {Philipp}}]{neufeld2006discovery}
{Neufeld}, D.~A., {Schilke}, P., {Menten}, K.~M., {et~al.} 2006, \aap, 454, L37

\bibitem[{{Neufeld} \& {Wolfire}(2016)}]{neufeld2016chemistry}
{Neufeld}, D.~A. \& {Wolfire}, M.~G. 2016, \apj, 826, 183

\bibitem[{{Ozeki} \& {Saito}(1995)}]{Ozeki1995}
{Ozeki}, H. \& {Saito}, S. 1995, \apjl, 451, L97

\bibitem[{{Pabst} {et~al.}(2019){Pabst}, {Higgins}, {Goicoechea}, {Teyssier},
  {Berne}, {Chambers}, {Wolfire}, {Suri}, {Guesten}, {Stutzki}, {Graf},
  {Risacher}, \& {Tielens}}]{pabst2019disruption}
{Pabst}, C., {Higgins}, R., {Goicoechea}, J.~R., {et~al.} 2019, \nat, 565, 618

\bibitem[{{Parsons} {et~al.}(2012){Parsons}, {Thompson}, {Clark}, \&
  {Chrysostomou}}]{Parsons2012}
{Parsons}, H., {Thompson}, M.~A., {Clark}, J.~S., \& {Chrysostomou}, A. 2012,
  \mnras, 424, 1658

\bibitem[{{Pauls} {et~al.}(1983){Pauls}, {Wilson}, {Bieging}, \&
  {Martin}}]{Pauls1983}
{Pauls}, A., {Wilson}, T.~L., {Bieging}, J.~H., \& {Martin}, R.~N. 1983, \aap,
  124, 23

\bibitem[{{Pety}(2005)}]{Pety2005}
{Pety}, J. 2005, in SF2A-2005: Semaine de l'Astrophysique Francaise, ed.
  F.~{Casoli}, T.~{Contini}, J.~M. {Hameury}, \& L.~{Pagani}, 721

\bibitem[{{Polehampton} {et~al.}(2007){Polehampton}, {Baluteau}, {Swinyard},
  {Goicoechea}, {Brown}, {White}, {Cernicharo}, \&
  {Grundy}}]{polehampton2007lws}
{Polehampton}, E.~T., {Baluteau}, J.-P., {Swinyard}, B.~M., {et~al.} 2007,
  \mnras, 377, 1122

\bibitem[{{Polehampton} {et~al.}(2005){Polehampton}, {Menten}, {Br{\"u}nken},
  {Winnewisser}, \& {Baluteau}}]{polehampton2005far}
{Polehampton}, E.~T., {Menten}, K.~M., {Br{\"u}nken}, S., {Winnewisser}, G., \&
  {Baluteau}, J.~P. 2005, \aap, 431, 203

\bibitem[{{Prasad} \& {Huntress}(1980)}]{prasad1980model}
{Prasad}, S.~S. \& {Huntress}, W.~T., J. 1980, \apjs, 43, 1

\bibitem[{{Roshi} {et~al.}(2006){Roshi}, {De Pree}, {Goss}, \&
  {Anantharamaiah}}]{Roshi2006}
{Roshi}, D.~A., {De Pree}, C.~G., {Goss}, W.~M., \& {Anantharamaiah}, K.~R.
  2006, \apj, 644, 279

\bibitem[{{Salas} {et~al.}(2019){Salas}, {Oonk}, {Emig}, {Pabst}, {Toribio},
  {R{\"o}ttgering}, \& {Tielens}}]{Salas2019}
{Salas}, P., {Oonk}, J.~B.~R., {Emig}, K.~L., {et~al.} 2019, \aap, 626, A70

\bibitem[{{Salgado} {et~al.}(2017){Salgado}, {Morabito}, {Oonk}, {Salas},
  {Toribio}, {R{\"o}ttgering}, \& {Tielens}}]{Salgado2017}
{Salgado}, F., {Morabito}, L.~K., {Oonk}, J.~B.~R., {et~al.} 2017, \apj, 837,
  142

\bibitem[{{Sato} {et~al.}(2010){Sato}, {Reid}, {Brunthaler}, \&
  {Menten}}]{sato2010trigonometric}
{Sato}, M., {Reid}, M.~J., {Brunthaler}, A., \& {Menten}, K.~M. 2010, \apj,
  720, 1055

\bibitem[{{Sheffer} {et~al.}(2008){Sheffer}, {Rogers}, {Federman}, {Abel},
  {Gredel}, {Lambert}, \& {Shaw}}]{Sheffer2008}
{Sheffer}, Y., {Rogers}, M., {Federman}, S.~R., {et~al.} 2008, \apj, 687, 1075

\bibitem[{{Snow} \& {McCall}(2006)}]{SnowMcCall2006}
{Snow}, T.~P. \& {McCall}, B.~J. 2006, \araa, 44, 367

\bibitem[{{Sorochenko} \& {Tsivilev}(2000)}]{Sorochenko2000A}
{Sorochenko}, R.~L. \& {Tsivilev}, A.~P. 2000, Astronomy Reports, 44, 426

\bibitem[{{Stoerzer} {et~al.}(1995){Stoerzer}, {Stutzki}, \&
  {Sternberg}}]{Stoerzer1995}
{Stoerzer}, H., {Stutzki}, J., \& {Sternberg}, A. 1995, \aap, 296, L9

\bibitem[{{Tahani} {et~al.}(2016){Tahani}, {Plume}, {Bergin}, {Tolls},
  {Phillips}, {Caux}, {Cabrit}, {Goicoechea}, {Goldsmith}, {Johnstone}, {Lis},
  {Pagani}, {Menten}, {M{\"u}ller}, {Ossenkopf-Okada}, {Pearson}, \& {van der
  Tak}}]{Tahani2016}
{Tahani}, K., {Plume}, R., {Bergin}, E.~A., {et~al.} 2016, \apj, 832, 12

\bibitem[{{Thiel} {et~al.}(2019){Thiel}, {Belloche}, {Menten}, {Giannetti},
  {Wiesemeyer}, {Winkel}, {Gratier}, {M{\"u}ller}, {Colombo}, \&
  {Garrod}}]{Thiel2019}
{Thiel}, V., {Belloche}, A., {Menten}, K.~M., {et~al.} 2019, \aap, 623, A68

\bibitem[{{Thronson} \& {Harper}(1979)}]{Thronson1979}
{Thronson}, H.~A., J. \& {Harper}, D.~A. 1979, \apj, 230, 133

\bibitem[{{Ulich} \& {Haas}(1976)}]{ulich1976absolute}
{Ulich}, B.~L. \& {Haas}, R.~W. 1976, \apjs, 30, 247

\bibitem[{{Ungerechts} {et~al.}(1997){Ungerechts}, {Bergin}, {Goldsmith},
  {Irvine}, {Schloerb}, \& {Snell}}]{ungerechts1997chemical}
{Ungerechts}, H., {Bergin}, E.~A., {Goldsmith}, P.~F., {et~al.} 1997, \apj,
  482, 245

\bibitem[{{van der Tak} {et~al.}(2007{\natexlab{a}}){van der Tak}, {Black},
  {Sch{\"o}ier}, {Jansen}, \& {van Dishoeck}}]{van2007computer}
{van der Tak}, F.~F.~S., {Black}, J.~H., {Sch{\"o}ier}, F.~L., {Jansen}, D.~J.,
  \& {van Dishoeck}, E.~F. 2007{\natexlab{a}}, \aap, 468, 627

\bibitem[{{van der Tak} {et~al.}(2007{\natexlab{b}}){van der Tak}, {Black},
  {Sch{\"o}ier}, {Jansen}, \& {van Dishoeck}}]{vanderTak2007}
{van der Tak}, F.~F.~S., {Black}, J.~H., {Sch{\"o}ier}, F.~L., {Jansen}, D.~J.,
  \& {van Dishoeck}, E.~F. 2007{\natexlab{b}}, \aap, 468, 627

\bibitem[{{van der Tak} {et~al.}(2020){van der Tak}, {Lique}, {Faure}, {Black},
  \& {van Dishoeck}}]{vdTak2020}
{van der Tak}, F. F.~S., {Lique}, F., {Faure}, A., {Black}, J.~H., \& {van
  Dishoeck}, E.~F. 2020, Atoms, 8, 15

\bibitem[{{van Dishoeck} {et~al.}(1996){van Dishoeck}, {Beaerda}, \& {van
  Hemert}}]{vanDishoeck1996}
{van Dishoeck}, E.~F., {Beaerda}, R.~A., \& {van Hemert}, M.~C. 1996, \aap,
  307, 645

\bibitem[{{van Dishoeck} \& {Black}(1986)}]{van1986comprehensive}
{van Dishoeck}, E.~F. \& {Black}, J.~H. 1986, \apjs, 62, 109

\bibitem[{{Vejby-Christensen} {et~al.}(1997){Vejby-Christensen}, {Andersen},
  {Heber}, {Kella}, {Pedersen}, {Schmidt}, \& {Zajfman}}]{Vejby1997}
{Vejby-Christensen}, L., {Andersen}, L.~H., {Heber}, O., {et~al.} 1997, \apj,
  483, 531

\bibitem[{{Viti} {et~al.}(2000){Viti}, {Williams}, \& {O'Neill}}]{Viti2000}
{Viti}, S., {Williams}, D.~A., \& {O'Neill}, P.~T. 2000, \aap, 354, 1062

\bibitem[{{Walker} {et~al.}(2016){Walker}, {Kalinauskaite}, {McCarthy},
  {Trappe}, {Murphy}, {Helldner}, {Pantaleev}, \& {Flygare}}]{Walker2016}
{Walker}, G.~W., {Kalinauskaite}, E., {McCarthy}, D.~N., {et~al.} 2016, in
  Society of Photo-Optical Instrumentation Engineers (SPIE) Conference Series,
  Vol. 9914, \procspie, 99142V

\bibitem[{{Wallstr{\"o}m} {et~al.}(2013){Wallstr{\"o}m}, {Biscaro}, {Salgado},
  {Black}, {Cherchneff}, {Muller}, {Bern{\'e}}, {Rho}, \&
  {Tielens}}]{Wallstrom2013}
{Wallstr{\"o}m}, S. H.~J., {Biscaro}, C., {Salgado}, F., {et~al.} 2013, \aap,
  558, L2

\bibitem[{{Walmsley} {et~al.}(2000){Walmsley}, {Natta}, {Oliva}, \&
  {Testi}}]{Walmsley2000}
{Walmsley}, C.~M., {Natta}, A., {Oliva}, E., \& {Testi}, L. 2000, \aap, 364,
  301

\bibitem[{{Welty} {et~al.}(2020){Welty}, {Sonnentrucker}, {Snow}, \&
  {York}}]{Welty2020}
{Welty}, D.~E., {Sonnentrucker}, P., {Snow}, T.~P., \& {York}, D.~G. 2020,
  \apj, 897, 36

\bibitem[{{Wiesemeyer} {et~al.}(2018){Wiesemeyer}, {G{\"u}sten}, {Menten},
  {Dur{\'a}n}, {Csengeri}, {Jacob}, {Simon}, {Stutzki}, \&
  {Wyrowski}}]{Wiesemeyer2018}
{Wiesemeyer}, H., {G{\"u}sten}, R., {Menten}, K.~M., {et~al.} 2018, \aap, 612,
  A37

\bibitem[{{Wyrowski} {et~al.}(1997){Wyrowski}, {Walmsley}, {Natta}, \&
  {Tielens}}]{Wyrowski1997}
{Wyrowski}, F., {Walmsley}, C.~M., {Natta}, A., \& {Tielens}, A.~G.~G.~M. 1997,
  \aap, 324, 1135

\bibitem[{{Young} {et~al.}(2012){Young}, {Becklin}, {Marcum}, {Roellig}, {De
  Buizer}, {Herter}, {G{\"u}sten}, {Dunham}, {Temi}, {Andersson}, {Backman},
  {Burgdorf}, {Caroff}, {Casey}, {Davidson}, {Erickson}, {Gehrz}, {Harper},
  {Harvey}, {Helton}, {Horner}, {Howard}, {Klein}, {Krabbe}, {McLean}, {Meyer},
  {Miles}, {Morris}, {Reach}, {Rho}, {Richter}, {Roeser}, {Sandell}, {Sankrit},
  {Savage}, {Smith}, {Shuping}, {Vacca}, {Vaillancourt}, {Wolf}, \&
  {Zinnecker}}]{young2012early}
{Young}, E.~T., {Becklin}, E.~E., {Marcum}, P.~M., {et~al.} 2012, \apjl, 749,
  L17

\bibitem[{{Zapata} {et~al.}(2011){Zapata}, {Schmid-Burgk}, \&
  {Menten}}]{Zapata2011}
{Zapata}, L.~A., {Schmid-Burgk}, J., \& {Menten}, K.~M. 2011, \aap, 529, A24

\bibitem[{{Ziurys} \& {Friberg}(1987)}]{Ziurys1987}
{Ziurys}, L.~M. \& {Friberg}, P. 1987, \apjl, 314, L49

\end{thebibliography}
%
\begin{appendix}
\section{\texorpdfstring{CH$_{2}$}{CH2} non-detections}\label{appendix:ch2_nondetections}
In this appendix we list the rms noise levels for the sources toward which we do not detect any o-CH$_{2}$.  

\begin{table}
\small{
     \centering
     \caption{Onsala 20~m telescope non-detections.}
     \begin{tabular}{lllrcc}
          \hline \hline
         Source & \multicolumn{1}{c}{$\alpha_{{\rm J2000}}$}  & \multicolumn{1}{c}{$\delta_{{\rm J2000}}$}&  \multicolumn{1}{c}{$\upsilon_\text{LSR}$}  & Line & rms\tablefootmark{*} \\  
         
         &
         & 
         &  \multicolumn{1}{c}{[kms$^{-1}$]}  &  \multicolumn{1}{c}{[GHz]} & \multicolumn{1}{c}{[mK]}\\ \hline
          Orion & 05:35:16.96 & -05:22:02.7 & +9.7 & 68 & 40 \\
          KL/S (1) & & & & 69 & 33\\
          Orion  & 05:35:14.28 & -05:22:27.5 & +9.0 & 70 & 24 \\ 
          KL/S (2) & & & & 68 & 53\\
          & & & & 69 & 48\\
          Orion  & 05:35:14.28 & -05:23:16.5 & +9.0 & 70 & 23 \\
          KL/S (3) & & & & 68 & 55\\
          & & & & 69 & 43\\
          Orion & 05:35:16.46 & -05:23:22.7 & +9.0 & 70 & 27 \\
          KL/S (4) & & & & 68 & 61\\
          & & & & 69 & 37\\
          Orion  & 05:35:13.10 & -05:23:56.0 & +9.0 & 70 & 17 \\ 
          KL/S (5)& & & & 68 & 40 \\
          & & & & 69 & 30\\ 
          Orion  & 05:35:24.96 & -05:22:32.7 & +9.0 & 68 & 30\\ 
          KL/S (6)& & & & 69 & 25\\
          Orion & 05:35:25.30 & -05:24:34.0 & +11.0 & 70 & 17 \\ 
          Bar (1) & & & & 68 & 34 \\
          & & & & 69 & 25\\
          Orion  & 05:35:22.80 & -05:25:01.0 & +11.0 & 68 & 55\\
          Bar (2)& & & & 69 & 38\\
          Orion & 05:35:25.40 & -05:25:25.0 & +11.0 & 70 & 18 \\
          Bar (3) & & & & 68 & 41 \\
          & & & & 69 & 30\\
          Orion  & 05:35:23.10 & -05:24:35.0 & +11.0 & 70 & 23 \\
          Bar (4) & & & & 68 & 47 \\
          & & & & 69 & 41\\
          Orion & 05:35:20.81 & -05:25:17.1 & +11.0 & 68 & 38\\
          Bar (5) & & & & 69 & 28\\
          W51~E & 19:23:44.00 & 14:30:30.00 & +55.3 & 68 & 21.  \\
          & & & & 69 & 18\\
          W51~M & 19:23:42.00 & 14:30:36.00 & +58.4 & 68 &  24 \\
          & & & & 69 & 21 \\
          W51~N & 19:23:40.00 & 14:31:10.00 & +60.0 & 68 & 15  \\
          & & & & 69 & 12\\
          W49~N & 19:10:13.55 & 09:06:14.70 & +8.8 & 68 & 26 \\
          & & & & 69 & 21 \\
          W43 & 18:47:36.89 & -01:55:30.20 & +91.1 &  68 & 26 \\
          & & & & 69 & 19 \\
          DR21 & 20:39:02.00 & 42:19:42.00 & -2.5 & 68 & 11\\
          & & & & 69 & 7  \\
          W75~N & 20:38:36.45 & 42:37:35.10 & +7.5 & 68 & 9 \\ 
          & & & & 69 & 8    \\
          S140 &  22:19:11.53 & 63:17:46.90  & -12.3 & 68 & 34  \\ & & & & 69 & 25   \\
          \hline 
     \end{tabular}

     \label{tab:detection_limits}}
\end{table}          
          
\addtocounter{table}{-1}   
\begin{table}
\small{
     \centering
     \caption{Continued.}

     \begin{tabular}{lllrcc}
          \hline \hline
         Source & \multicolumn{1}{c}{$\alpha_{{\rm J2000}}$}  & \multicolumn{1}{c}{$\delta_{{\rm J2000}}$} &  \multicolumn{1}{c}{$\upsilon_\text{LSR}$}  & Line & rms\tablefootmark{*} \\  
         
         &   &   &  \multicolumn{1}{c}{[kms$^{-1}$]}  & [GHz] & \multicolumn{1}{c}{[mK]}\\ \hline
          W3(OH) & 02:27:04.10 & 61:52:22.00 & -46.0 & 70& 7 \\
          & & & & 68 & 13   \\
          & & & & 69 & 8    \\
          S233 & 05:35:51.19 & 35:44:12.90 & -17.5 &  70& 10\\
          & & & & 68 & 17   \\ 
          & & & & 69 & 15   \\
          S235 & 05:40:53.30 & 35:41:49.00 & -57.0 &70 & 19  \\
          & & & & 68 & 37   \\
          & & & & 69 & 30   \\
          CRL618 & 04:42:53.62 & 36:06:53.30 & -26.0 & 70 & 60 \\
          & & & & 68 & 100   \\ 
          & & & & 69 & 80   \\
          CRL2688 & 21:02:18.27 & 36:41:37.00 & -32.0 & 70 & 12 \\ & & & & 68 & 20   \\
          & & & & 69 & 18   \\
          NGC7023 & 21:01:36.90 & 68:09:48.00 & 2.5 & 70& 12\\
          & & & & 68 & 24   \\
          & & & & 69 & 22   \\
          NGC7027 & 21:07:01.75 & 42:14:10.00 & 8.9 &70 & 6  \\
          & & & & 68 & 11   \\ 
          & & & & 69 & 9    \\
          NGC7538 & 23:13:37.20 & 61:30:00.00 & -64.9 &70 & 19  \\
          & & & & 68 & 38   \\
          & & & & 69 & 28   \\
          Cas A & 23:23:24.90 & 58:50:03.30 & -2600.0 &  70 &  8 \\ 
          & & & & 68 & 15   \\ 
          & & & & 69 & 12   \\
          IC~443 & 06:18:02.70 &  22:39:03.00 & -4.4 &  70 & 10 \\ 
          & & & & 68 & 16   \\
          & & & & 69 & 14   \\  
                     \hline
     \end{tabular}
     \tablefoot{Columns are: (left to right) source name, J2000 coordinates of observed position, assumed LSR velocity, line ID, and $1\sigma$ rms noise level. \tablefoottext{*}{On the $T_{\text{mb}}$ scale the rms noise is quoted for a spectral resolution of 0.97~km~s$^{-1}$.} All the sources are SFRs, except for NGC7027 which is a planetary nebula, CRL618 and CRL2688, which are proto-planetary nebulae and Cas A and IC~443, which are supernova remnants. NGC7023 is a reflection nebula with a PDR.}
     \label{tab:detection_limits2} 
     }
\end{table}

\section{\texorpdfstring{CH$_{2}$}{CH2} 70~GHz HFS stacking}\label{appendix:hfs_stacking}
\begin{figure}
    \centering
    \includegraphics[width=0.4\textwidth]{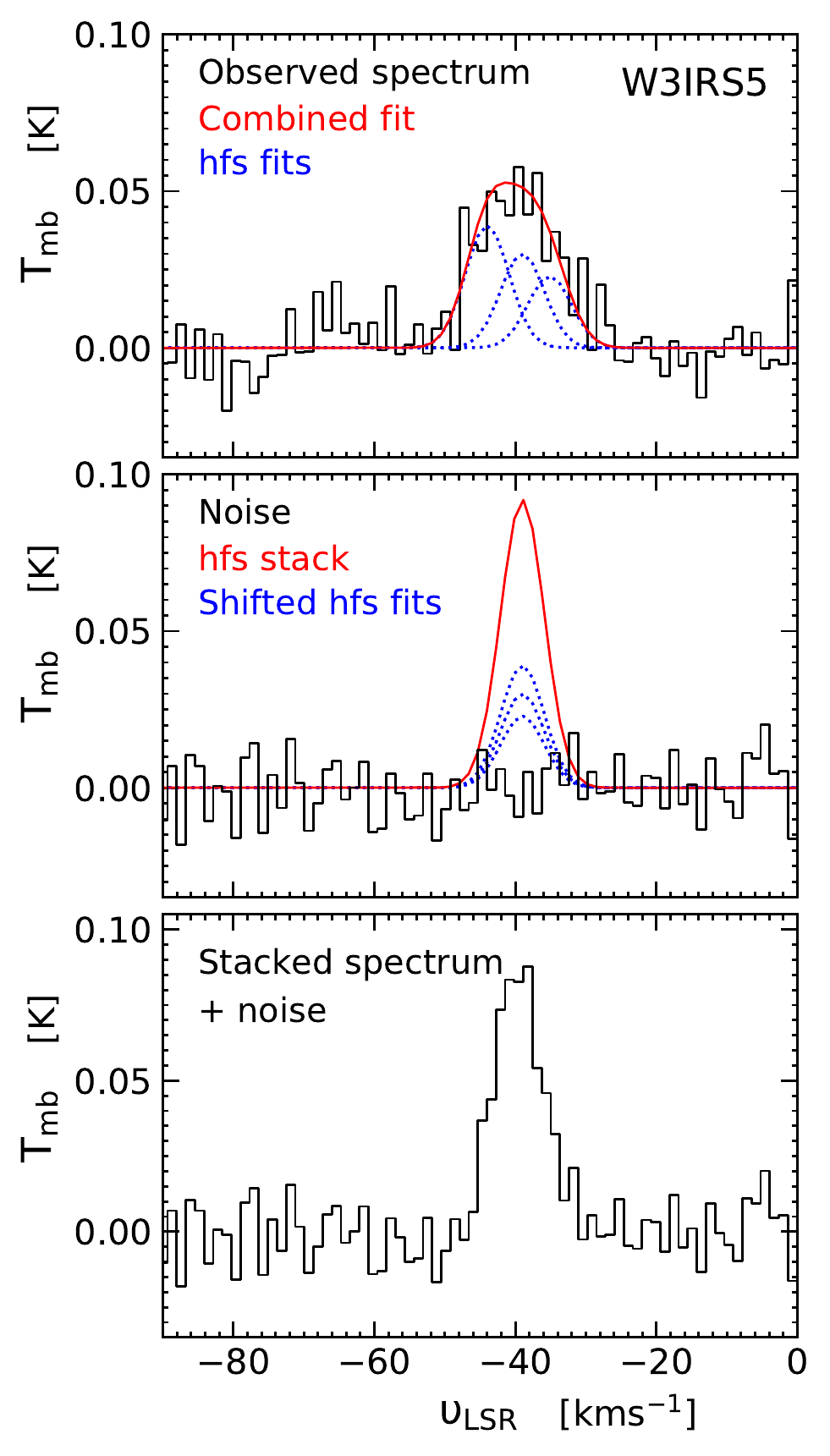}
    \caption{Top: Observed o-CH$_{2}$ 70~GHz spectrum towards W3~IRS5 (black) alongside the HFS decomposed fits (dotted blue) and the combined fit (red). Middle: The HFS stacked profile (red) with the fit to each HFS component shifted to the systemic velocity of the source (dotted blue) and the Gaussian noise profile (black). Bottom: Resultant stacked spectrum with the addition of noise (black). }
    \label{fig:stacking}
\end{figure}

As mentioned in Sec.~\ref{subsec:results_othersources} and \ref{subsec:comparison_with_crrl}, we stack the HFS-decomposed model fits of each of the individual HFS components of the 70~GHz transition of \cht, prior to comparing its line profile with the observed line profiles of the CRRLs. Furthermore, in order to carry out a more objective comparison we incorporated the expected noise of the (stacked and averaged) 70~GHz spectra by including an additive Gaussian noise term. As an example, the different steps that are a part of this exercise are illustrated for the 70~GHz spectrum observed toward W3~IRS5 in Fig.~\ref{fig:stacking}.

\section{Observed recombination lines}\label{appendix:summary_RRL}
In this appendix we present the observed spectra of the different RRLs. As mentioned in Sect.~\ref{sec:observations}, there are two observing setups used for these observations. The first setting observed only the H-, He-, and CRRLs for a principal quantum number of 75, near 15~GHz. This setup was used to carry out observations toward the different Orion positions and W3~IRS5 (displayed in Fig.~\ref{fig:W3IRS5_RRL}). The second setup was used for the remainder of the sources toward which o-CH$_{2}$ was observed, utilising a 5~GHz bandwidth from 12.5-17.5~GHz, which covered RRLs with principal quantum numbers between 80 and 72. Subsequently, for these sources, the RRL profiles for the different quantum numbers are stacked and averaged; based on these profiles, the contributions of the He RRL (modelled using a Gaussian fit) is subtracted to yield the CRRL profile used in the analysis. 
\begin{figure*}
\includegraphics[width=0.31\textwidth]{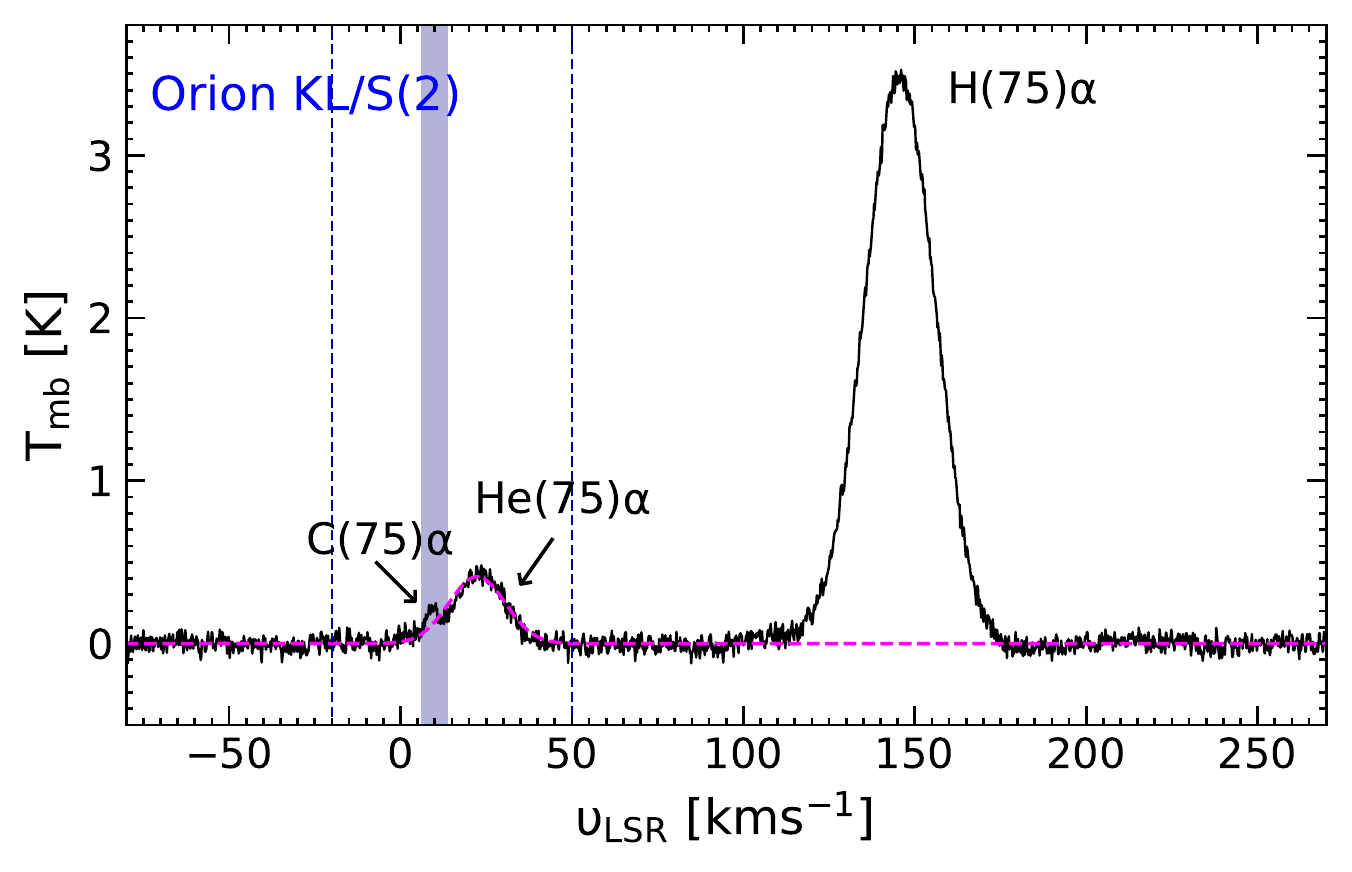}\quad
\includegraphics[width=0.31\textwidth]{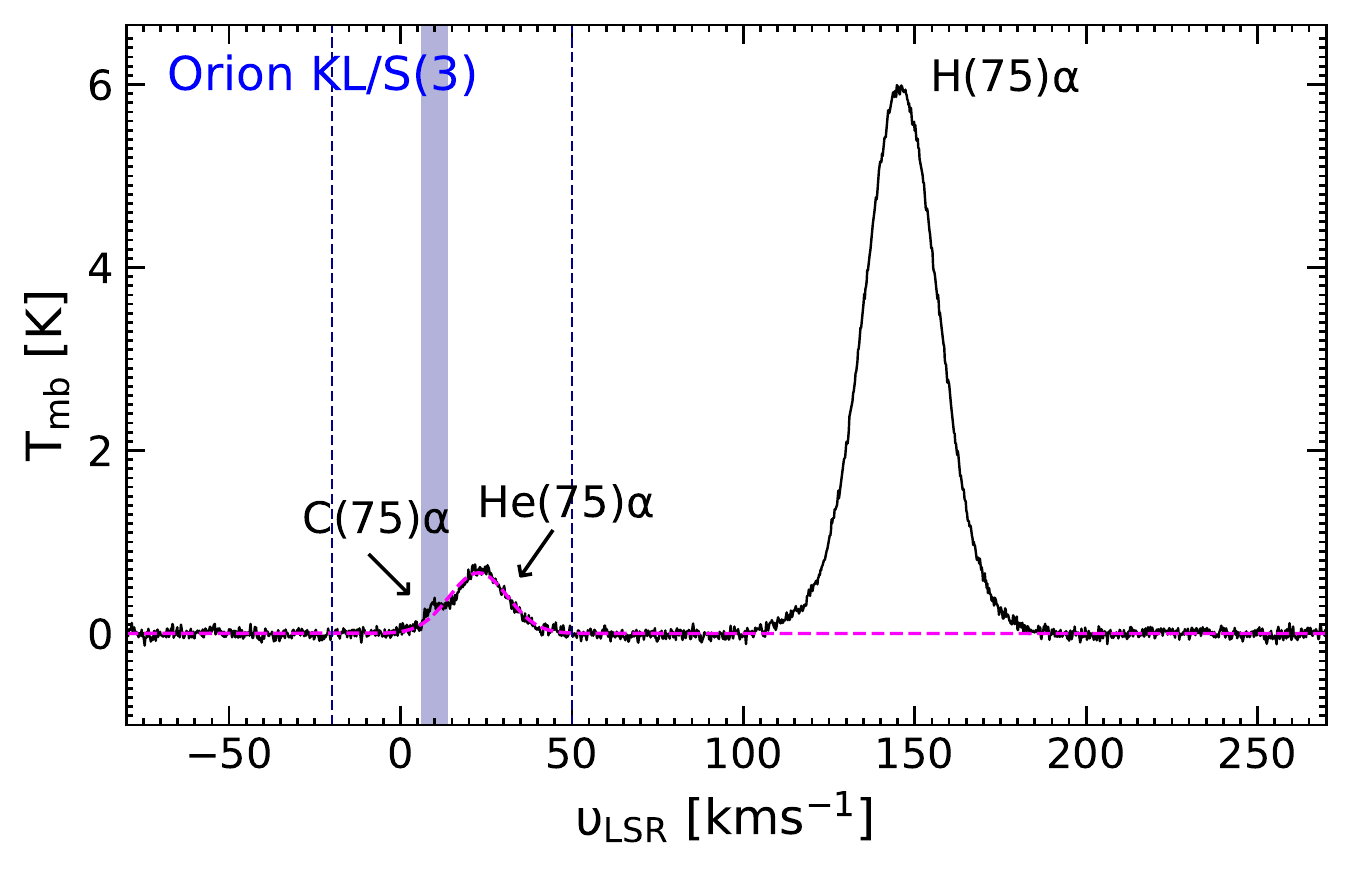}\quad
\includegraphics[width=0.31\textwidth]{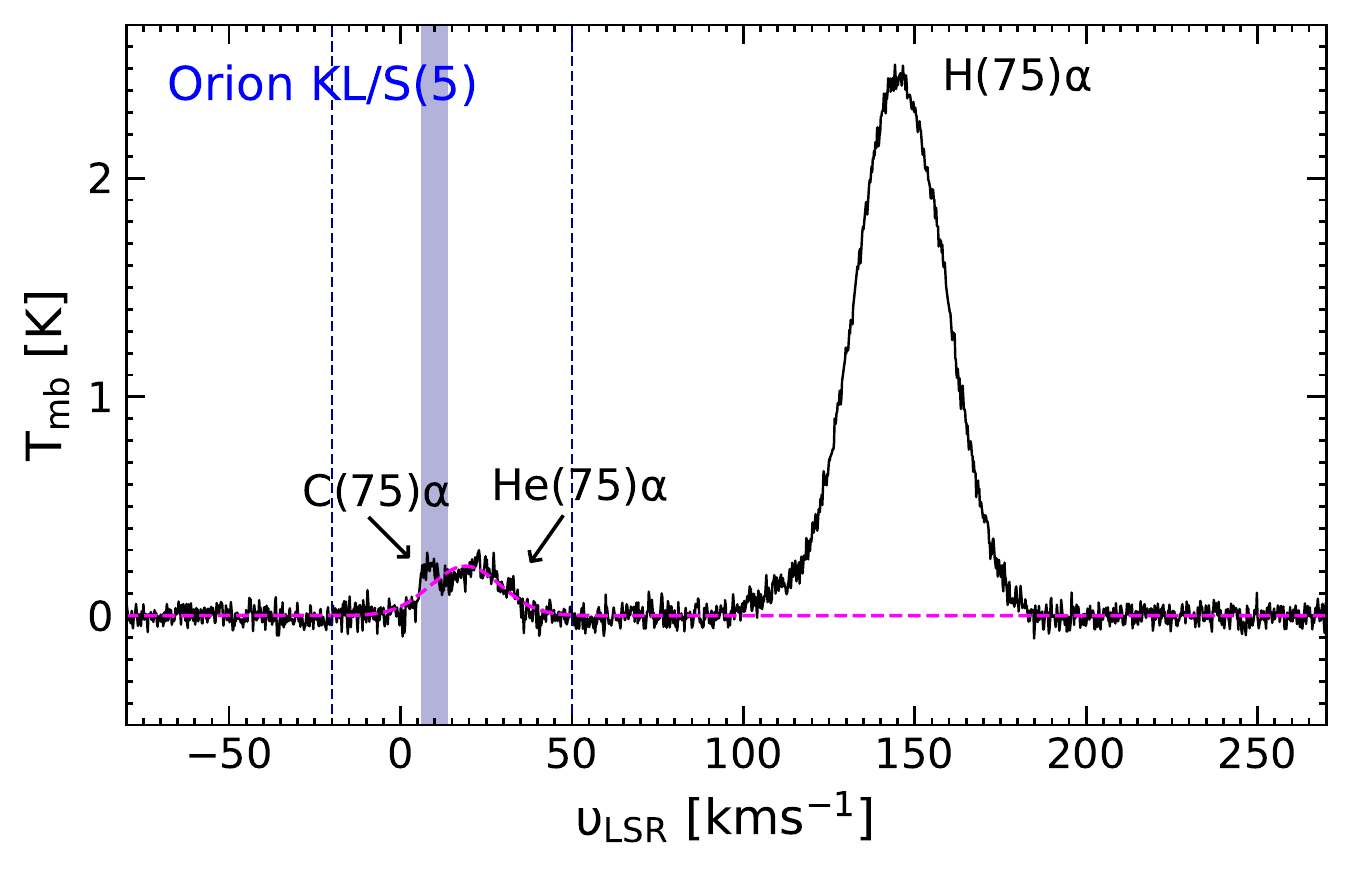}\quad
\includegraphics[width=0.32\textwidth]{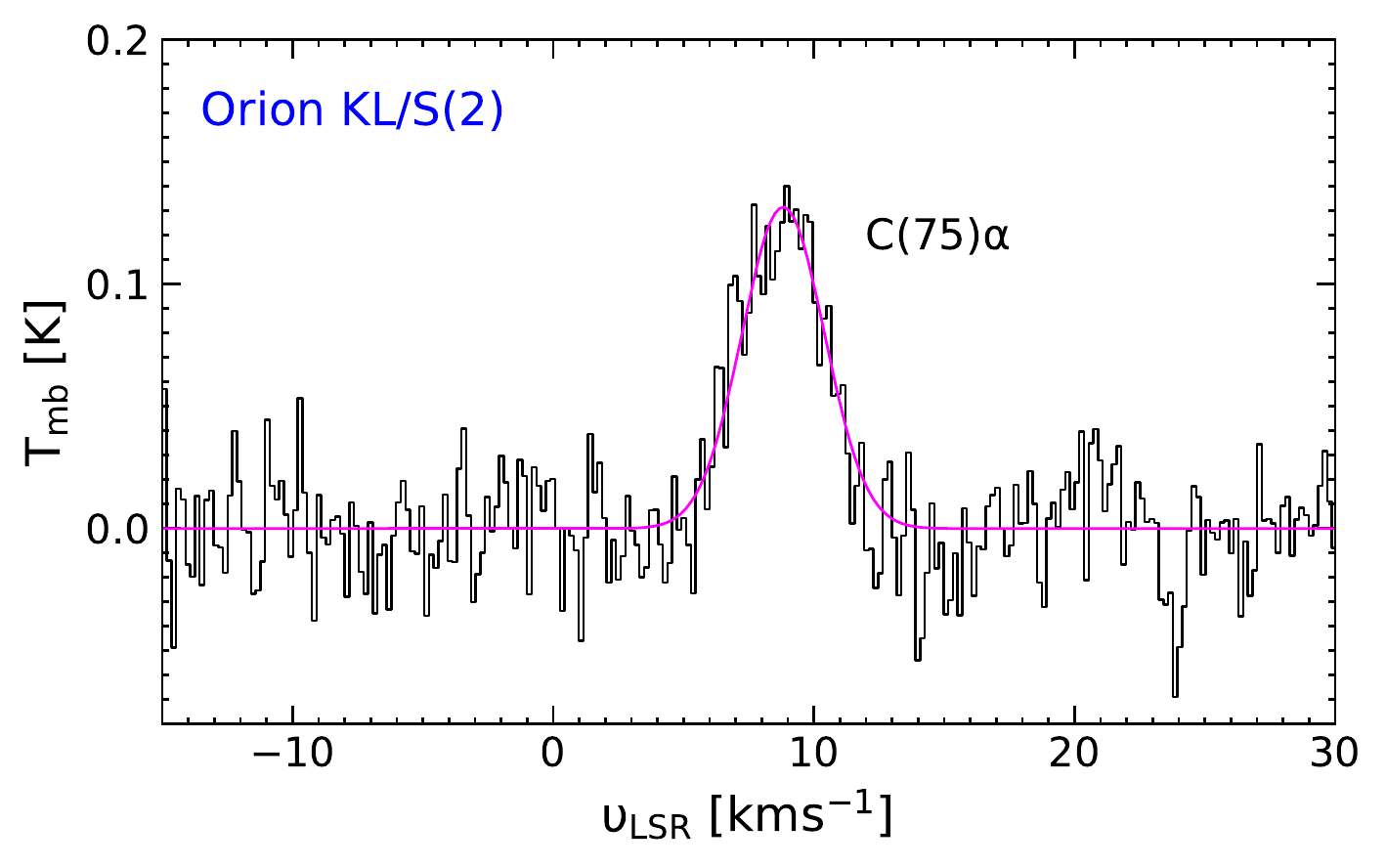}\quad
\includegraphics[width=0.32\textwidth]{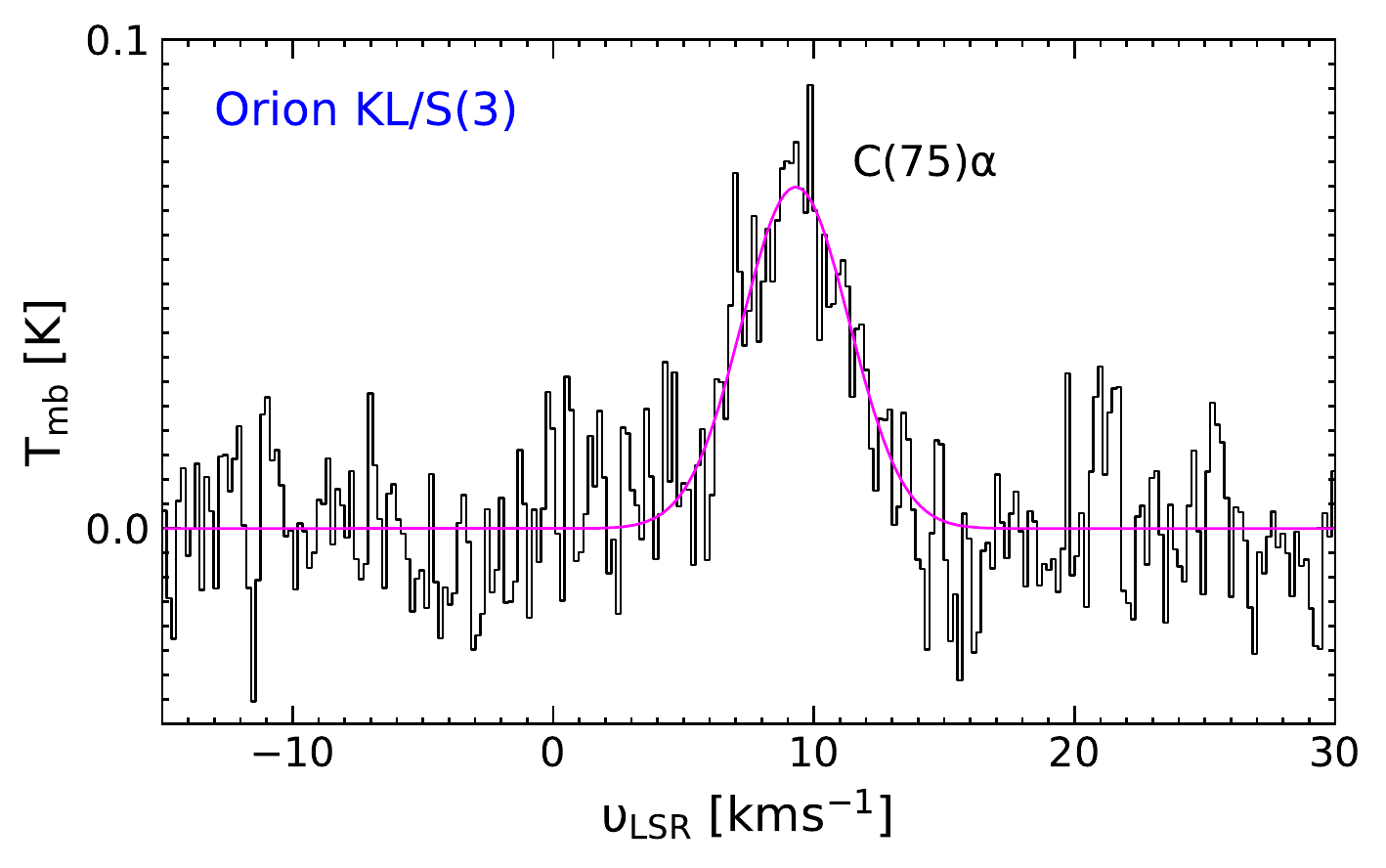}\quad
\includegraphics[width=0.32\textwidth]{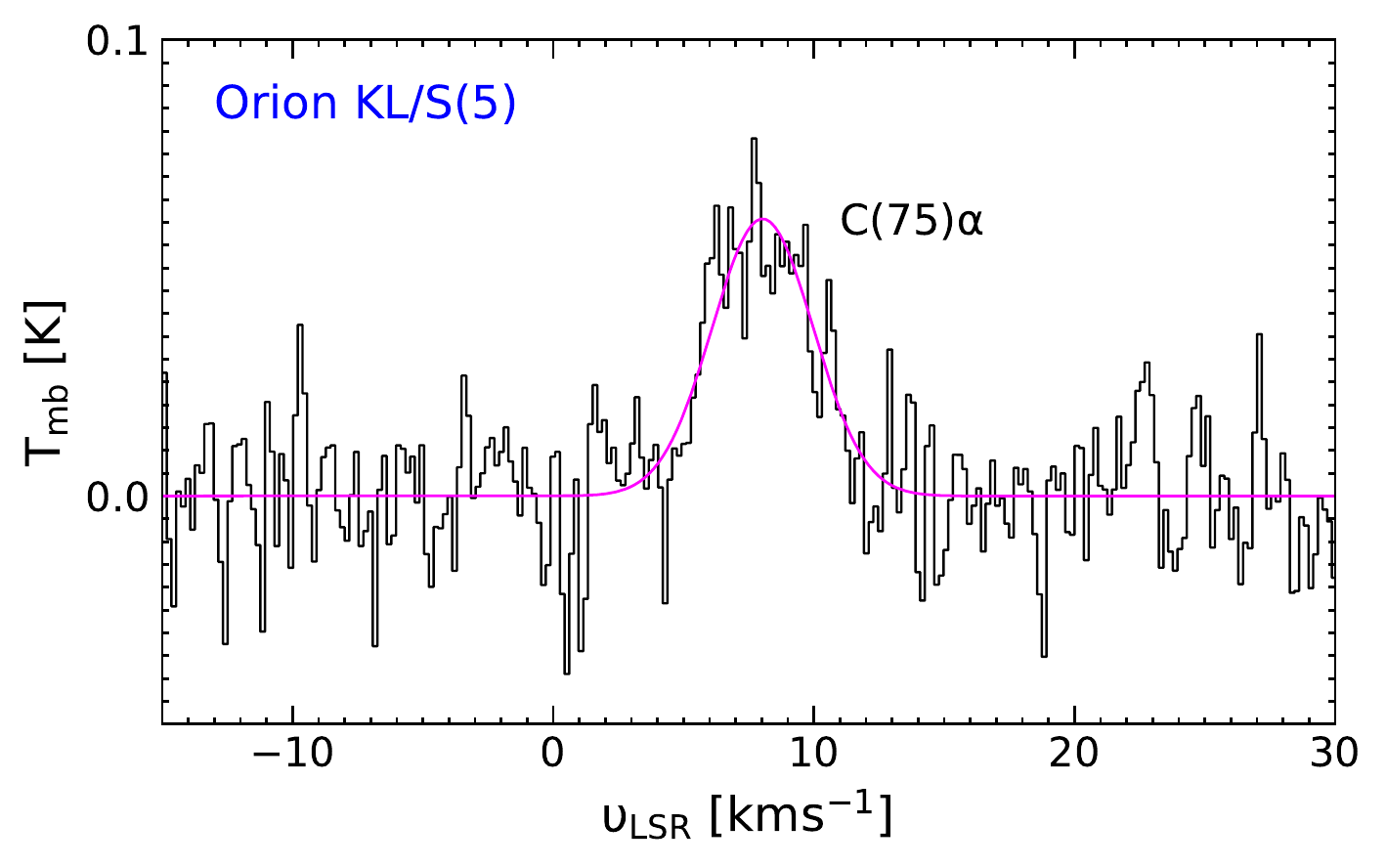}\quad
\caption{Top: H-, He-, and CRRLs toward different Orion KL/S OSO pointing positions, with the Gaussian fit to the He-RRL displayed by the dashed magenta curve. Bottom: Resulting CRRL with the Gaussian fit to the profile displayed in magenta. The velocity scale is given with respect to the CRRL.}
\label{fig:OrionKLS6_H-CRRL}
\end{figure*}

\begin{figure*}
\centering
\includegraphics[width=0.31\textwidth]{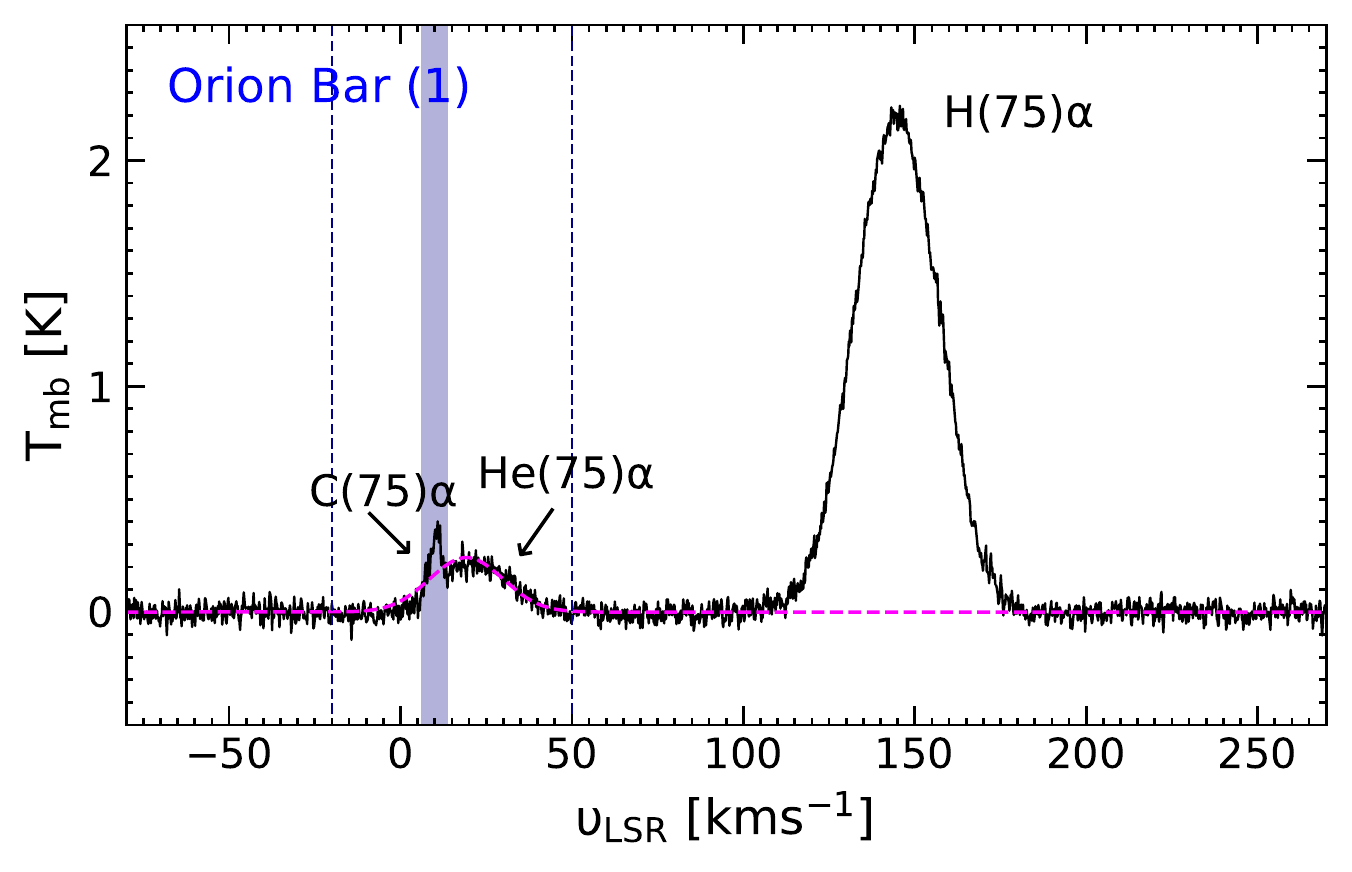}\quad
\includegraphics[width=0.31\textwidth]{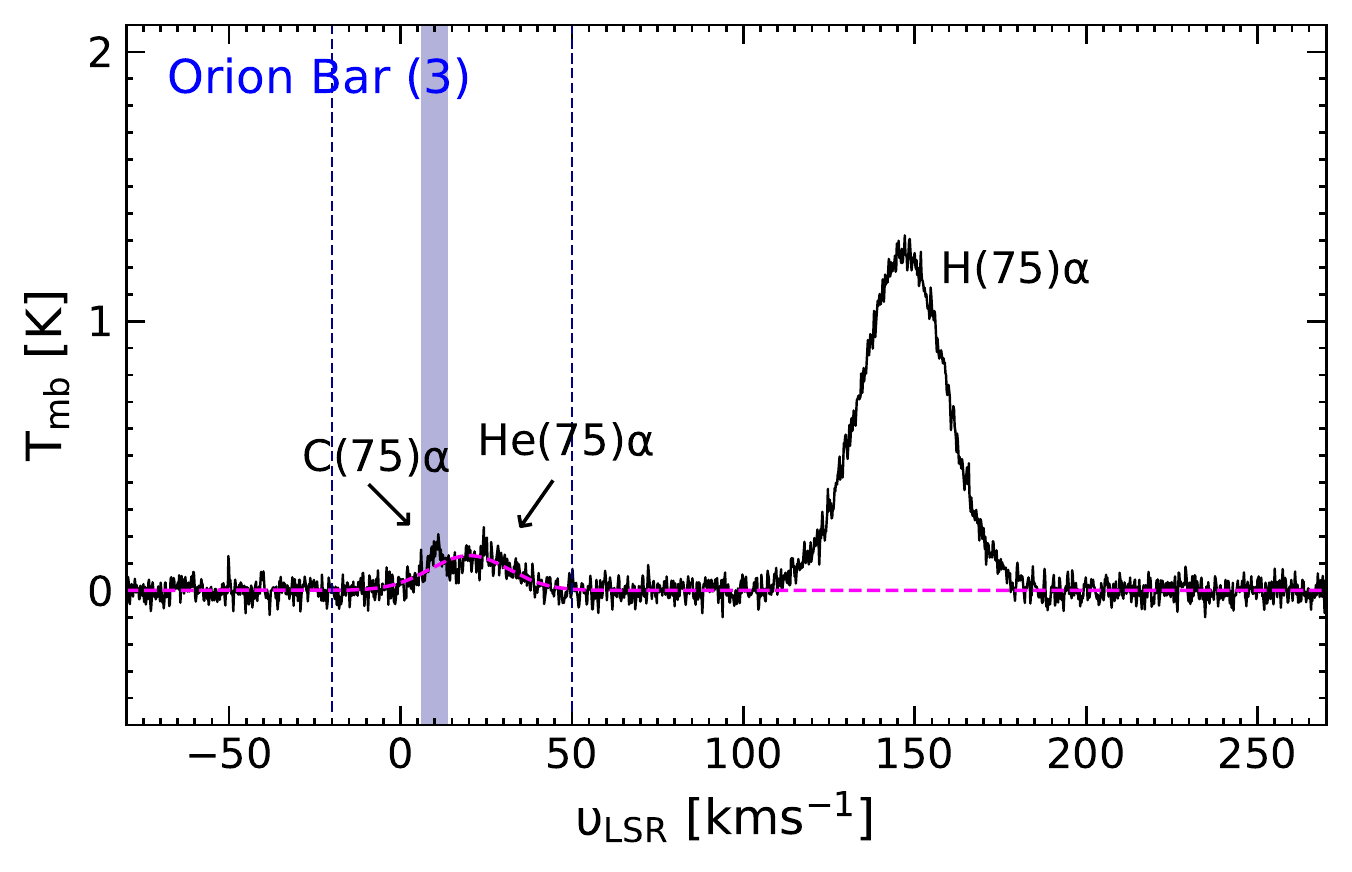}\quad
\includegraphics[width=0.31\textwidth]{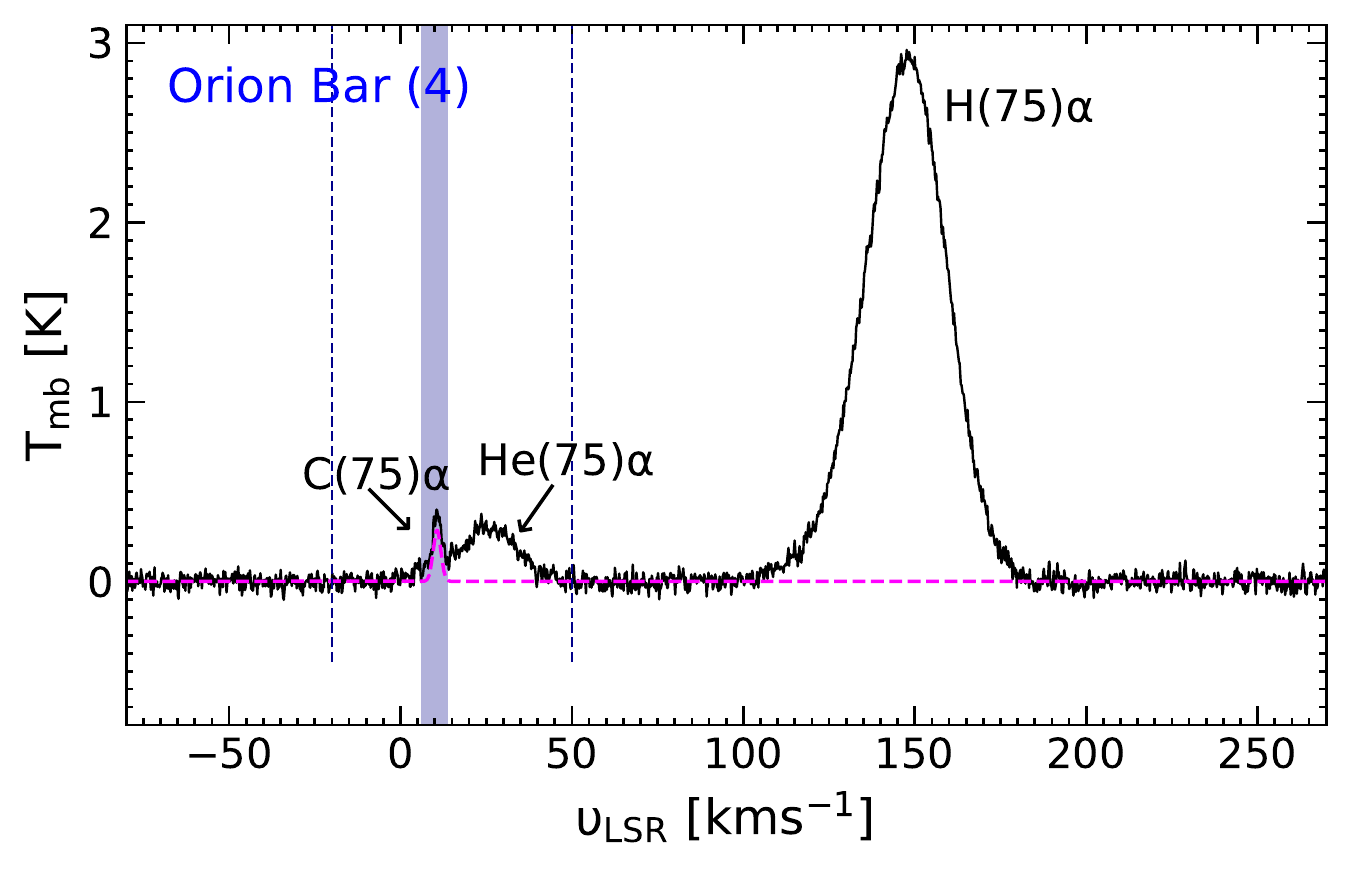}\quad
\includegraphics[width=0.32\textwidth]{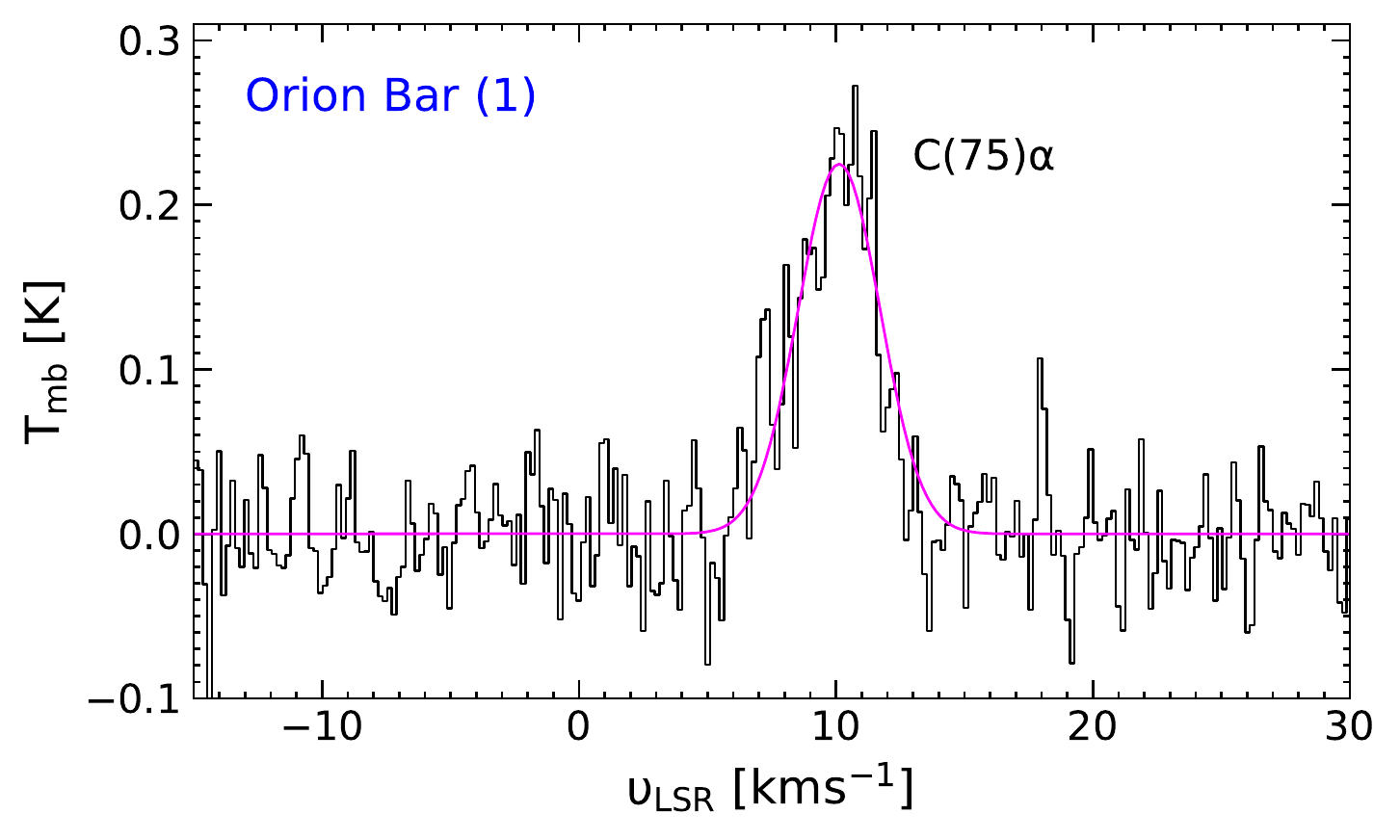}\quad
\includegraphics[width=0.32\textwidth]{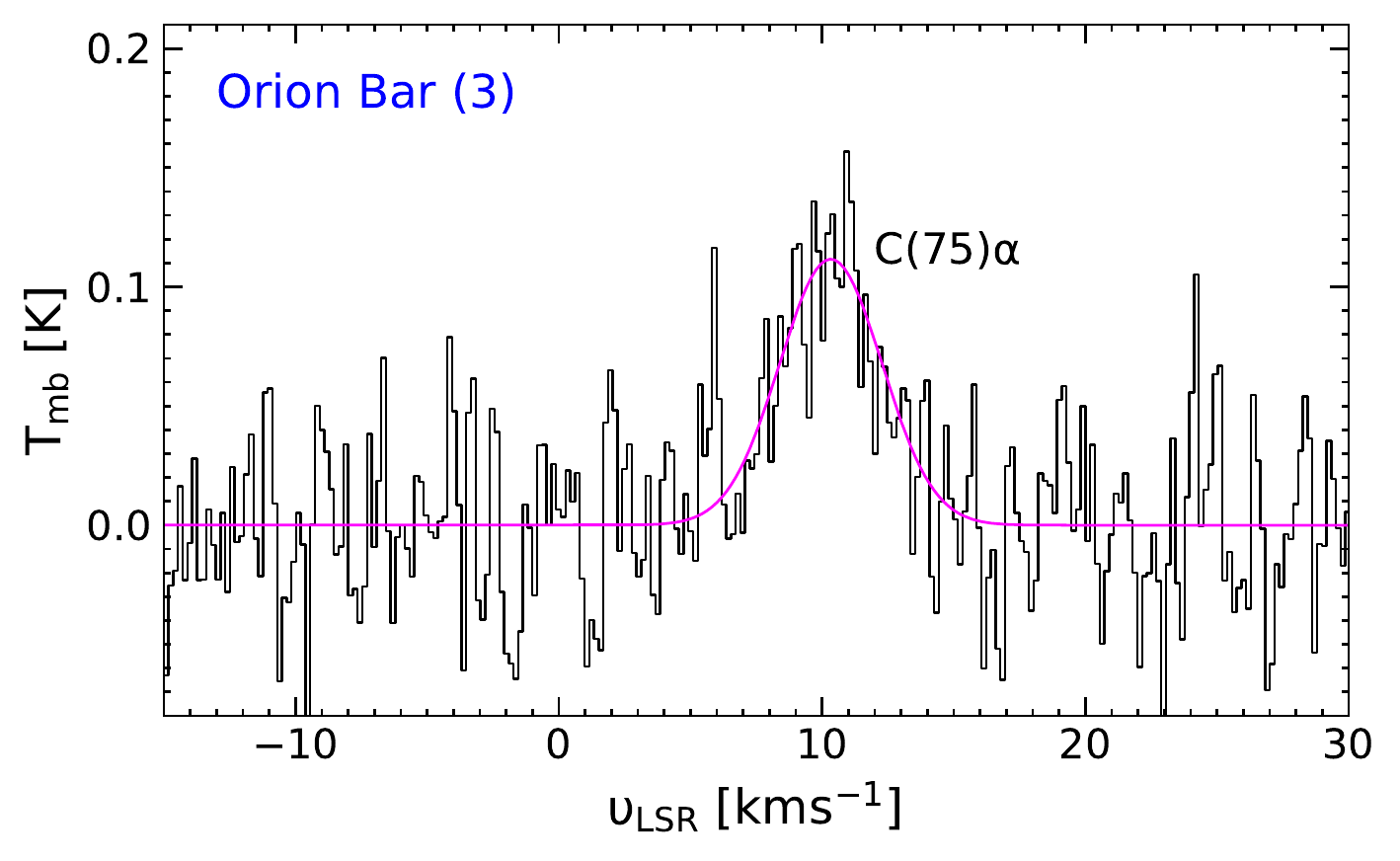}\quad
\includegraphics[width=0.32\textwidth]{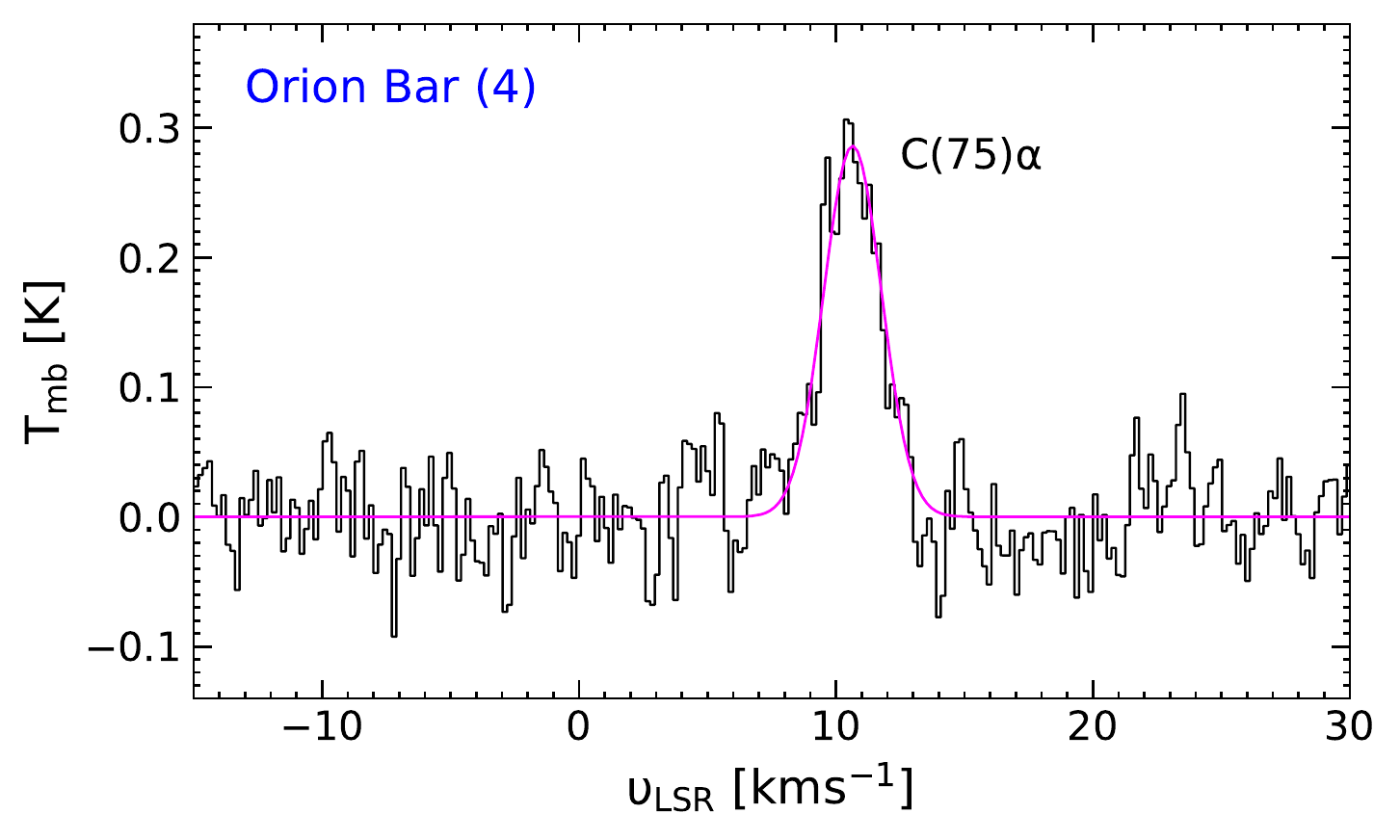}\quad
\caption{Same as Fig.~\ref{fig:OrionKLS6_H-CRRL}, but toward the Orion Bar OSO pointing positions.}
\label{fig:OrionBar_H-CRRL}
\end{figure*}

\begin{figure*}
    \centering
    \includegraphics[width=0.95\textwidth]{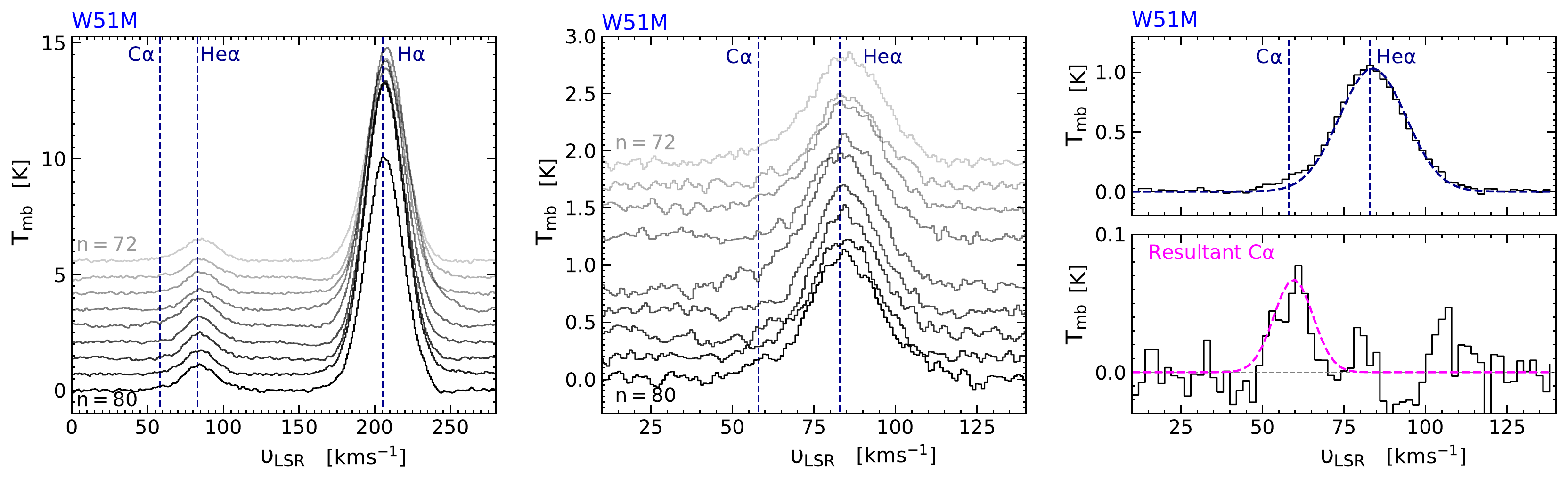}\quad 
    \includegraphics[width=0.95\textwidth]{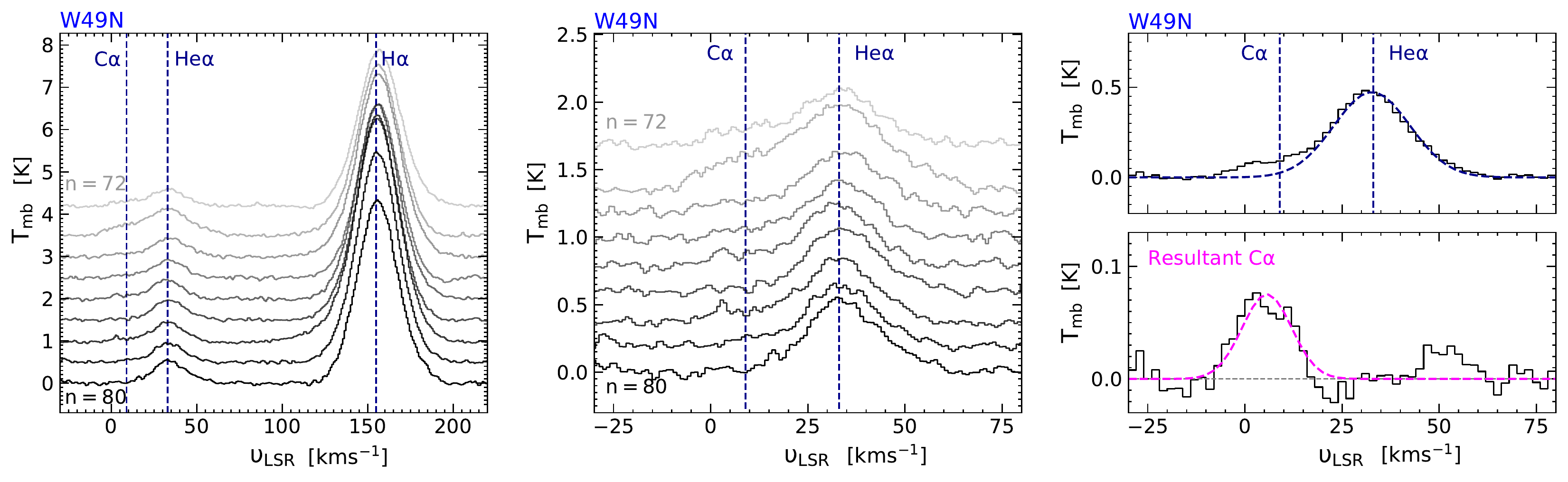}\quad 
    \includegraphics[width=0.95\textwidth]{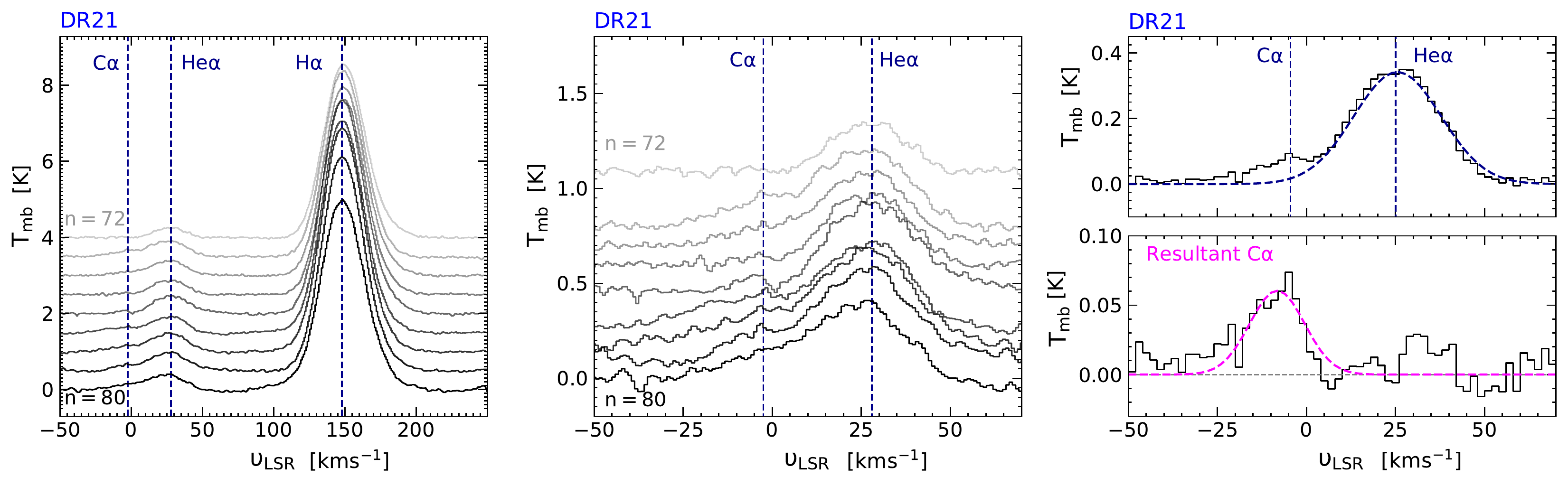}\quad 
    \includegraphics[width=0.95\textwidth]{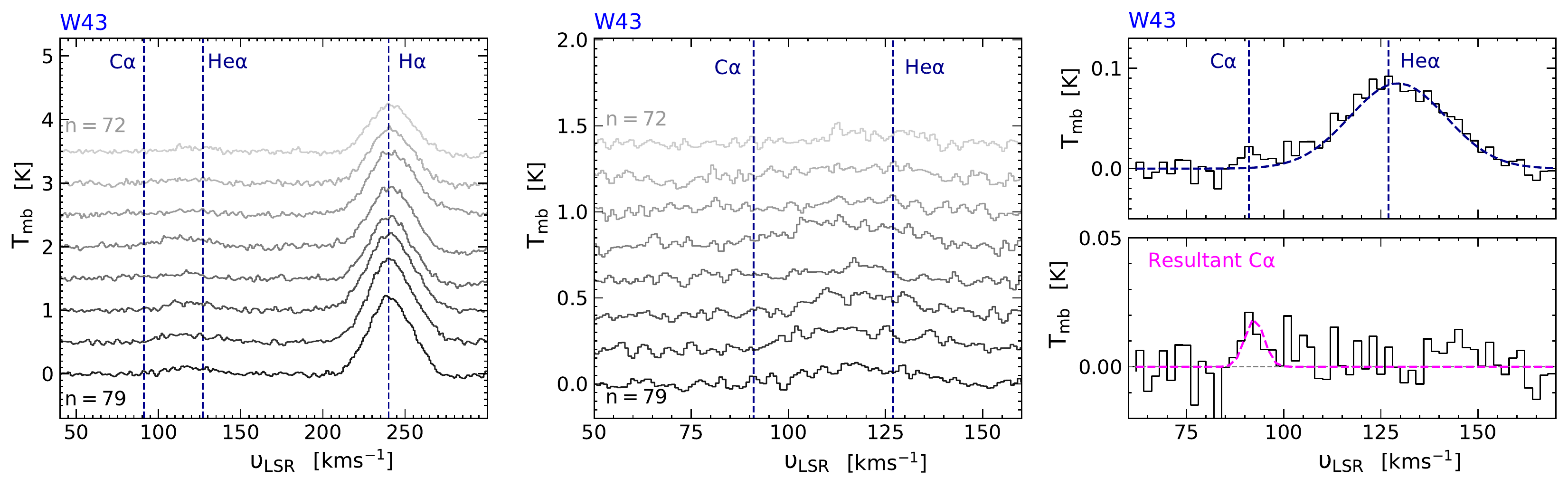}
    \caption{Left: H-, He-, and CRRLs observed with the Effelsberg 100~m telescope for $\alpha$ lines with principal quantum number, $n$, from 72 (top) to 80 (bottom) (except for W43) in steps of $\Delta n = 1$. The positions of the different lines are marked by dashed dark blue lines. The spectra are uniformly offset by 0.5~K on the $T_{\text{mb}}$. Centre: Zoomed in view of the He-, and CRRL. The offsets are not uniform but varied for the ease of viewing with a typical offset of 0.2~K on the $T_{\text{mb}}$ scale. Right: Average He-, and CRRL profile (top) with the Gaussian fit to the He RRL presented by the dashed dark blue curve. The resulting spectrum of the CRRL is displayed below with its fit given by the dashed magenta Gaussian. For all spectra, the velocity scale is appropriate for the CRRL. The different rows (from top to bottom) present these results for W51~M, W49~N, DR21, and W43, respectively.  }
    \label{fig:other_crrls}
\end{figure*}

\begin{figure*}
    \centering
    \includegraphics[width=0.4\textwidth]{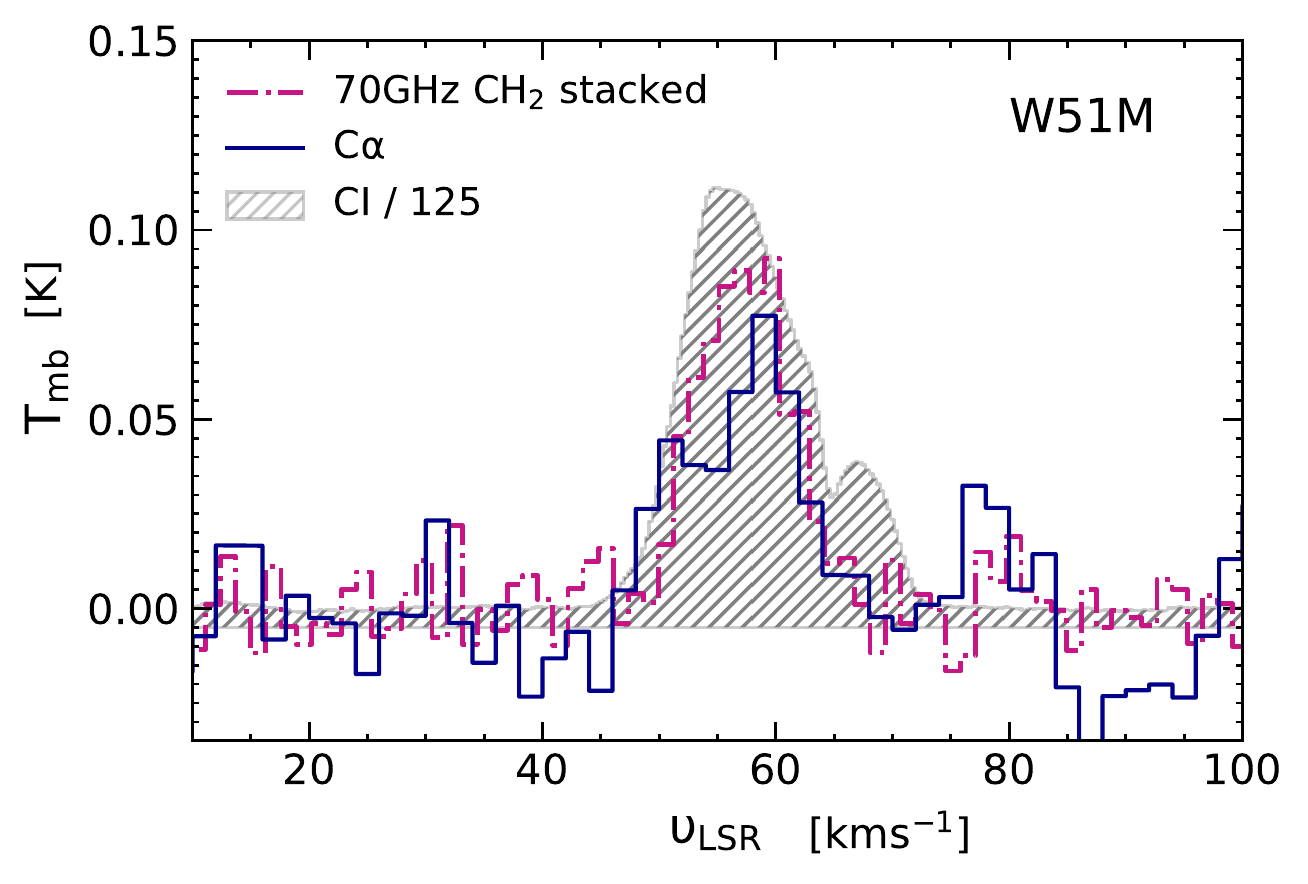}\quad
    \includegraphics[width=0.4\textwidth]{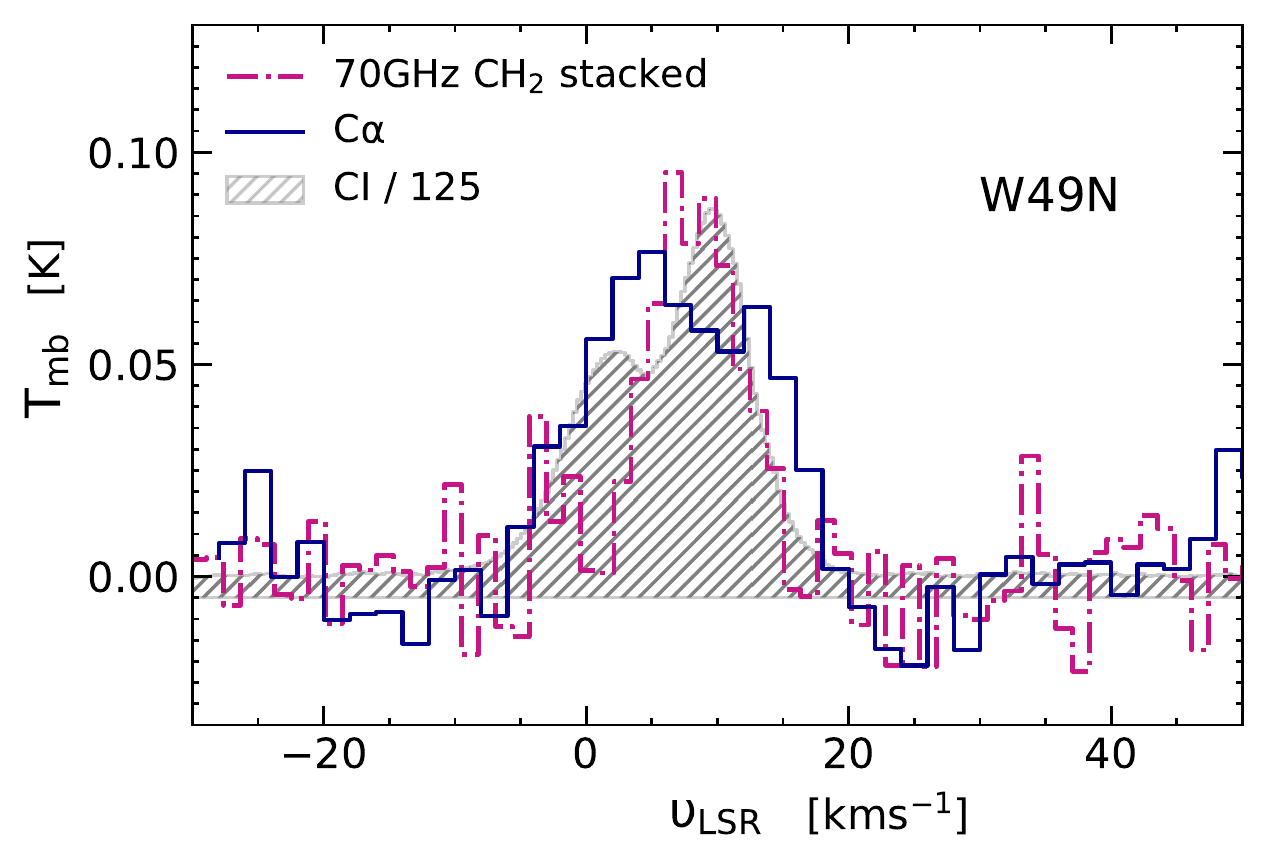}\quad
    \includegraphics[width=0.4\textwidth]{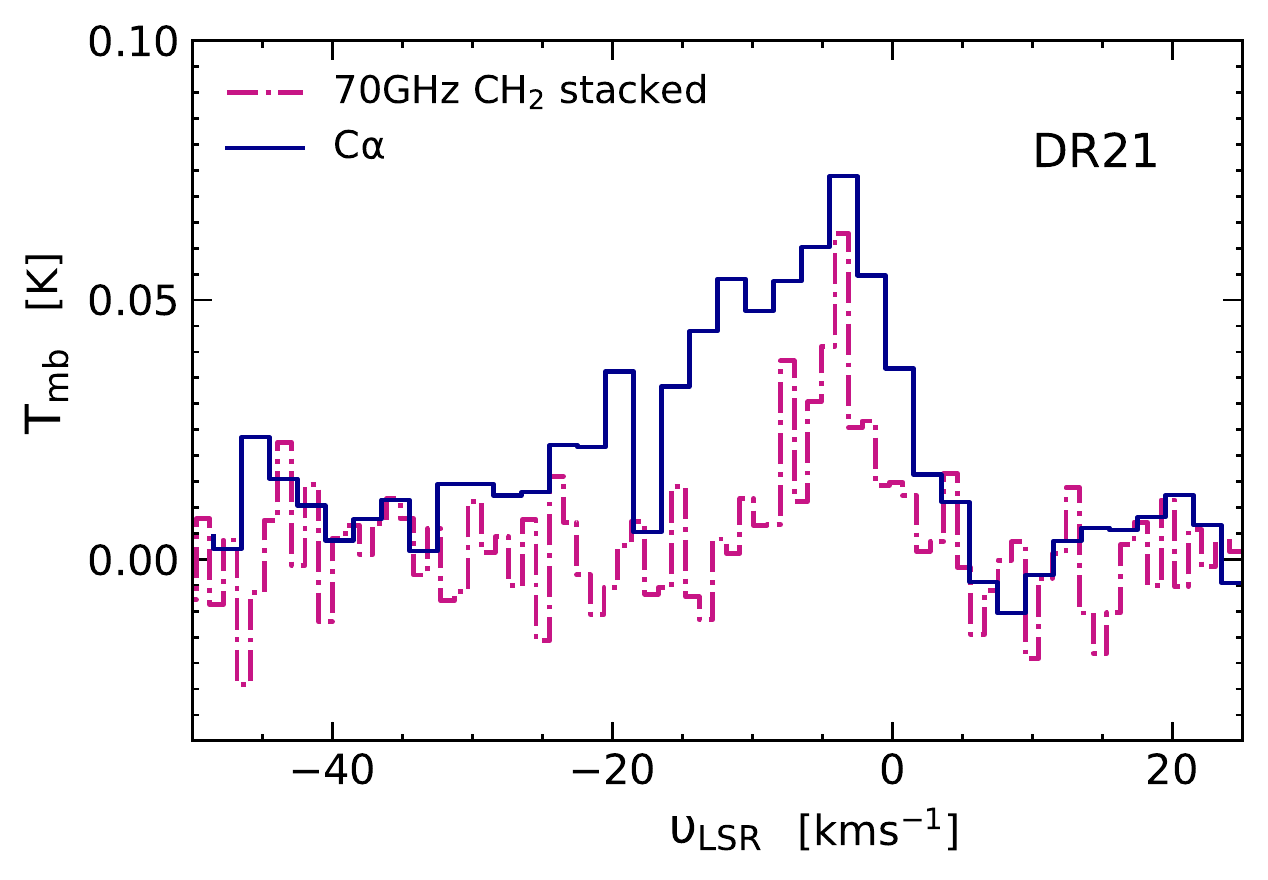}\quad
    \includegraphics[width=0.4\textwidth]{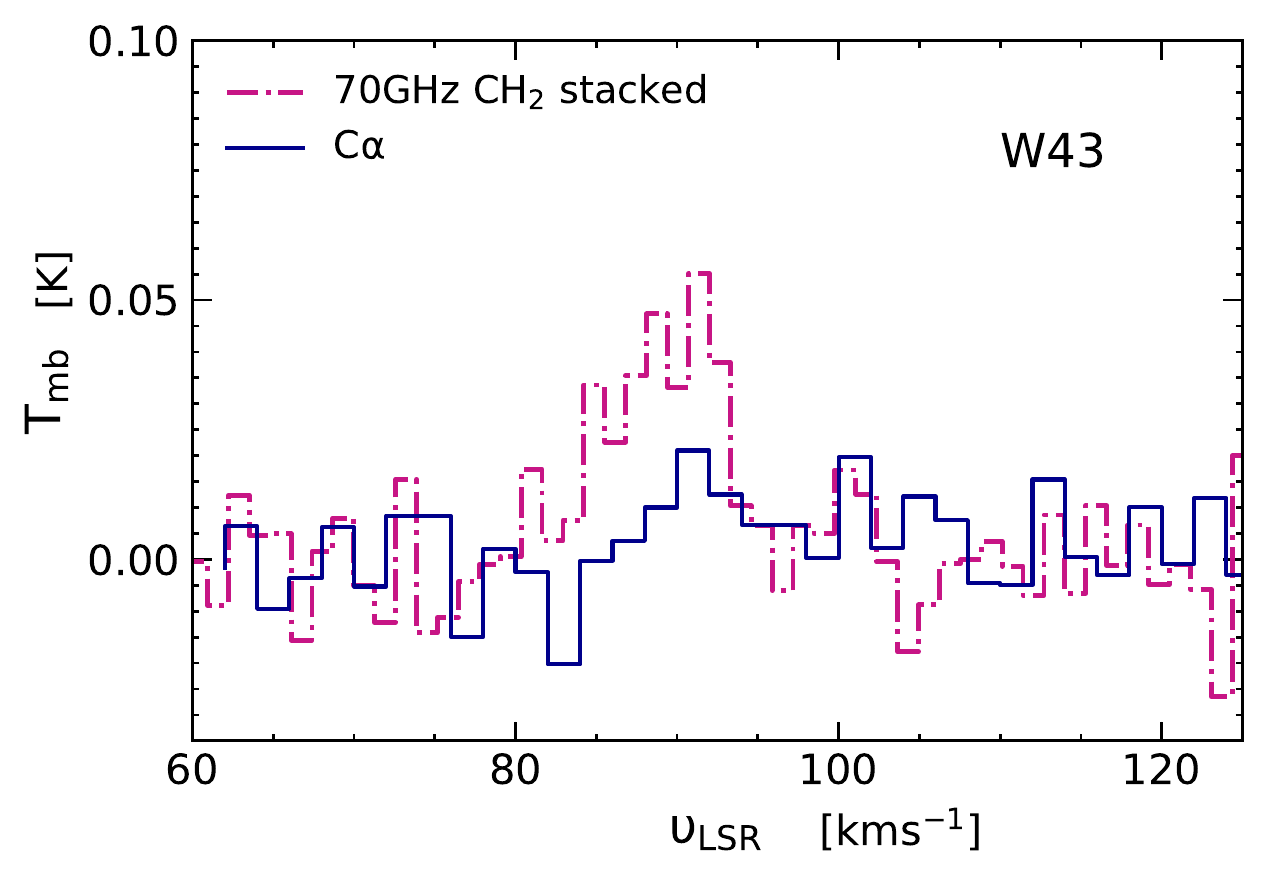}
    \caption{Decomposed CRRL profile in blue alongside the 70~GHz HFS stacked CH$_{2}$ line profile displayed by the dashed-dotted violet curve toward (clock-wise from top) W51~M, W49~N, DR21, and W43, respectively. The stacked profile of the 70~GHz CH$_{2}$ transition was obtained from the HFS decomposition model. The hatched grey regions in the W51~M and W49~N spectra display the line profiles of the $^{3}P_{1}-{}^{3}P_{0}$ transition of C{\tiny I} at 492.160~GHz scaled down by a factor of 125 on the $T_{\text{mb}}$ scale.}
    \label{fig:RRL_comparison_plots}
\end{figure*}
\begin{table}
\centering 
\caption{Best-fit Gaussian parameters for the CRRLs.}
    \begin{tabular}{lrrr}
    \hline \hline
    \multicolumn{1}{c}{Source} & \multicolumn{1}{c}{$\upsilon$} & \multicolumn{1}{c}{$\Delta \upsilon$} & \multicolumn{1}{c}{$T_{\text{mb}}$} \\
    & \multicolumn{1}{c}{[km~s$^{-1}$]} & \multicolumn{1}{c}{[km~s$^{-1}$]} & \multicolumn{1}{c}{[mK]} \\
    \hline 
         Orion KL/S (2) & 8.8(0.1) & 3.7(0.2) & 140.0(5.4) \\
         Orion KL/S (3) & 9.3(0.1) & 4.8(0.3) & 74.4(3.3)\\
         Orion KL/S (5) & 8.1(0.1) & 4.6(0.3) & 64.7(3.3)\\
         Orion Bar (1) & 10.1(0.1) & 3.7(0.2) & 239.4(8.5)\\
         Orion Bar (3) & 10.3(0.2) & 4.6(0.5) & 118.8(9.1)\\
         Orion Bar (4) & 10.6(0.1) & 2.6(0.1) & 304.6(11.1)\\ 
         W3~IRS5 & -38.4(0.1) & 6.8(0.4) & 131.4(6.6)\\
         W51~M & 59.8(1.0) & 6.1(1.8) & 69.0(3.7)\\
         W49N & 5.6(0.6) & 6.7(1.5) & 79.0(3.7)\\
         DR21 & -8.2(2.6) & 7.7(1.5)& 60.5(6.1) \\ 
         W43 & 92.4(0.6) & 2.4(1.0) & 18.6(3.6) \\
         \hline
    \end{tabular}
    
    \label{tab:CRRL_params}
\end{table}
\end{appendix}

\end{document}